\documentclass[fleqn,usenatbib]{mnras}

\usepackage{newtxtext,newtxmath}

\usepackage[T1]{fontenc}
\usepackage{ae,aecompl}

\usepackage{graphicx}	
\usepackage{amsmath}	
\usepackage{amssymb}	

\title[IFU observations of NGC 1566]{NGC 1566: analysis of the nuclear region from optical and near-infrared Integral Field Unit spectroscopy}

\author[Patr\'icia da Silva et al.]{
Patr\'icia da Silva\thanks{Contact e-mail: \href{mailto:p.silva2201@gmail.com}{p.silva2201@gmail.com}}, 
J. E. Steiner\thanks{\href{mailto:joao.steiner@iag.usp.br}{joao.steiner@iag.usp.br}},
R. B. Menezes\thanks{\href{mailto:robertobm@astro.iag.usp.br}{robertobm@astro.iag.usp.br}}
\\
Instituto de Astronomia, Geof\'isica e Ci\^encias Atmosf\'ericas, Departamento de Astronomia, Universidade de S\~ao Paulo, 05508-090, SP, Brazil\\
}

\date{}

\pubyear{2017}

\begin{document}
\label{firstpage}
\pagerange{\pageref{firstpage}--\pageref{lastpage}}
\maketitle

\begin{abstract}
We analysed the centre of NGC 1566, which hosts a well-studied active galactic nucleus (AGN), known for its variability. With the aid of techniques such as Principal Component Analysis Tomography, analysis of the emission-line spectra, channel maps, Penalized Pixel Fitting and spectral synthesis applied to the optical and near-infrared data cubes, besides the analysis of \textit{Hubble Space Telescope} images, we found that: (1) the AGN has a Seyfert 1 emission, with a very strong featureless continuum that we described as a power law with spectral index of 1.7. However, this emission may come not only from the AGN [as its point spread function (PSF) is broader than the PSF of the broad-line region (BLR)], but from hot and young stars, the same ones that probably account for the observed $\sigma$-drop. (2) There is a correlation between redshift and the full width at half-maximum of the BLR emission lines. With a simple model assuming gravitational redshift, we described it as an emitting ring with varying emitting radii and small inclination angles. (3) There is an H \textsc{ii} region close to the AGN, which is composed of many substructures forming an apparent spiral with a velocity gradient. (4) We also detected a probable outflow coming from the AGN and it seems to contaminate the H \textsc{ii} region emission. (5) We identified an H$_2$ rotating disc with orientation approximately perpendicular to this outflow. This suggests that the rotating disc is an extension of an inner torus/disc structure, which collimates the outflow emission, according to the Unified Model.
\end{abstract}

\begin{keywords}
galaxies: active -- galaxies: individual: NGC 1566 -- galaxies: kinematics and dynamics -- galaxies: nuclei -- galaxies: Seyfert.
\end{keywords}



\section{Introduction}

Active galactic nuclei (AGNs) are characterized by their spectral emission, which cannot be attributed only to stars. The study of these objects is very important, as they can provide relevant information about the origin and evolution of galaxies. NGC 1566 is a galaxy that has a very well-studied active nucleus, notorious for its variability. It is a nearly face-on grand design spiral galaxy with morphological type SAB(s)bc (like the Milky Way). It is the brightest galaxy of the Dorado group and it presents a small bar of 1.7 kpc (32$\arcsec\!\!$.5) of length \citep{dev73,hackwell,comeron}. Observed in the near-infrared (NIR), the orientation of this bar is north-south. The galaxy's distance is about 10.8 Mpc (NASA Extralagactic Database - NED). In the radio (8.6 GHz), NGC 1566 shows a compact nuclear emission and a 'blob' at 3$\arcsec$ north of the centre \citep{morg99}. This galaxy also has an outer stellar formation ring, 10 kpc from the nucleus, and two other similar rings, between 1 and 3 kpc from the nucleus \citep{ag04}. The spiral arms have a strong star formation, with dust features in their inner lanes \citep{gw74}; they have many H \textsc{ii} regions, whose luminosity function, based on H$\alpha$ relative flux scale, is well described by a power law \citep{comte82}.

The NGC 1566 nucleus has a Seyfert emission \citep{devdev61,shobb} that was later classified as type 1, based on optical spectra observations, that showed that the H$\alpha$ and H$\beta$ profiles have broad components with strong asymmetry towards the red \citep{osmer74}. In the literature one can find that  4.2$\lesssim$ H$\alpha$/H$\beta$ $\lesssim$4.7 \citep{osmer74,hp,martin74}, but broad-line region (BLR) photoionization models, assuming that it is composed by a high temperature region ($T_e$= 15000K) and a low temperature region ($T_e$= 10000K), resulted in an H$\alpha$/H$\beta$ ratio of about 3.2 \citep{joly}.

In the NIR, the nucleus presents Br$\gamma$ emission with a broad and a narrow component, besides evidence of a dust torus with temperature of 1000K \citep{smajic}, which was determined by \textit{K}-band continuum fit, from a combination of stellar spectra, a power law and a blackbody curve. In the mid-infrared, both continuum and spectral structures between 10 and 18 $\mu$m (from the emission of silicates) are consistent with the thermal emission from a dust torus with clumpy morphology \citep{Tomp09}.
 
NGC 1566's nuclear spectrum in X-rays (0.5-195 keV) is well reproduced by models that consist of a sum of partially or completely transmitted nuclear emission, its reflection on the accretion disc and the reprocessed emission from the dust torus, followed by a strong emission of Fe K$\alpha$ \citep{kawa}. \citet{ehle} verified that the nucleus' luminosity, in the spectral region between 0.1 and 2.4 keV, is $L_x$ = $10^{41}$ erg s$^{-1}$, the nuclear spectrum being well fitted by a power law with spectral index of 2.3.

This galaxy is known by the variability of its nuclear activity. Variations were observed both in the \textit{B}-band nuclear apparent magnitude - from 13.5 to 14.6 mag \citep{dev73} - and in the nuclear spectrum, in which an H$\beta$ intensity decrease relative to [O \textsc{iii}] was observed, in a time-scale of years \citep{pg70}. This variability was also observed in H$\alpha$ and H$\beta$ profiles and in the non-stellar continuum of the nuclear emission, which varied from Seyfert 1.9 to 1.2 in four months. The explanation for this rapid increase of the Balmer lines and of the broad component of H$\alpha$ (7000 km s$^{-1}$) is a recurrent outburst in the nuclear region \citep{alloin}.

The variability of the nuclear spectrum was also observed from X-rays to IR, the IR variation having a delay of some months to a year relative to the optical and ultraviolet emission \citep{clavel}. 

\citet{ag04} detected, based on deviations from the Satoh model fitted to the kinematic optical data of ionized gas, an apparent outflow located at the edges of the galaxy's bar. Besides that, other non-circular motions were detected in some regions (10$\arcsec$ from the nucleus). Such motions probably represent an inflow of gas to the nucleus, which could be a feeding mechanism of the AGN. \citet{davies} analysed optical integral field observations centred on the nucleus of NGC 1566 and verified that, in ther inner regions (where the ionization parameter is log $U$ $\sim$ 0), the environment is radiation pressure dominated, which can lead to outflows that can affect areas far beyond the radius of influence of the black hole. 

\citet{hp} found evidences of a decrease of the [O \textsc{iii}]/H$\beta$ ratio and of an increase of the [N \textsc{ii}]/[O \textsc{ii}] ratio towards the nucleus. This can be due to an increase of oxygen abundance, which cools the gas, leading to a lower electronic temperature and thus to a decrease of the [O \textsc{iii}] flux.

\citet{beckman} analysed photometric (\textit{V}, \textit{R} and \textit{I} filters) and optical spectral data and verified that the nucleus and the spiral arms of NGC 1566 have bluer colours than the other parts of the galaxy. The bluer colour of the nucleus is probably due to the Seyfert nuclear activity. The \textit{M/L} ratio values, obtained up to the distance of 13.5 kpc from the nucleus, can be explained by taking into account only the main-sequence stellar populations, without the need  to assume the presence of giant stars or dark matter. Based on the observed values of \textit{V-R} and \textit{V-I}, it was possible to infer the presence of O-, B- and A-type stars in the nucleus.

\textit{K}-band data in the NIR, observed with the Spectrograph for Integral Field Observations in the Near Infrared (SINFONI) and with Naos-Conica, both instruments of the Very Large Telescope, showed a circular emission of H$_2$ molecular gas somewhat off-centred along the east-west direction and with  a radius of less than 1$\arcsec$ \citep{mezcua}. Additional SINFONI data in the \textit{K}-band, analysed by \citet{smajic}, showed that the NGC 1566 circumnuclear region is composed of molecular gas, dust and old stellar populations. The molecular gas and the stellar populations are rotating around the nucleus, although there is a disturbance in the molecular gas velocity field towards the nucleus. A star-forming region was observed southwest from the nucleus, with [SFR]= 2.6 $\times 10^{-3}$ M$_{\sun}$ yr$^{-1}$. Using both SINFONI and Atacama Large Millimeter Array (ALMA) 350 GHz data, \citet{smajic} verified that there is a cold and warm molecular gas spiral that may indicate feeding of the AGN. A molecular gas spiral towards the nucleus was also observed by  \citet{comb14}, using ALMA data with spatial resolution of 25 pc. That spiral is very similar to the dust spiral observed in the \textit{Hubble Space Telescope}'s (\textit{HST}) extinction images. \citet{comb14} proposed that the black hole must be influencing the gas dynamics, in order to reverse gravitation torques. These torques are directing the gas towards the nucleus, which may result in the feeding of the AGN. The authors also observed a nuclear disc of dense molecular gas, with deficiency of H \textsc{ii} regions and atomic gas. This molecular gas displays a well-behaved rotation, without the presence of outflows, feedback or feeding of the AGN. 

In X-rays, an extended emission was discovered around the nucleus, certainly associated with the nuclear activity \citep{elvis}. According to \citet{ehle}, this emission is similar to the extended radio emission, suggesting a link between the hot gas and the magnetic field.

\begin{figure*}
\begin{center}
  \includegraphics[scale=0.39]{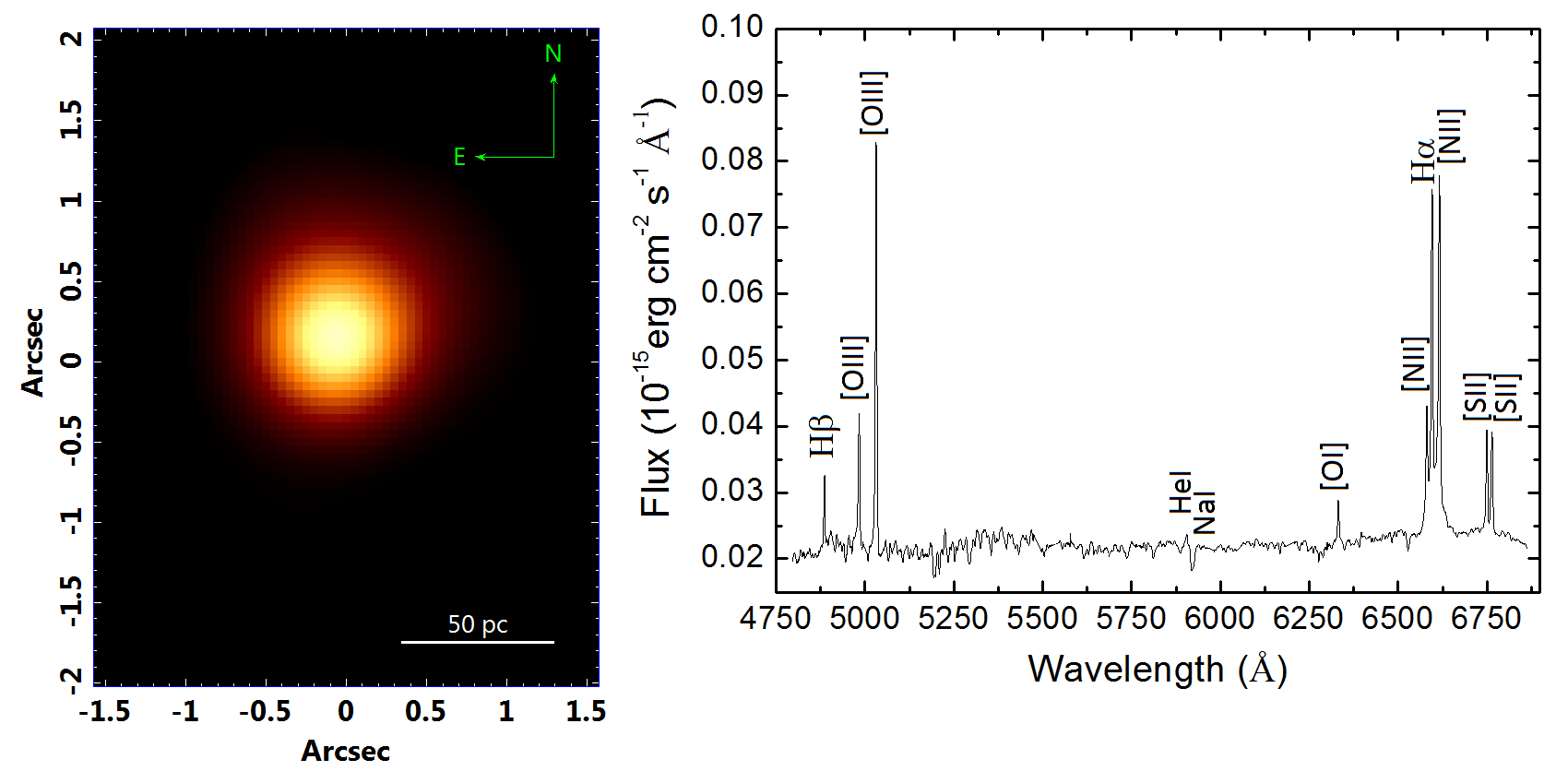}
\caption{Image of the GMOS data cube collapsed along the spectral axis (left) and its average spectrum (right).\label{fig1}}
\end{center}
\end{figure*}

\begin{figure*}
\begin{center}
  \includegraphics[scale=0.35]{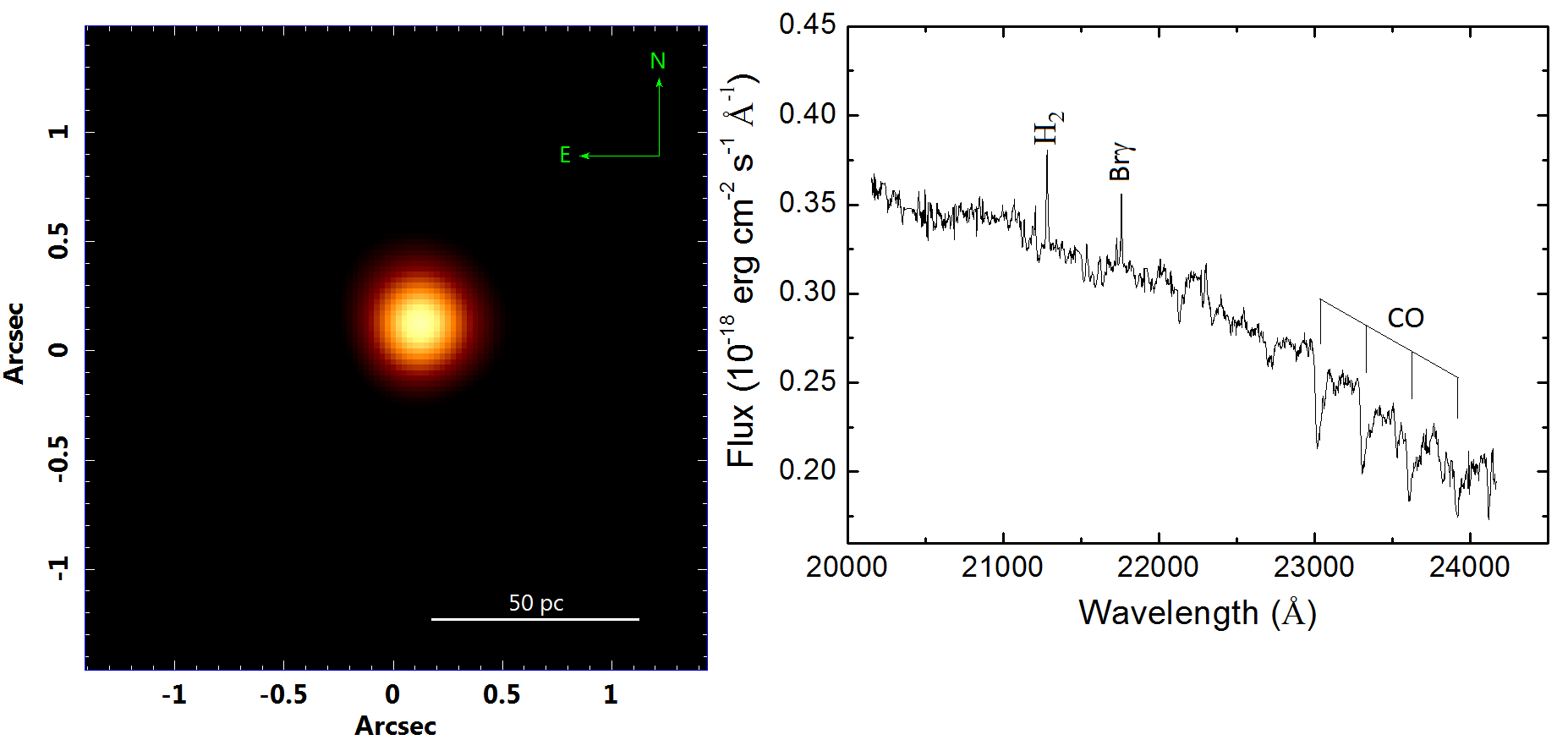}
\caption{Image of the \textit{K-}band SINFONI data cube collapsed along the spectral axis (left) and its average spectrum (right).\label{bandaK}}

  \includegraphics[scale=0.35]{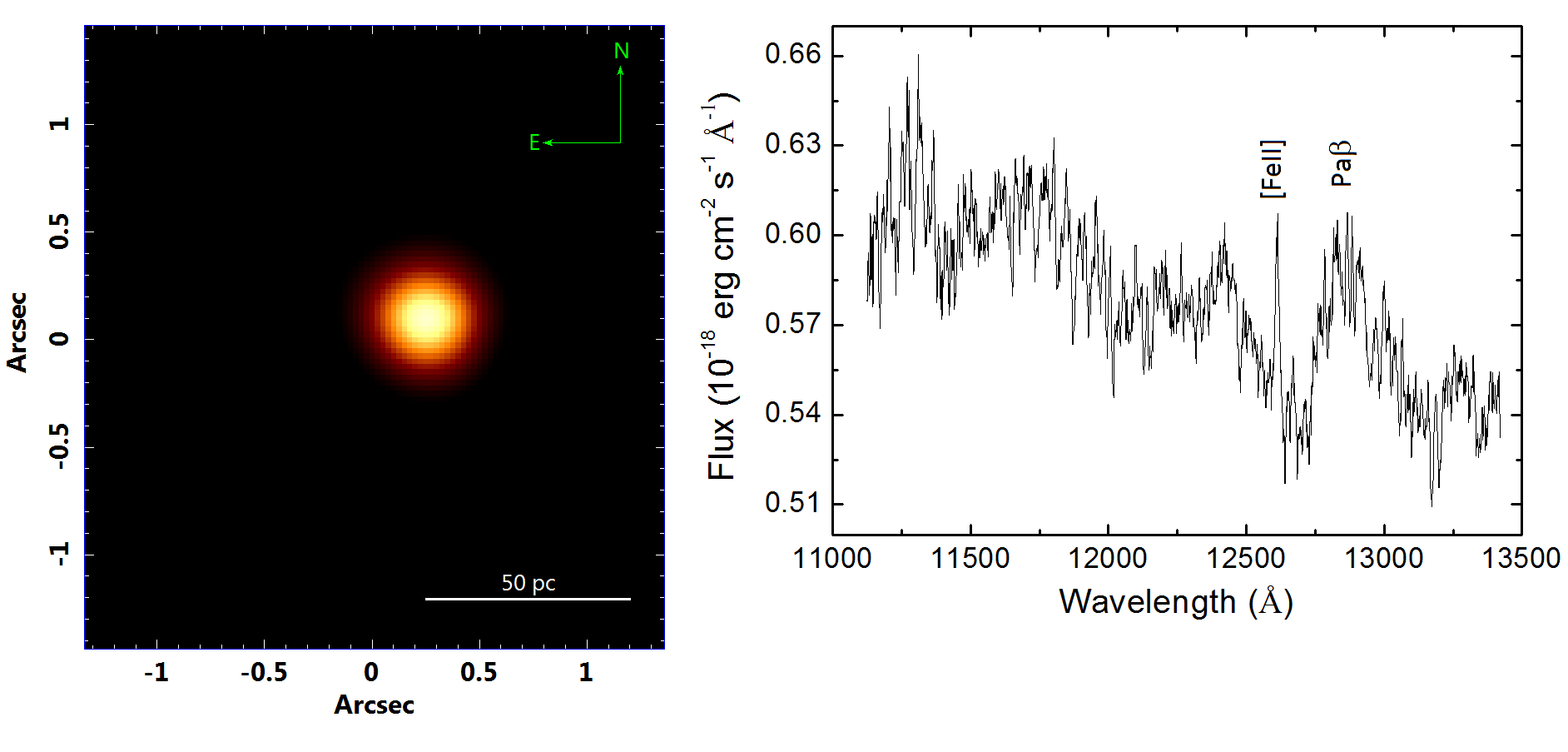}
\caption{Image of the \textit{J-}band SINFONI data cube collapsed along the spectral axis (left) and its average spectrum (right).\label{bandaJ}}
\end{center}
\end{figure*}

A stellar kinematic study performed by \citet{bottema} revealed that the stellar velocity dispersion is smaller at the edges of the galaxy and increases towards the nucleus, where it reaches a constant value of 115 km s$^{-1}$. \citet{smajic} observed a stellar rotation around the nucleus of NGC 1566. The authors also observed a slight decrease in the velocity dispersion in the nucleus. The black hole's mass, estimated from the $M-\sigma$ relation, with the stellar velocity dispersion value equal to 100 km s$^{-1}$ \citep{nelson}, is 8.3$\times10^{6}$ M$_{\sun}$ \citep{woo}.

The goal of this work is to analyse the nuclear region of NGC 1566, using a data cube observed in the optical with the Integral Field Unit (IFU) of the Gemini Multi-Object Spectrograph (GMOS), installed at the Gemini-South telescope. This study was complemented with a re-analysis of the SINFONI data observed in the \textit{K} and \textit{J} bands, obtained from the SINFONI data archive, and with \textit{HST} images of this galaxy, obtained from the \textit{HST} data archive. This work is focused on the properties of: the emission-line spectrum, the AGN's featureless continuum, the stellar kinematics and the ionized and molecular gas kinematics. 

The sections of this paper are divided as follows: section \ref{sec2} describes the conditions of the observations and presents a brief description of the data reduction and treatment. Section  \ref{sec3} shows  the results obtained with our first analysis tool, Principal Component Analysis (PCA) Tomography, applied to the optical data cube, and also shows the characterization of the featureless continuum detected in this galaxy. In section \ref{sec4}, we present the analysis of the \textit{HST} images. In section \ref{sec5}, we calculate the emission-line ratios of the two sources found in the nuclear region of NGC 1566. The gas kinematics (ionized and molecular) and the stellar kinematics are analysed in Sections \ref{sec6} and \ref{sec7}, respectively. In section \ref{sec8} we discuss the results of this work and, in section \ref{sec9}, we present the main conclusions. In Appendix \ref{sinteseespectral} we show the results of the spectral synthesis applied to the GMOS data cube.
 
 \section{observations, reduction and data treatment} \label{sec2}

\subsection{Optical data}

The optical observations of the nucleus of NGC 1566 were made on 2013 October 10, using the IFU of GMOS, installed on the Gemini-South telescope, in one-slit mode. The observation program was GS-2013B-Q-3. The field of view (FOV) has 5$\arcsec$ $\times$ 3$\arcsec\!\!$.5 and three exposures of 910 s were taken, with spectral resolution of R = 4340. The grating used for the observations was R831+G5322, with a central wavelength of 5850\AA.

The data reduction was made in \textsc{iraf} environment  and consisted of the determination of trim, bias subtraction, cosmic ray rejection with the \textsc{lacos} routine \citep{van01}, spectra extraction, correction of pixel-to-pixel gain variations (with response curves obtained with GCAL-flat images), correction of fiber-to-fiber gain variations and of asymmetric illumination patterns of the instrument (with response maps obtained with twilight flat images), wavelength calibration (using the CuAr lamp images), sky subtraction, atmospheric extinction correction, telluric absorptions removal, flux calibration and data cube construction. 

Three data cubes were constructed, with spatial pixels (spaxels) of 0$\arcsec\!\!$.05 $\times$ 0$\arcsec\!\!$.05 and a spectral coverage of 4790-6260\AA. The full width at half-maximum (FWHM) of the point spread function (PSF) of the reduced data cubes, estimated with the image of the red wing of the broad component of H$\alpha$, is 0$\arcsec\!\!$.75.

Following the process described in \citet{rob1} and \citet{rob2}, the treatment of the data cubes was performed after the reduction, with scripts written in Interactive Data Language (\textsc{idl}). This treatment consisted, first of all, of the differential atmospheric refraction correction. The three data cubes were, then, combined in the form of a median. After that, a Butterworth spatial filtering \citep{gwoods}, the instrumental fingerprint removal (using PCA Tomography - see the next section) and the Richardson-Lucy deconvolution 
\citep{rich,lucy} were applied. This last step required a variation law of the FWHM of the PSF with the wavelength, which was estimated from the standard star data cube. After the deconvolution, with 10 interactions, the FWHM of the image of the red wing of H$\alpha$'s broad component was 0$\arcsec\!\!$.66. 

Fig.~\ref{fig1} shows the image of the data cube collapsed along the spectral axis, herewith its average spectrum. Possible broad components of the H$\alpha$ and H$\beta$ lines can be observed, together with a strong emission of [O \textsc{iii}]$\lambda$5007 and other narrow emission lines.

With the goal of characterizing the emission-line spectra, we performed a stellar continuum subtraction, using the synthetic stellar spectra provided by the spectral synthesis (see Appendix \ref{sinteseespectral}). This procedure resulted in a data cube mainly with gas emission. 

The optical data cube of NGC 1566 described above was used in this work to analyse: the emission-line spectrum of the ionized gas, the AGN featureless continuum emission, the current stellar populations (with spectral synthesis), the stellar kinematics and the ionized gas kinematics. 

\begin{figure*}
\begin{center} 
  \includegraphics[width=1.0\linewidth,height=0.4\linewidth]{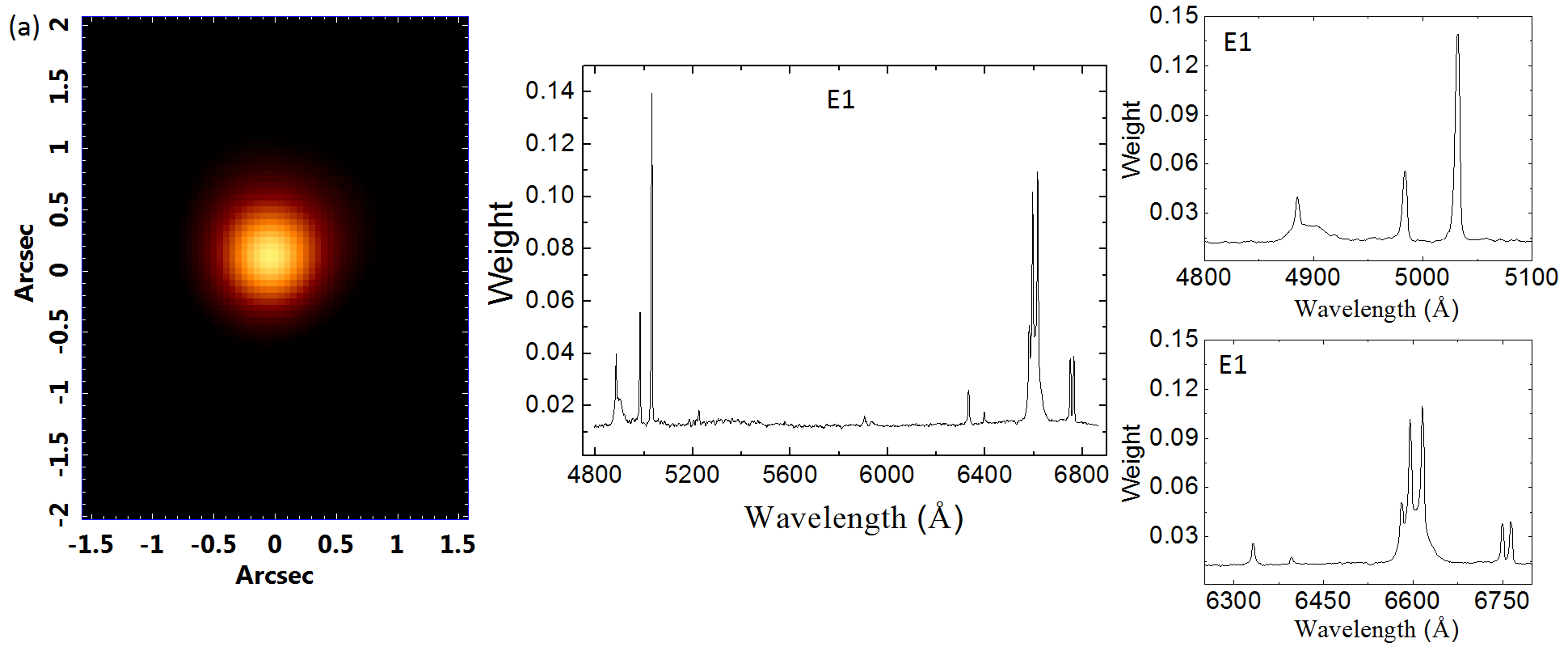}
  
  \vspace{0.5cm}
  
  \includegraphics[width=1.0\linewidth,height=0.4\linewidth]{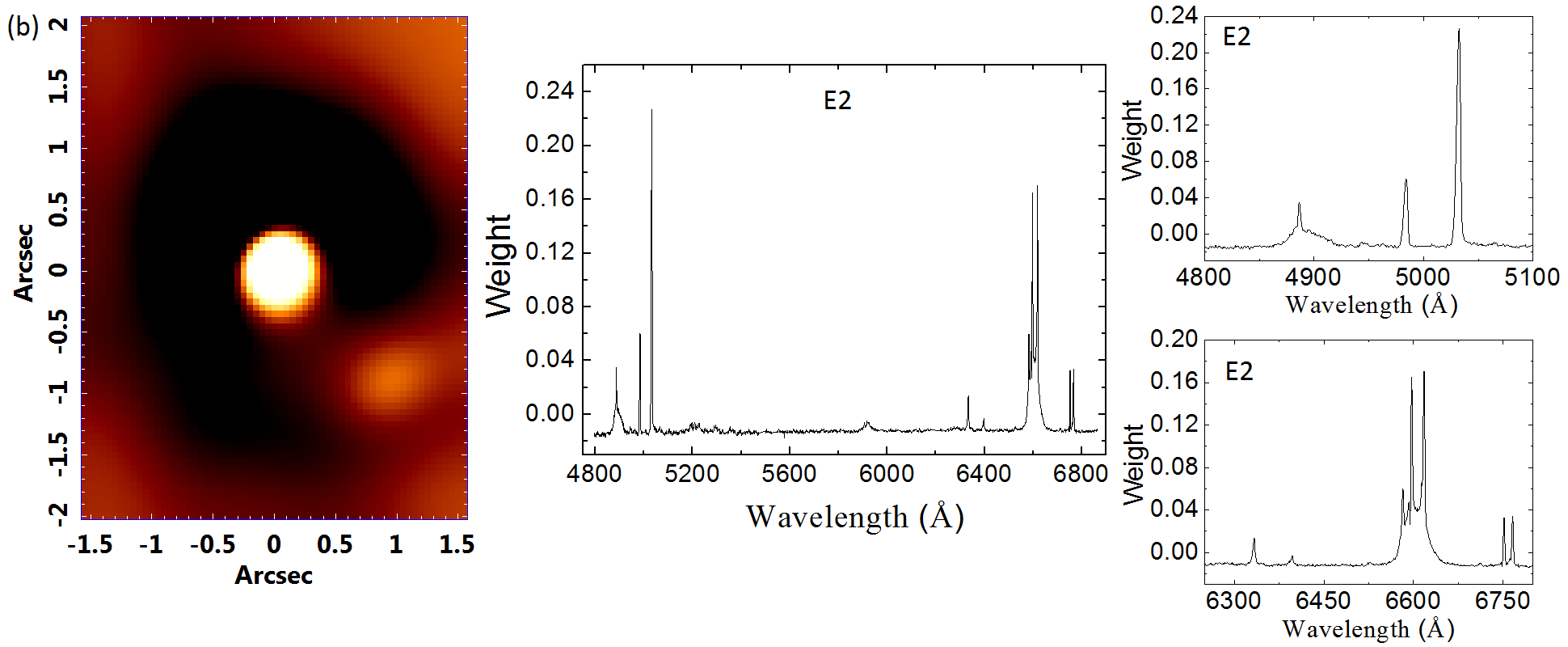}
  
    \vspace{0.5cm}
    
  \includegraphics[width=1.0\linewidth,height=0.4\linewidth]{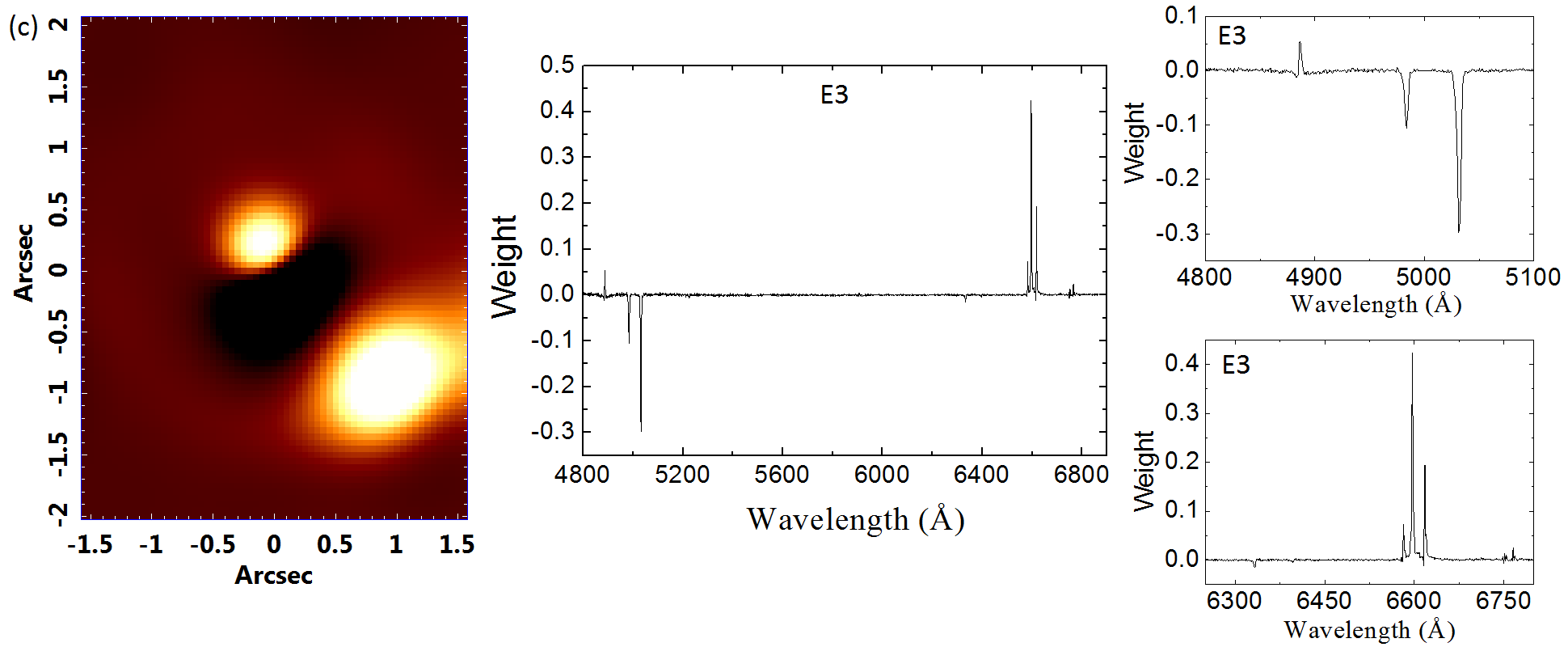}
  
  \caption{Tomograms and eigenspectra obtained with PCA Tomography applied to the GMOS data cube of NGC 1566. (a) Tomogram and eigenspectrum 1, (b) tomogram and eigenspectrum 2, (c) tomogram and eigenspectrum 3 and (d) tomogram and eigenspectrum 4. The eigenspectra are not redshift corrected. \label{fig2}}
\end{center}
\end{figure*}

\begin{figure*}
\begin{center}
 \includegraphics[width=1.0\linewidth,height=0.4\linewidth]{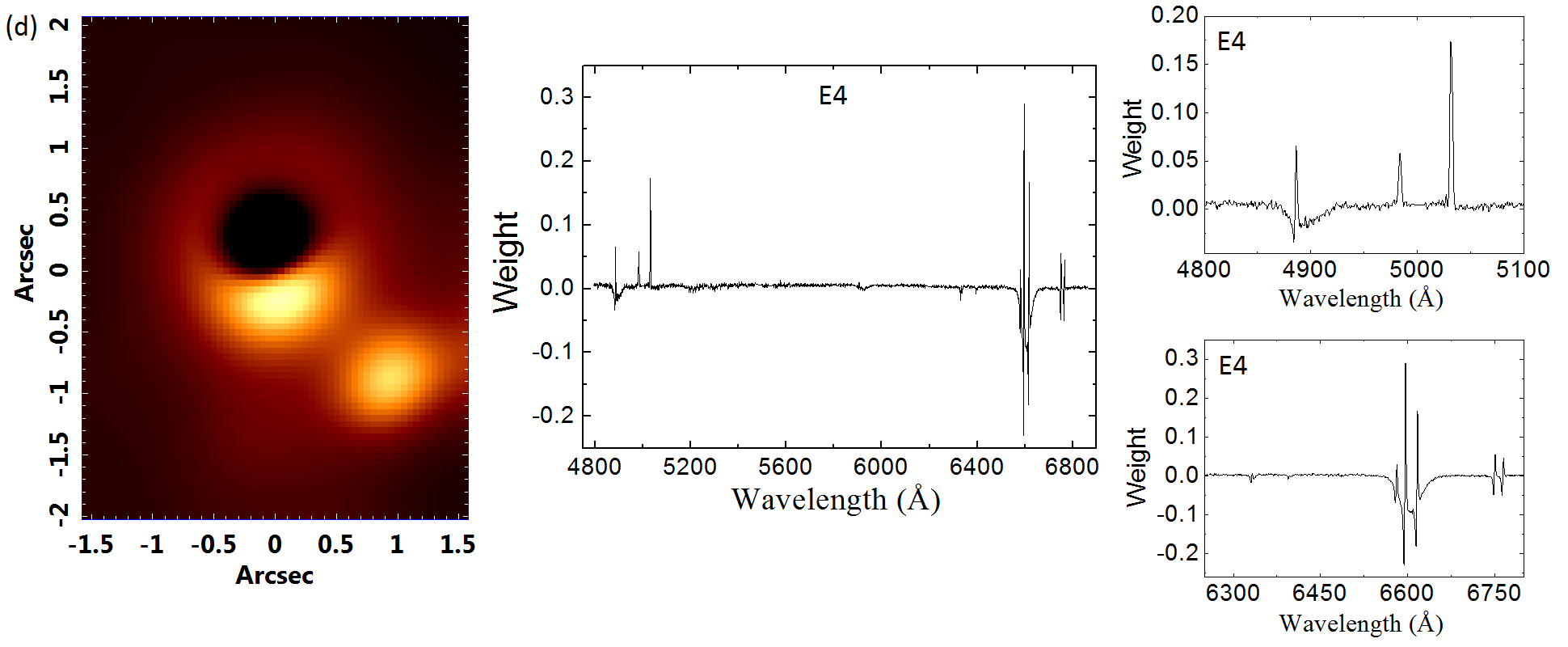}
  \contcaption{\label{fig22}}

\end{center}
\end{figure*}

\begin{figure}
\begin{center}
  \includegraphics[scale=0.45]{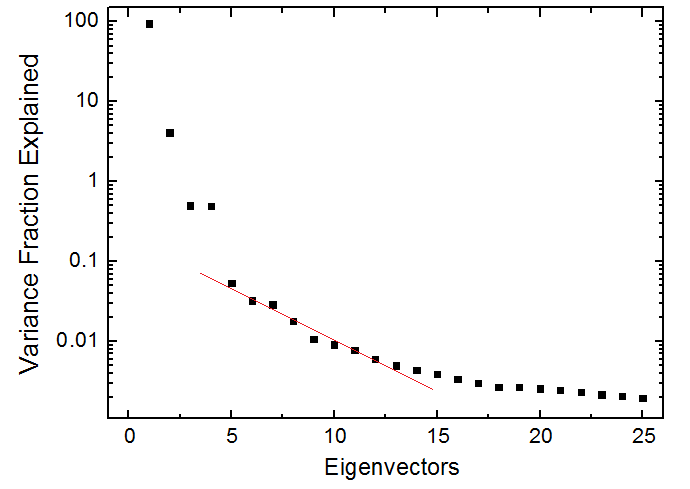}
  \caption{'Scree test'. The red line shows that, from eigenvector E5 onwards, the decrease rate of variance fraction explained becomes constant. This shows that the noise is relevant in the data from this eigenvector onwards, making interpretation difficult. \label{fig3}}
\end{center}
\end{figure}

\subsection{Near-infrared data}

The  observations in the NIR, obtained from the SINFONI data archive, were made on 2008 October 2 (\textit{J} band) and on 2008 November 1 (\textit{K} band). The observation program was 082.B-0709(A) (PI: Beckert). The fore optics of 0.1$\arcsec$ was used, which resulted in a FOV of 3$\arcsec\!\!$.2 $\times$ 3$\arcsec\!\!$.2. Two 300 s observations were made for the \textit{J} and \textit{K} bands, with central wavelengths of 12500 and 22000\AA, respectively.

The data reduction was made with the \textsc{gasgano} software\footnote[1]{\href{http://www.eso.org/sci/software/gasgano.html}{http://www.eso.org/sci/software/gasgano.html}} and included bad pixel correction, flat-field correction, spatial rectification (using distortion-fibre images), wavelength calibration, sky subtraction and data cube construction. At the end, data cubes with spaxels of 0$\arcsec\!\!$.05 were obtained. The FWHM of PSF of the \textit{J-} and \textit{K-}band data cubes were $\sim$ 0$\arcsec\!\!$.41 and $\sim$ 0$\arcsec\!\!$.40, respectively. These estimates were taken from images of Pa$\beta$ and Br$\gamma$ broad component red wings.

After the reduction, a data treatment similar to the one used for the optical data cubes was applied -see \citet{rob1,rob2}. First, a differential atmospheric refraction correction was performed and the resulting data cubes were combined in the form of a mean. Then, a spatial re-sampling was carried out, resulting in spaxels of 0$\arcsec\!\!$.025. After that, a Butterworth spatial filtering was applied, the instrumental fingerprint  was removed and the data cube was deconvolved using the Richardson-Lucy procedure. The estimate of the PSFs of the data cubes, required for the deconvolution, was obtained from images of the red wings of Pa$\beta$ and Br$\gamma$ broad components. Using these same images, we verified that the FWHM of the PSFs of the treated data cubes, in the \textit{J} and \textit{K} bands, was $\sim$ 0$\arcsec\!\!$.33.

Fig.~\ref{bandaK} shows the collapsed image and the mean spectrum of the data cube in the \textit{K} band. The CO molecular bands are very prominent in this spectrum. Fig.~\ref{bandaJ} shows the collapsed image and mean spectrum of the data cube in the \textit{J} band. One may notice that Pa$\beta$ is the only emission line clearly visible in this spectrum, which is considerably noisier than the one in the \textit{K} band. 

The stellar continuum of the data cubes in the \textit{J} and \textit{K} bands was subtracted, as well as it was done for the optical data cube, to allow a more detailed analysis of the emission-line spectrum. In the \textit{K} band, the subtraction was made using synthetic stellar spectra provided by the Penalized Pixel Fitting (pPXF) method \citep{cappellari}. This procedure involves the stellar continuum fit, using spectra of a given base (for more details, see section \ref{sec7}). On the other hand, the stellar continuum subtraction in the \textit{J}-band data cube was performed simply by subtracting splines fitted to the spectra (with masked emission lines), since, in this case, there is no suitable stellar spectra base to be used.  

The NIR data cubes of the central region of NGC 1566 were used to analyse the emission-\textit{}line spectrum, the ionized gas kinematics in some areas of the FOV and also the molecular gas kinematics (using the H$_2\lambda$21218 line, in the \textit{K} band).

\begin{figure*}
\begin{center}
 \includegraphics[scale=0.5]{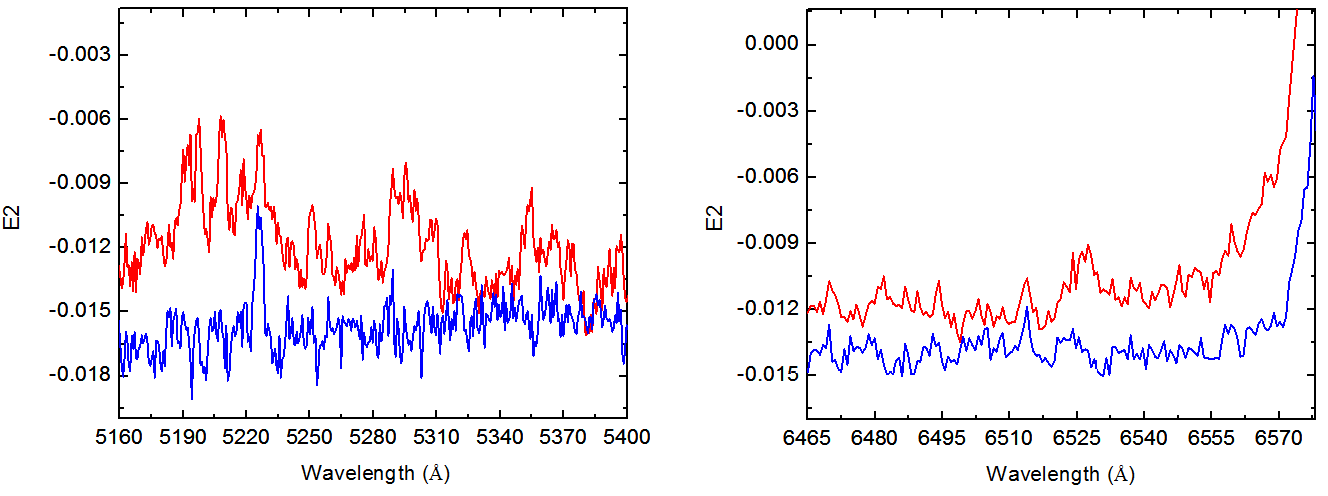}
  \caption{Two wavelength ranges of eigenspectrum E2 (not redshift corrected): in red, the original eigenspectrum E2 and, in blue, the same eigenspectrum obtained from the data cube after the removal of the featureless continuum contamination, whose determined spectral index is 1.7.\label{fig10}}
\end{center}
\end{figure*}

\section{PCA Tomography} \label{sec3}

After the treatment, PCA Tomography \citep{steiner} was applied to the optical data cube. PCA is an orthogonal linear transformation of coordinates that passes the data to a new system in which the coordinates are the eigenvectors of the covariance matrix, in order of the explained variance. In the case of PCA Tomography (which consists of PCA applied to data cubes), the variables of the system are the spectral pixels, while the observables are the spaxels. As a consequence, the eigenvectors of PCA Tomography are linear combinations of the spectral pixels and are very similar to the spectra. Because of that, we call them eigenspectra. Tomograms are the projections of the spaxels on the eigenvectors, being, then, images that indicate the degree of correlation between each spaxel and the associated eigenspectrum. It is important to analyse eigenspectra and tomograms simultaneously for a complete interpretation of the data. The PCA Tomography analysis reveals a large variety of phenomena in the data cube, many of them hard to detect with traditional methods \citep{ricci11,menezes13,ricci14}.

\subsection{Eigenvectors and tomograms}

The first set of tomogram/eigenvector (Fig.~\ref{fig2}a) represents the characteristics associated with most of the data cube variance. This eigenspectrum resembles the data cube's average spectrum, which reveals the data redundancy (see Fig.~\ref{fig1}). However, this eigenspectrum displays less structures compatible with the absorption lines than the average spectrum, indicating that it is less related to the stellar population's emission than the average spectrum. It is possible to see correlations compatible with the emission lines coming from the central region (bright) of the tomogram (correlated to the eigenspectrum). Structures compatible with the broad components of H$\alpha$ and H$\beta$ lines are evident, with a strong emission of [O \textsc{iii}]$\lambda$5007 and with the emission of the typical lines of partial ionization regions ([O \textsc{i}]$\lambda$6300, [S \textsc{ii}]$\lambda\lambda$6716, 6731 and [N \textsc{ii}]$\lambda\lambda$6548, 6584). All these characteristics reveal the existence of an AGN in the bright region of the associated tomogram. The variance fraction explained by this eigenvector is about 95 per cent. 

The second eigenvector (see Fig.~\ref{fig2}b) is correlated to the main emission lines of this spectral region and to the broad components of H$\alpha$ e H$\beta$. There is, also, an anticorrelation with many stellar absorption lines in this spectral region. Based on that, it is possible to infer that the tomogram's bright regions represent the main areas from where the emission lines are emitted. At the same time, one may notice that the stellar absorptions are more accentuated in the tomogram's dark region (anticorrelated to the eigenspectrum) than in the bright regions. One of the bright regions, Region 1, is centred on the galaxy's nucleus (centre of the FOV), while the second region, Region 2, is centred at $\Delta_x =$ 1$\arcsec$ and $\Delta_y=$ -1$\arcsec$. This eigenvector explains about 4 per cent of the data cube variance.

Eigenvector E3 is correlated to the low ionization lines (H$\beta$, H$\alpha$, [N \textsc{ii}]$\lambda\lambda$6548, 6584, [S \textsc{ii}]$\lambda\lambda$6716, 6731) and anticorrelated to the higher ionization lines ([O \textsc{iii}]$\lambda\lambda$4959, 5007). So, the tomogram's bright regions (see Fig.~\ref{fig2}c), which are correlated to eigenvector E3, are regions of lower ionization and the tomogram's central dark region emits the higher ionization lines. In other words, this set of tomogram/eigenvector is differentiating the regions of high and low ionization and explains 0.5 per cent of the variance.

Eigenspectrum E4 (see Fig.~\ref{fig2}d) is anticorrelated to the broad components of the H$\alpha$ e H$\beta$ lines. With the exception of the [O \textsc{iii}]$\lambda\lambda$4959,5007 and [O \textsc{i}]$\lambda$6300 lines, there are anticorrelations to the blue wings of all the emission lines and correlations to the red wings of the same lines. In view of this, the set tomogram/eigenvector 4 represents certain phenomena: the kinematics of the low ionization lines, the location of the BLR and the emission of the higher ionization lines. The central bright region is the same one observed in tomogram 3, therefore, indicating again the location from where the higher ionization emission lines are emitted. In addition, eigenspectrum E4 also has correlations to the red wings of all other emission lines. Therefore, it is possible that at least part of this region is in redshift. The second bright area in the tomogram, which coincides with Region 2, is not correlated to the higher ionization lines (see tomogram 3), so it must be associated with the correlations to the red wings of the emission lines observed in eigenspectrum E4. Based on that, Region 2 should also be in redshift. The dark area, besides indicating the location of the BLR, as it is anticorrelated to the broad components of the H$\alpha$ and H$\beta$ lines, also coincides with the area where the gas is in blueshift. The possible correlation of the bright part of the central region to the red wings of the lines and the anticorrelation of the other part, which shows the blueshift area, suggests the existence of a kinematic phenomenon, being an outflow or a gas rotation. This eigenvector explains about 0.49 per cent of the variance.

It is not possible to interpret the data clearly from the set of tomogram/eigenvector 5 onwards. This is because noise becomes dominant as the variance fraction decreases. By summing the variance fractions of all other eigenvectors, we have 0.24 per cent; so, there is little information shrouded in a lot of noise. A way to see that is through the 'scree test'. Generally, when the rate of decrease of the explained variance fraction reaches a nearly constant value (which happens here from eigenvector E5 onwards - Fig.~\ref{fig3}), the noise dominates the eigenvectors, making their interpretation more difficult. But it is also possible to see that from eigenvector E15 onwards, the decreasing rate is flatter and there is little difference between the variance fractions of the eigenvectors, which shows that the noise is even more expressive in the data from this eigenvector onwards.

\begin{figure*}
\begin{center}

  \includegraphics[scale=0.35]
    {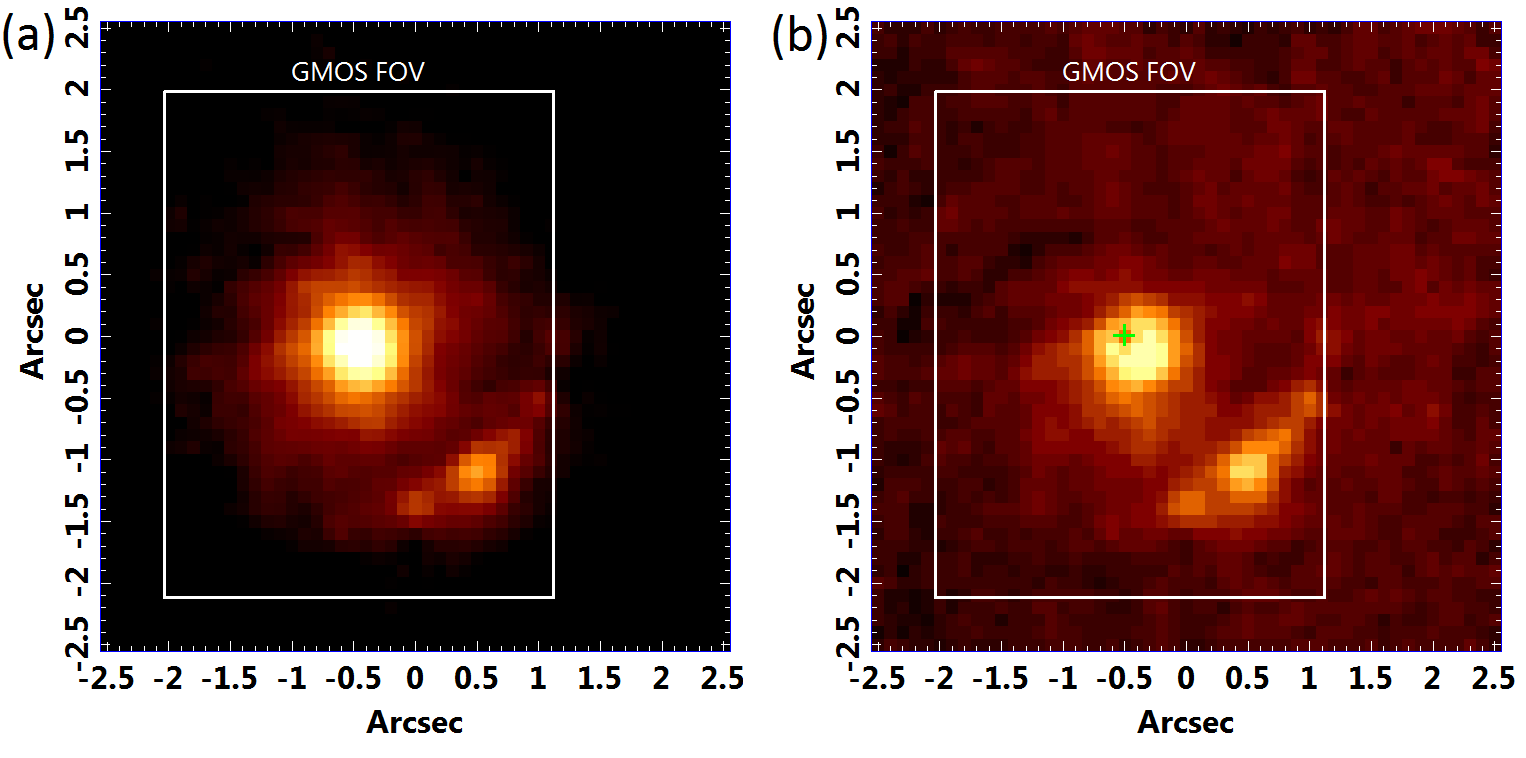}
  \caption{(a) Filter F658N image obtained with WFPC2 of the \textit{HST}, at the wavelength 6562\AA, without the stellar emission (represented by the filter F555W image multiplied by a constant). In this image, it is possible to see the central region, which corresponds to the AGN, and the H \textsc{ii} region, according to Fig.~\ref{fig4}. A square representing the size of GMOS/IFU FOV was added to the image. (b) H$\alpha$/F555W ratio image, with the green cross representing the emission peak in the filter F555W. The size of the cross represents the position uncertainty of 2$\sigma$.\label{hst_ha}}
  
\end{center}
\end{figure*}

\begin{figure*}
\begin{center}

  \includegraphics[scale=0.35]
  {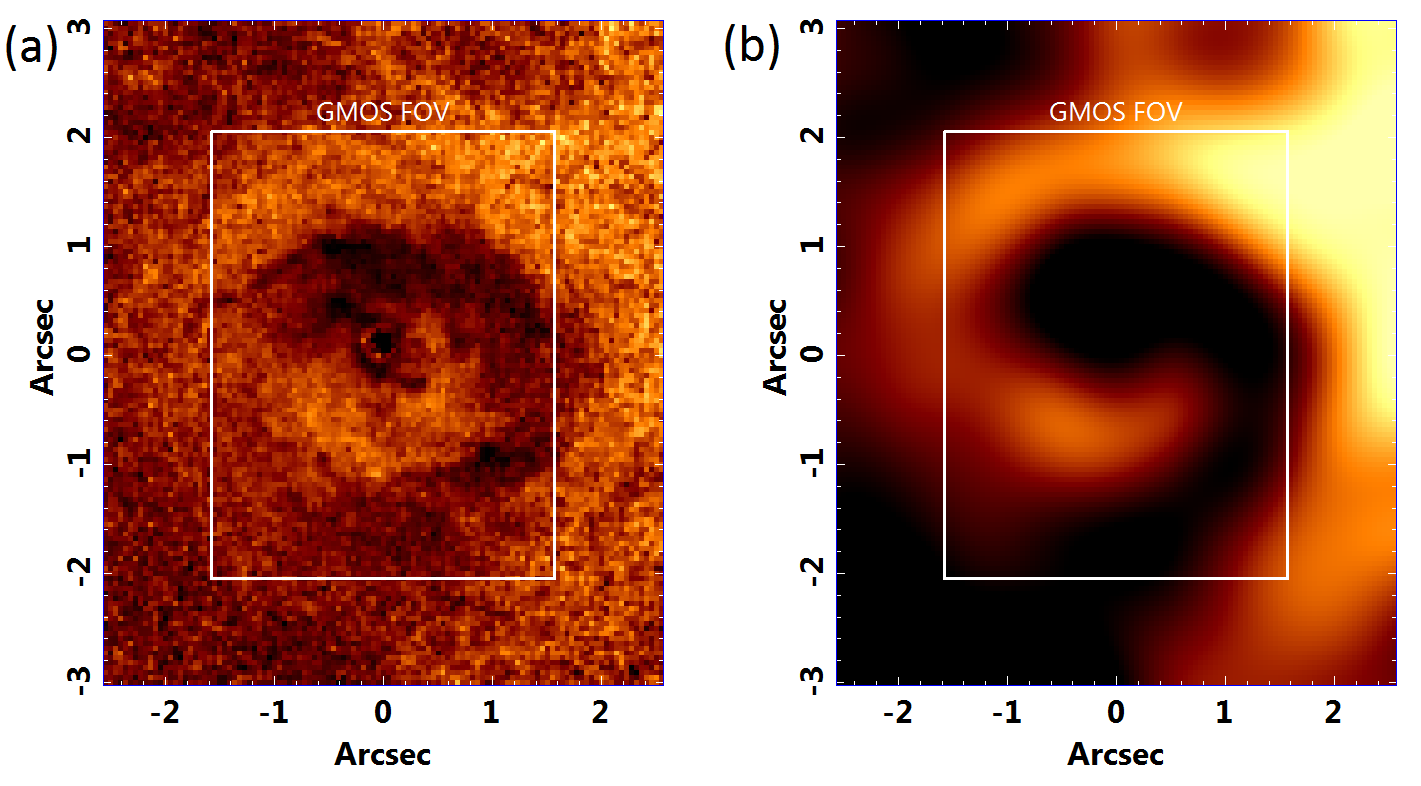}
  \caption{(a)\textit{HST} image corresponding to the subtraction F555W-F814W (equivalent to \textit{V-I}). (b)The same image convolved with the PSF of the optical data cube. Both images have the indication of the size of the GMOS/IFU FOV. The bright areas indicate regions where spectra are redder and the dark areas indicate regions where spectra are bluer.\label{hst_vi}}
  
\end{center}
\end{figure*}

\subsection{Featureless continuum}\label{secfeatcont}

Great part of the AGN emission occurs in the form of a featureless continuum, which behaves mathematically as a power law,

\begin{equation}
F_{\nu}= A{\nu}^{-\alpha},
\end{equation}

\noindent where the exponent $\alpha$ is called spectral index. 

Eigenspectrum 2 (Fig.~\ref{fig2}b) reveals that the stellar absorption lines are less intense in the AGN region than in its surroundings. We believe that this is due to the obfuscation of these lines by the featureless continuum emitted by the AGN. In order to find an appropriate spectral index to describe this emission, synthetic data cubes containing power laws at the AGN position were built, with different spectral indexes and different multiplicative constants. All the others values in the synthetic data cubes were assumed to be 0. After that, each frame of the synthetic data cube was convolved with a PSF, with FWHM given by the variation law estimated from the data cube of the standard star used during the data reduction. This variation law requires a value for FWHM at a reference wavelength. We took the wavelength of H$\alpha$ as reference and we also convolved the synthetic data cubes assuming different values for the reference FWHM. The convolved synthetic data cubes were then subtracted from the original one. Lastly, PCA Tomography was applied to the resulting data cubes. Fig.~\ref{fig10} shows the eigenvector E2 obtained from the PCA Tomography of the data cube before the subtraction of the featureless continuum, as well as the same eigenvector obtained with the PCA Tomography of the data cube that revealed the best subtraction, which was achieved assuming a featureless continuum with a spectral index of 1.7. The reference FWHM (at the wavelength of H$\alpha$) for the variation law of the PSF convolved with the synthetic data cube that best removed the absorption lines obfuscation in the central region (FWHM $\sim$ 0$\arcsec\!\!$.75) is higher than the FWHM of the PSF (estimated from the image of the broad component of H$\alpha$) of the treated data cube of NGC 1566 (FWHM= 0$\arcsec\!\!$.66). This suggests that the detected emission, with the approximate form of a power law, is not coming from the AGN alone (see Appendix \ref{sinteseespectral}).

\section{Analysis of \textit{HST} images}\label{sec4}

In order to analyse with greater detail the circumnuclear structures in this galaxy, we compared our data with some \textit{HST} archive images, observed with the Wide-Field Planetary Camera 2 (WFPC2) using the broad-band F555W and F814W filters \citep{erwin,georgiev}, which correspond, approximately, to the \textit{V} and \textit{I} bands, respectively. The images were taken on 1995 August 30 and three exposures of 100 s were made for each filter. We also retrieved, from the \textit{HST} data archive, three 600-s exposures, taken with WFPC2, on 2009 September 9, using the narrow-band F658N filter.

\begin{figure}
\begin{center}

  \includegraphics[scale=0.5]
  {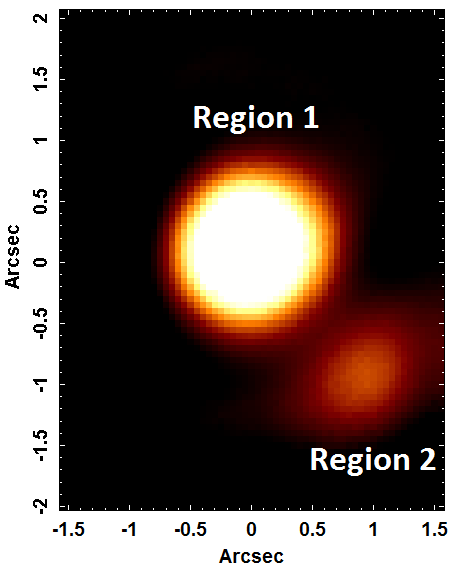}
  \caption{Image of the data cube collapsed along the spectral region of the [N \textsc{ii}]$\lambda$6548, H$\alpha$ and [N \textsc{ii}]$\lambda$6584 lines. In this figure, the two main emitting regions are indicated: Region 1 (AGN), centred on the nucleus, and Region 2 (H \textsc{ii} region), centred on $\Delta_x =$ 1$\arcsec$ and $\Delta_y=$ -1$\arcsec$. \label{fig4}}
  
\end{center}
\end{figure}

An H$\alpha$ image was obtained with the F658N filter, after the subtraction of the image in the F555W filter (multiplied by a constant), which represents the stellar emission. The result obtained after rotating the subtracted image, so that it had the same GMOS data orientation presented here, is shown in Fig.~\ref{hst_ha}(a). We can see that, besides the central emitting region (Region 1), there are other significant emitting regions southwest from the nucleus, which correspond to Region 2, observed in the optical data cube (see Fig.~\ref{fig4}). An interesting detail is that these emitting regions, located at about 1$\arcsec\!\!$.4 from the nucleus, are placed along a structure that is similar to a spiral arm in the central region of NGC 1566. In section \ref{sec5}, we see that Region 1 represents the AGN, which has Seyfert emission, and Region 2 is an H \textsc{ii} region close to the nucleus.

Fig.~\ref{hst_ha}(b) corresponds to the ratio between the H$\alpha$ and F555W filter images (H$\alpha$/F555W). It shows exactly the same structures depicted in Fig.~\ref{hst_ha}(a), and it is possible to visualize in greater detail that Region 2 has many emitting areas. Furthermore, we noticed that the point corresponding to the emitting peak in the F555W filter image is not coincident with the emitting peak in the H$\alpha$ image. Another characteristic that we cannot see very well in Fig.~\ref{hst_ha}(a) is the extended emission to the south-west, which coincides with the outflow direction observed in the [O \textsc{iii}] images (see section \ref{sec6}). 

An image equivalent to \textit{V}-\textit{I} was made by subtracting, in scale of magnitude, the images in the F555W and F814W filters (see Fig.~\ref{hst_vi}a). It is easy to see that the bright areas, which represent the redder spectral emission, are arranged in a pattern around the central region that closely resembles a spiral. By convolving this \textit{HST} image with the GMOS PSF of the data analysed here, we obtain the image shown in Fig.~\ref{hst_vi}(b), which reveals clearly the spiral structure mentioned before. We also notice that the bluer regions, with a similar spiral pattern, include the central region (due to the featureless continuum emission from the AGN and to the presence of young stellar populations there) and the H \textsc{ii} region (possibly due to the presence of young stellar populations at the spiral arm). 

\section{AGN and the H II region} \label{sec5}

\begin{figure*}
\begin{center}

  \includegraphics[scale=0.46]
  {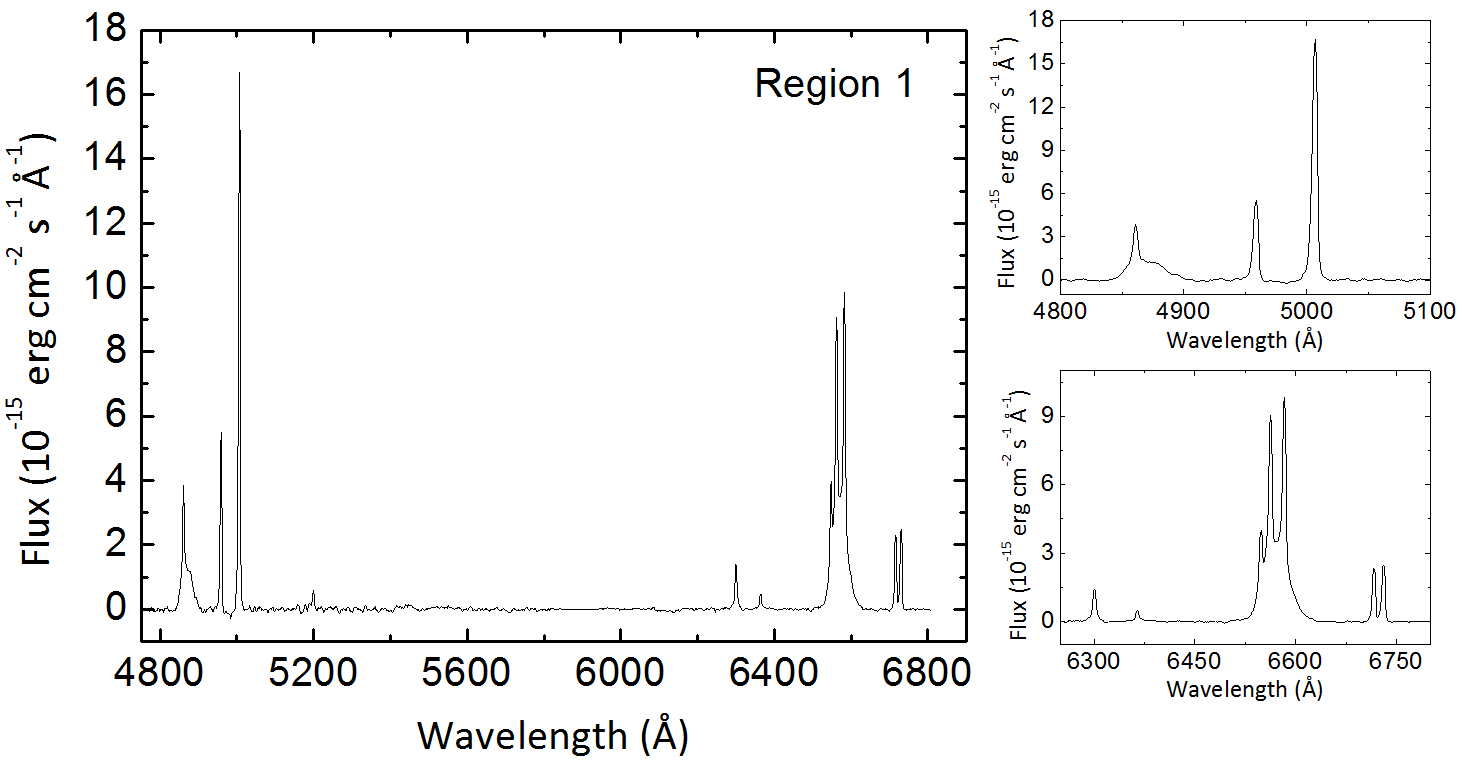}
  \caption{Spectrum of the NGC 1566 nucleus (Region 1: AGN) extracted from the GMOS/IFU data cube.\label{fig5}}
  
\end{center}
\end{figure*}

\begin{figure*}
\begin{center}

  \includegraphics[scale=0.46]{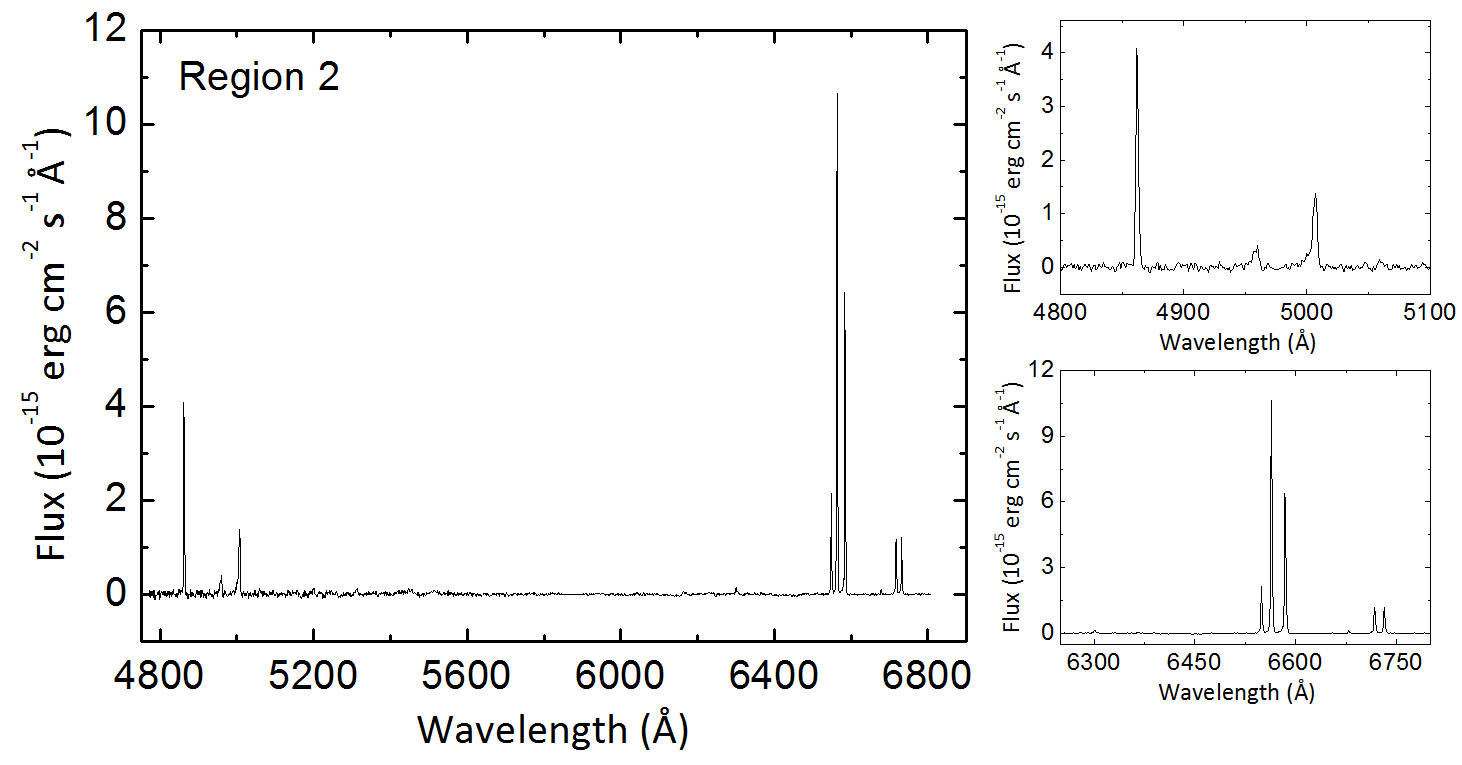}
  \caption{Spectrum of the emitting region found close to the nucleus of NGC 1566 (Region 2: H \textsc{ii} region) extracted from the GMOS/IFU data cube.\label{fig6}}
  
\end{center}
\end{figure*}

\begin{figure*}
\begin{center}

  \includegraphics[scale=0.3]
  {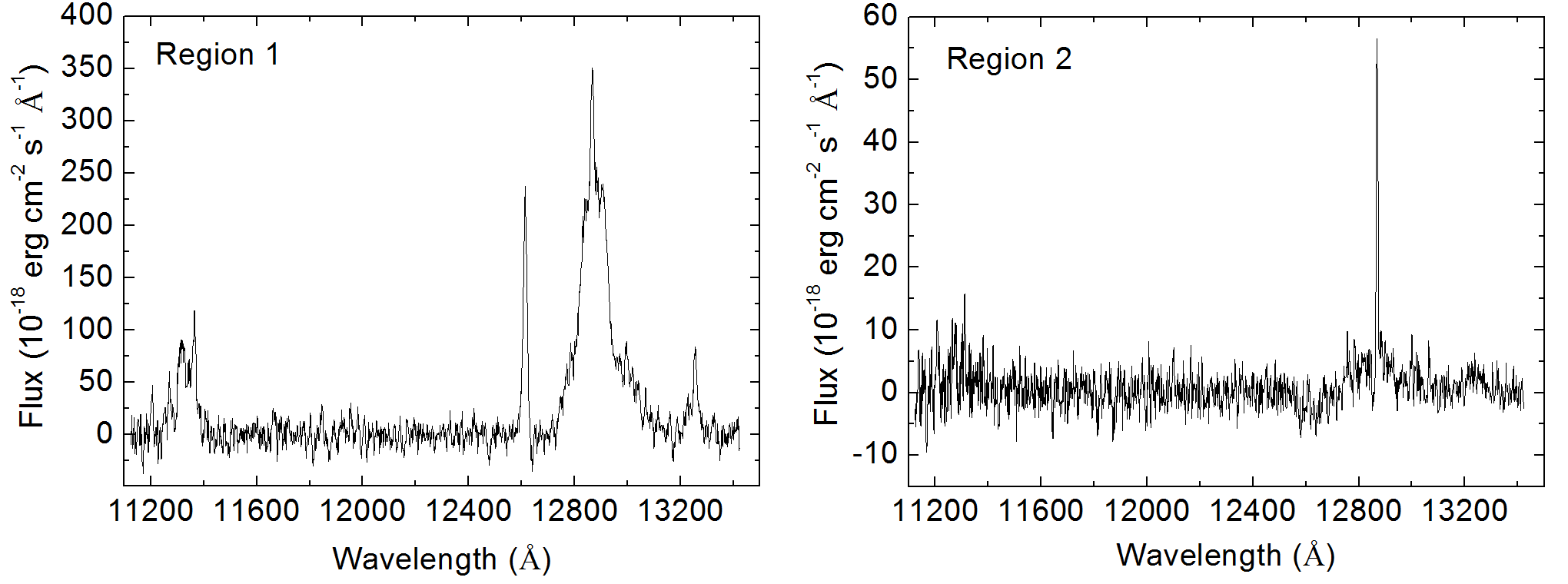}
  \caption{Spectrum of the NGC 1566 nucleus (Region 1: AGN) and of the H \textsc{ii} region found close to the nucleus (Region 2: H \textsc{ii} region) extracted from the \textit{J}-band SINFONI data cube.\label{espectrosj}}
  
\end{center}
\end{figure*}

\begin{figure*}
\begin{center}

  \includegraphics[scale=0.3]
  {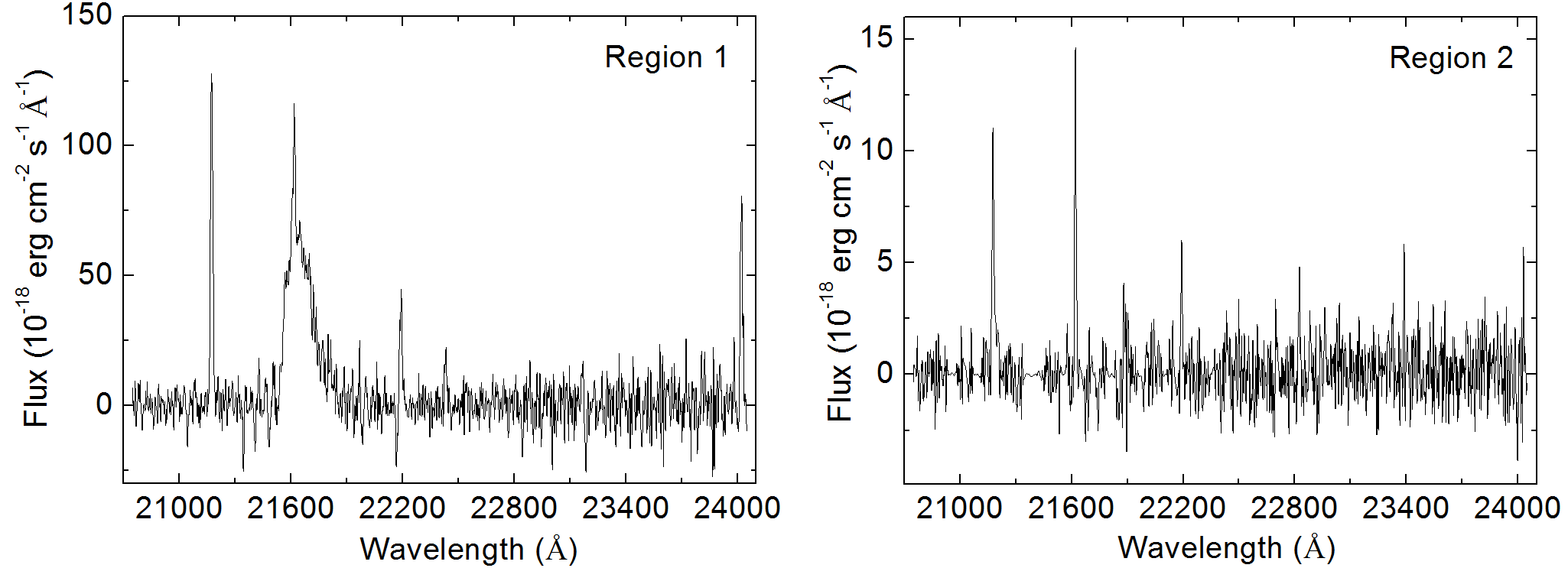}
  \caption{Spectrum of the NGC 1566 nucleus (Region 1: AGN) and of the H \textsc{ii} region found close to the nucleus (Region 2: H \textsc{ii} region) extracted from the \textit{K}-band SINFONI data cube.\label{espectrosk}}
  
\end{center}
\end{figure*}

\begin{figure*}
\begin{center}

  \includegraphics[scale=0.23]
  {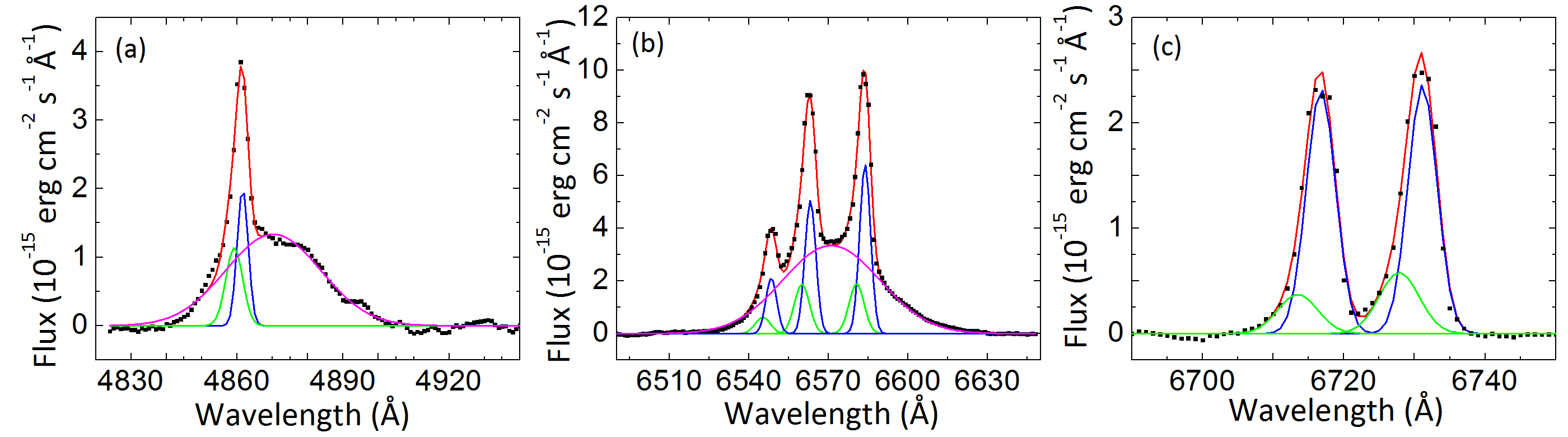}
  \caption{Gaussian fits of the following lines: (a) H$\beta$, (b) [N \textsc{ii}]$\lambda$6548, H$\alpha$ and [N \textsc{ii}]$\lambda$6584 and (c) [S \textsc{ii}]$\lambda\lambda$6716,6731. The black points correspond to the observed data. The blue and green lines represent the narrow components of the emission lines. The pink curves are the broad components of the H$\alpha$ and H$\beta$ lines and the red curves represent the final fits of the Gaussian compositions. \label{fig7}}
  
\end{center}
\end{figure*}

\begin{figure*}
\begin{center}

  \includegraphics[scale=0.5]{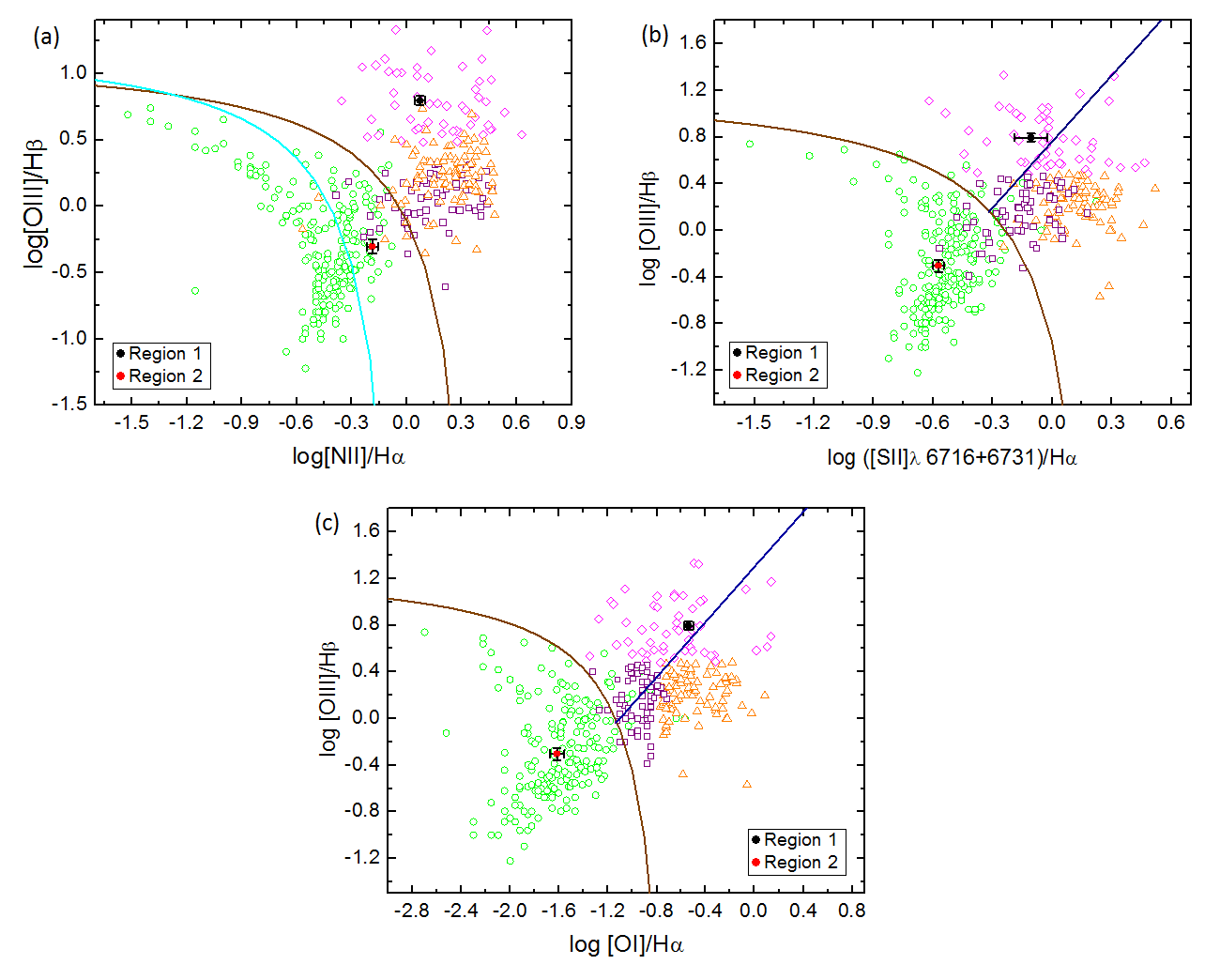}
  \caption{Diagnostic diagrams of the main emission-line ratios in the optical data cube of NGC 1566. The central emitting region (Region 1) is represented by the black point and Region 2, by the red square. All other points correspond to the objects analysed by \citet{ho}: the green circles are H \textsc{ii} regions, the orange triangles are LINERs (low-ionization nuclear emission-line regions), the violet squares are transition objects and the pink diamonds represent Seyfert galaxies. The brown line in the diagrams (a)-(c) shows the maximum limit of ionization by a starburst obtained by \citet{kewley1}. The cyan line represents the division between H \textsc{ii} regions and AGNs made by \citet{kauffmann} and the blue line represents the division between Seyfert galaxies and LINERs obtained by \citet{kewley2}.\label{fig9}}
  
\end{center}
\end{figure*}

\subsection{Analysis of optical spectra}

In order to make a detailed analysis of the emission lines of the data cube, a stellar continuum subtraction was performed. After applying the spectral synthesis (see Appendix \ref{sinteseespectral}), the resulting synthetic stellar spectra were subtracted from the original data cube, which resulted in a data cube with mainly emission lines (gas data cube). 

The PCA Tomography results (see section \ref{sec3}) show two distinct emitting regions (see Fig.\ref{fig4}): one centred on the nucleus (Region 1) and the other (Region 2) centred on $\Delta_{x}= 1\arcsec$ and $\Delta_y= -1\arcsec$, whose projected distance from the nucleus is about 73.3 $\pm$ 1.9 pc (1$\arcsec\!\!$.40 $\pm$ 0$\arcsec\!\!$.04). In order to study the spectral properties of the second emitting region, we extracted a spectrum from a circular region of the gas data cube, centred on $\Delta_{x}= 1\arcsec$ and $\Delta_y= -1\arcsec$, whose radius (0$\arcsec\!\!$.3) is equal to half of the PSF's FWHM. We likewise extracted a spectrum with the same radius from Region 1 (see Figs ~\ref{fig5} and ~\ref{fig6}). The two extracted spectra were corrected for the interstellar extinction, using the values of the H$\alpha$/H$\beta$ ratio (considering only the narrow components of the lines) and the extinction law of \citet{cardelli}.

The spectrum of Region 1 shows, clearly, H$\alpha$ and H$\beta$ broad components. In order to separate the blended lines of this spectrum, first of all, we fitted the [S \textsc{ii}]$\lambda$6716, 6731 lines with a sum of two Gaussian sets (set 1 and set 2). In each set, we assumed a velocity ($V$) and a width ($\sigma$) for the Gaussians. Thus, each [S \textsc{ii}] line was fitted with two Gaussian functions with different values of $V$ and $\sigma$. The resulting fits, shown in Fig.~\ref{fig7}(c), reproduced the [S \textsc{ii}] lines with good precision. The Gaussian widths (FWHM) of each set, corrected for the instrumental resolution, are 207 $\pm$ 10 and 290 $\pm$ 44 km s$^{-1}$. After that, we fitted the [N \textsc{ii}]$\lambda$6548;6584 + H$\alpha$ lines with another Gaussian sum (Fig.~\ref{fig7}b). In this case, we used two sets (blue and green Gaussians - see Fig.~\ref{fig7}b) of three Gaussians (two Gaussians for each line), each set having the same values of $V$ and $\sigma$ obtained in the [S \textsc{ii}] fit. In other words, the [S \textsc{ii}] lines were taken as a template to determine the main kinematic parameters of the Gaussians used to fit the other lines. A broad Gaussian was added to represent the H$\alpha$ broad component. One may notice that the Gaussians describe the [N \textsc{ii}]+H$\alpha$ lines very well. The Gaussian representing the H$\alpha$ broad component has a luminosity of (5.6 $\pm$ 0.5) $\times$ 10$^5 L_{\sun}$ and its FWHM is indicated in Table ~\ref{tableredshiftgrav}. The H$\beta$ line was fitted with a sum of two narrow Gaussians, with the same kinematic parameters provided by the fit of the [S \textsc{ii}] lines, and with a broad Gaussian, with the same width and velocity obtained from the fit of the H$\alpha$ broad component (Fig.~\ref{fig7}a). Although the H$\beta$ line has a more irregular profile relative to the others, due to the lower signal-to-noise (S/N) of this spectral region, the main characteristics of this line were well reproduced. The velocities ($V$ parameter) relative to the central wavelength of each line were: for the broad component, 375 $\pm$ 6 km s$^{-1}$; for set 1, 15 $\pm$ 6 km s$^{-1}$; and, for set 2, -132 $\pm$ 75 km s$^{-1}$.  

From the Gaussian fits obtained for the spectrum of Region 1 we know that the H$\alpha$ and H$\beta$ line profiles are compatible, since the same parameters were used to fit those lines. Another way to prove this is by subtracting, from the original spectrum, the Gaussian fits representing the narrow components of H$\beta$ and H$\alpha$ + [N \textsc{ii}]$\lambda\lambda$6548;6584. This procedure allowed us to isolate the H$\alpha$ and H$\beta$ broad components. Fig.~\ref{hahblarga} shows the overlap of the profiles of those broad components, in the velocity space. In this figure, the H$\alpha$ flux was divided by 2.81 (which is the minimum value for the H$\alpha$/H$\beta$ ratio, which occurs in the absence of extinction, assuming case B of recombination, with density of 10$^6$ cm$^{-3}$ and temperature equal to 10$^4$ K) for a better visualization. We notice that the ratio between the two broad components is, in fact, compatible with 2.81, which indicates that there is little extinction internal to the AGN's BLR in this galaxy. All the observed extinction is associated with the narrow components of the lines and, consequently, with the narrow-line region (NLR), and had already been corrected, based on H$\alpha$/H$\beta$ values. The widths of the broad components are compatible at 1$\sigma$ level - see Table \ref{tableredshiftgrav}. This result is consistent with that obtained by fitting these lines with Gaussian sums, as explained before.

\begin{figure}
\begin{center}

  \includegraphics[scale=0.3]{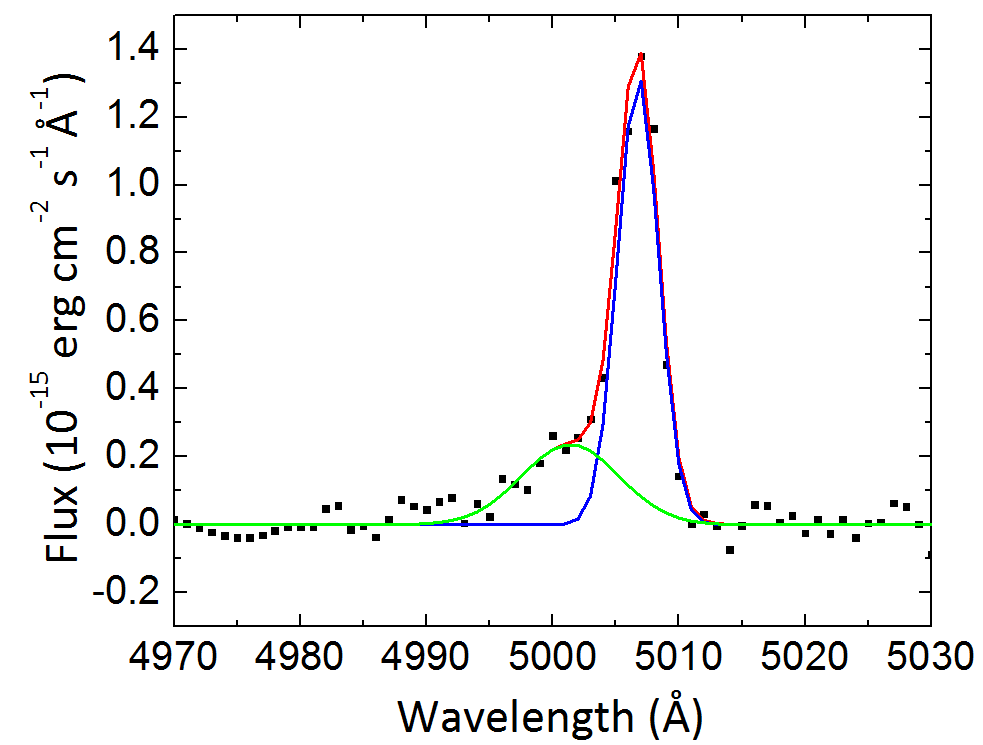}
  \caption{Region 2 [O \textsc{iii}]$\lambda$5007 line Gaussian fit. This was the only line in the region's spectrum that presented considerable contamination coming from the AGN. The black points are the observed data. The green line represents the contamination due to the AGN emission; the blue line, the [O \textsc{iii}]$\lambda$5007 line emitted by the H \textsc{ii} region; and the red curve is the total fit, which consists of the sum of the Gaussian fits. \label{fig8}}
  
\end{center}
\end{figure}

Region 2 spectrum does not have blended lines that require Gaussian fits for a decomposition. However, the [O \textsc{iii}]$\lambda$5007 line presents a probable contamination by the AGN emission, which appears as a prominent blue wing. In order to determine the [O \textsc{iii}]$\lambda$5007 flux without the contamination, we fitted it as a sum of two Gaussians: one representing the contamination and the other the emission of [O \textsc{iii}]$\lambda$5007 coming from Region 2 (see Fig.~\ref{fig8}). The values obtained for the FWHM of the Gaussians representing the [O \textsc{iii}]$\lambda$5007 line and the contamination by the AGN are 227 $\pm$ 12 and 544 $\pm$ 110 km s$^{-1}$, respectively. The FWHM of all other lines is 55 $\pm$ 10 km s$^{-1}$. The velocity ($V$ parameter) relative to the central wavelength of the [O \textsc{iii}]$\lambda$5007 line, for the narrow component, is -4 $\pm$ 5 km s$^{-1}$  and, for the Gaussian representing the contamination, is -326 $\pm$ 63 km s$^{-1}$. The velocity of all other lines is 51 $\pm$ 5 km s$^{-1}$.   

We calculated the following emission-line ratios for the two emitting regions: [O \textsc{iii}]$\lambda$5007/H$\beta$, [N \textsc{ii}]$\lambda$6584/H$\alpha$, ([S \textsc{ii}]$\lambda$6716+$\lambda$6731)/H$\alpha$, [O \textsc{i}]$\lambda$/H$\alpha$ and [S \textsc{ii}]$\lambda$6716/[S \textsc{ii}]$\lambda$6731 (see Table~\ref{tableratio}). Only the narrow components of the lines of the AGN spectrum (Region 1) were taken into account and, in the case of the Region 2 spectrum, the contamination of the [O \textsc{iii}]$\lambda$5007 line by the AGN was not considered. The resulting diagnostic diagrams (Figs ~\ref{fig9}a-c) show the points representing the two emitting regions of NGC 1566 and also the points corresponding to the objects analysed by \citet{ho}. One can see that Region 1 has emission-line ratios typical of a Seyfert galaxy, while Region 2, taking into account the criteria of \citet{kewley1}, can be classified as an H \textsc{ii} region (diagrams b and c of Fig.~\ref{fig9}) or as a transition object (Fig.~\ref{fig9}a). We believe that Region 2 is an H \textsc{ii} region contaminated by the emission from the AGN. Although the contamination was more significant in the [O \textsc{iii}]$\lambda$5007 line and was removed, there can still be contamination in other lines, more difficult to identify, and also to remove. 

We verified from the previous Gaussian fits that the velocity of the H \textsc{ii} region relative to the centre (AGN) is 36 $\pm$ 8 km s$^{-1}$.

\subsection{Analysis of NIR spectra}

As mentioned above, the subtraction of the stellar continuum of the \textit{K}-band data cube used the synthetic stellar spectra obtained with the pPXF method. In the \textit{J}-band data cube, this procedure was carried out by fitting and subtracting splines from the stellar continuum. We then extracted the resulting emission-line spectra from circular regions, with radius equal to half of the FWHM of the PSF of the data cubes, centred on Regions 1 and 2 of the gas cubes obtained in the \textit{J} and \textit{K} bands. The extracted spectra are shown in Figs ~\ref{espectrosj} and ~\ref{espectrosk}. 

\begin{figure}
\begin{center}

  \includegraphics[scale=0.3]{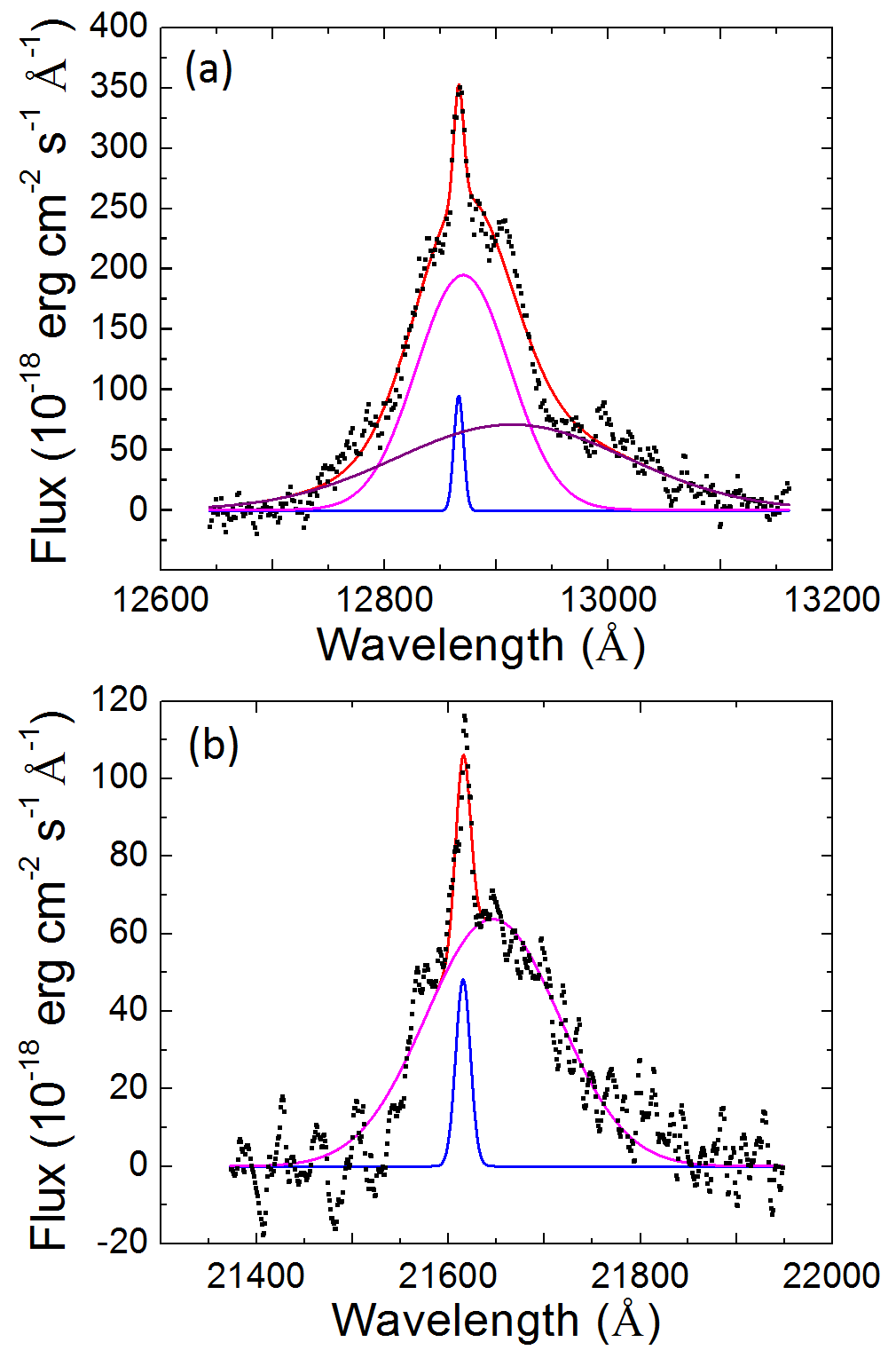}
  \caption{(a) Gaussian fit of the Pa$\beta$ line of Region 1 spectrum. Besides one narrow and one broad component, the fit detected a very broad component (in purple) which does not appear in the fits of all the other lines of Region 1. (b) Gaussian fit of the Br$\gamma$ line of Region 1. \label{ajustepb}}
  
\end{center}
\end{figure}

\begin{figure}
\begin{center}

  \includegraphics[scale=0.3]{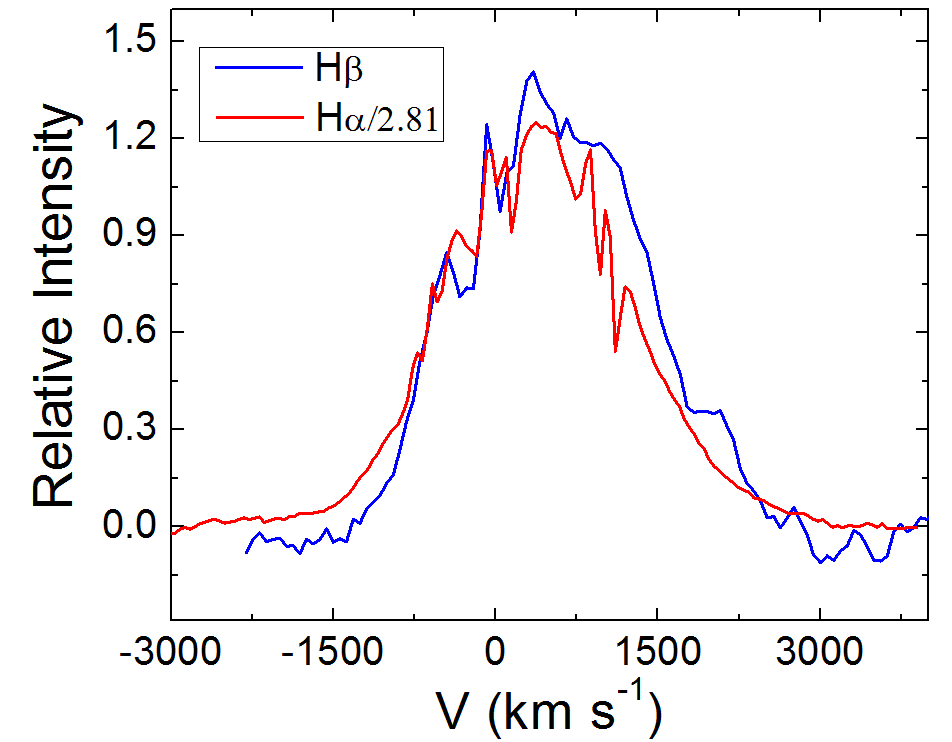}
  \caption{Broad components of the H$\beta$ (in blue) and H$\alpha$ (in red) lines in the velocity space. The H$\alpha$ line flux was divided by 2.81 (minimum value of the H$\alpha$/H$\beta$ ratio, assuming the case B of recombination, with $T=10^4$ K and density of 10$^6$ cm$^{-3}$, which indicates absence of extinction) so that the lines are in the same scale. \label{hahblarga}}
  
\end{center}
\end{figure}

Region 1 spectrum of the \textit{J}-band data cube shows a very broad Pa$\beta$ line and a narrow [Fe \textsc{ii}]$\lambda$12570 line. We fitted the Pa$\beta$ line with a sum of three Gaussians. Fig.~\ref{ajustepb}(a) shows the resulting fitting. Unlike the other fitted lines, the narrow component of the Pa$\beta$ line was satisfactorily represented by only one Gaussian curve (in blue). The broad component corresponds to the sum of two Gaussians (in magenta, the broad component; and in purple, the very broad component, in Fig.~\ref{ajustepb}a) with different FWHMs (see Table \ref{tableredshiftgrav}). The broader Gaussian (in purple) does not appear in the Region 1 spectra of the other data cubes. The FWHM of the  narrow component is 231 $\pm$ 14 km s$^{-1}$ and its velocity is 3 $\pm$ 5 km s$^{-1}$. The velocities of the magenta and purple broad components (see Fig.~\ref{ajustepb}a) are 98 $\pm$ 6 km s$^{-1}$ and 1107 $\pm$ 10 km s$^{-1}$, respectively, and their FWHM are shown in Table \ref{tableredshiftgrav}. 

Region 2 spectrum of the \textit{J}-band data cube shows only one visible emission line: narrow Pa$\beta$, with FWHM of 168 $\pm$ 12 km s$^{-1}$ and velocity of 64 $\pm$ 5 km s$^{-1}$.

\begin{table}
\centering
\caption{Emission-line ratios of Regions 1 (AGN) and 2 (H \textsc{ii} region).}
\label{tableratio}
\begin{tabular}{lll}
\hline
Emission-line ratios                          & Region 1        & Region 2          \\ \hline
{[}O \textsc{iii}{]}$\lambda$5007/H$\beta$              & 6.2 $\pm$ 0.5   & 0.50 $\pm$ 0.06   \\
{[}N \textsc{ii}{]}$\lambda$6584/H$\alpha$              & 1.18 $\pm$ 0.08 & 0.65 $\pm$ 0.05   \\
({[}S \textsc{ii}{]}$\lambda$ 6716+6731)/H$\alpha$      & 0.78 $\pm$ 0.16 & 0.267 $\pm$ 0.017 \\
{[}O \textsc{i}{]}$\lambda$6300/H$\alpha$               & 0.29 $\pm$ 0.03 & 0.024 $\pm$ 0.004 \\
{[}S \textsc{ii}{]}$\lambda$6716/{[}S \textsc{ii}{]}$\lambda$6731 & 0.9 $\pm$ 0.4   & 1.07 $\pm$ 0.08   \\ \hline
\end{tabular}
\end{table}

Region 1 spectrum of the \textit{K}-band data cube shows the H$_2\lambda\lambda$21218, 22234, 24084 and the Br$\gamma$ lines. Similarly to the procedure applied to Pa$\beta$, we fitted the Br$\gamma$ line with a sum of Gaussians, but, in this case, only two Gaussians were necessary: a narrow one, with FWHM of 250 $\pm$ 15 km s$^{-1}$ and $V$ parameter of 5 $\pm$ 5 km s$^{-1}$, and a broad one, with $V$ parameter of 432 $\pm$ 5 km s$^{-1}$ and FWHM shown in Table \ref{tableredshiftgrav}. Although the spectrum is very noisy in this region, the fitting results were satisfactory for our purposes. 

Region 2 spectrum of the \textit{K}-band data cube shows narrow H$_2$ and Br$\gamma$ lines, with velocity of 71 $\pm$ 7 km s$^{-1}$ and FWHM of 124 $\pm$ 12 km s$^{-1}$.

\subsection{Gravitational redshift}

\begin{figure}
\begin{center}

  \includegraphics[scale=0.31]{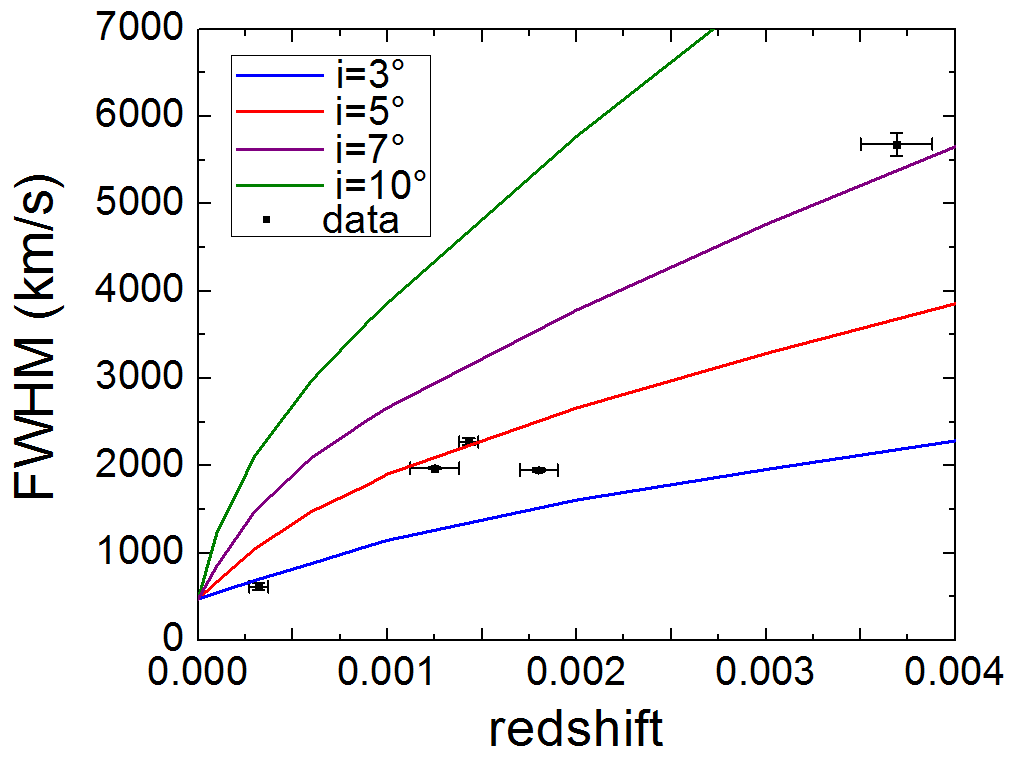}
  \caption{FWHM \textit{versus} redshift of all the broad components of the emission lines presented in Table \ref{tableredshiftgrav}. The curves correspond to the results obtained with a toy model, which considers the emission of a gas ring around a central black hole, with different radii and inclination angles. \label{zfwhm}}
   
\end{center}
\end{figure}

An important aspect to be noted in Table  \ref{tableredshiftgrav} is that there is an evident correlation between the width of the lines (FWHM) and their respective redshifts (see Fig.~\ref{zfwhm}). That can be explained by gravitational redshift: the closer to the AGN, the greater the redshift and the wider the lines, as long as the velocities have a Keplerian nature. The distinct values could be explained by the fact that the lines are emitted at different distances from the central object. This fact might be the result of different emissivities of the many regions of the BLR, or of the AGN's variability, since the spectra were taken in different epochs. 

The hypothesis to explain the red asymmetries in the permitted lines of AGN spectra as being due to the gravitational redshift has already been discussed many times in the literature. \citet{netzer} created a simple model of a rotating disc with velocity ($v$) between 5000 and 15 000 km s$^{-1}$, with some inclination angles and many radii, assuming gravitational redshift, and obtained line profiles shifted to the red. Further works also analysed the gravitational redshift influence in the AGN emission-line profiles, using more detailed models - see \citet{bon} and the references therein.

\begin{table*}
\centering
\caption{FWHMs, redshifts and the $R_{BLR}/R_{S}$ ratios of the broad components of the lines in Region 1 spectra.}
\label{tableredshiftgrav}
\begin{tabular}{llll}
\hline
Emission lines                & FWHM (km s$^{-1}$)           & Redshift ($\times 10^{-3}$) & $R_{BLR}/R_{S}$ \\ \hline
H$\beta$                      & 1950 $\pm$ 20  & 1.80 $\pm$ 0.10             & 279             \\
H$\alpha$                     & 1970 $\pm$ 10  & 1.25 $\pm$ 0.13             & 401             \\
Br$\gamma$                    & 2273 $\pm$ 40  & 1.43 $\pm$ 0.05             & 350             \\
Pa$\beta$ (broad)                    & 615 $\pm$ 40   & 0.32$\pm$ 0.05              & 1563            \\
 Pa$\beta$ (very broad)                           & 5681 $\pm$ 130 & 3.69 $\pm$ 0.19             & 136             \\ \hline
\end{tabular}
\end{table*}

Assuming that the redshift of the broad components of the observed lines is due to gravitational redshift, it is possible to determine the $R_{BLR}/R_S$ ratio from the following equation:

\begin{equation}
1+z = \left ( 1-\frac{R_s}{R_{BLR}} \right )^{-1/2}, \label{eqredgrav}
\end{equation}

\noindent where $z$ is the redshift of the line (in this case, the gravitational redshift), $R_{BLR}$ is the line emitting radius and $R_s$ is the black hole's Schwarzschild radius. The $R_{BLR}/R_S$ ratios calculated for all broad components of the emission lines of the AGN spectrum (which are, therefore, being emitted from the BLR) are shown in Table \ref{tableredshiftgrav}. In order to evaluate whether the AGN's variability can, in fact, result in the observed pattern of the broad components of the emission lines, we elaborated a toy model, assuming the existence of a circular ring of gas, emitter of a determined spectral line, rotating around the central black hole. Along this ring, the line is emitted with a Gaussian profile, with constant width and with the redshifts given by the combination of Keplerian values and gravitation redshift. The final observed line profile corresponds to the superposition of all the line emissions along the ring. The free parameters of this toy model are the ring's inclination relative to the line of sight, $i$, and the gravitational redshift, $z$, which is related to the gas' orbital velocity through the following expression (obtained from the equation \ref{eqredgrav}): 

\begin{equation}
1+z = \left ( 1-\frac{2v^2}{c^2} \right )^{-1/2}. \label{eqvel}
\end{equation}

To each combination of $z$ and $i$ considered in the model, the Gaussian width emitted along the ring was taken as being equal to the necessary value, so that the final profile is compatible with the observed one. The FWHM of all the resulting profiles were obtained with this toy model. The points corresponding to the observed parameters of the broad components, together with the traced curves from the toy model results, are shown in Fig.~\ref{zfwhm}. One may notice a correlation between $z$ and FWHM. The curves obtained from the toy model reproduce reasonably well the observed correlation, as long as the inclination is low ($i \sim 5 \degr$). It is important to emphasize that the different values of $z$ taken into account in the toy model are equivalent to different radii of the emitting gas ring. We are assuming that these different radii values where the line is emitted are a consequence of the AGN's variability. 

We also tried to reproduce the observed correlation (Fig.~\ref{zfwhm}) with a second toy model, assuming the existence of a spherical shell of gas, with only radial velocities, around the central black hole. In this case, however, all the obtained parameters were incompatible with the observed ones.

From the temporal variation of the profiles of the BLR lines, \citet{alloin85} estimated the BLR size as being less than 0.01 pc. If we assume that the BLR radius is half of this value, using the calculated values of the $R_{BLR}/R_s$ ratio (Table \ref{tableredshiftgrav}), we conclude that the upper limit of the black hole's mass is between 1.3 and 4 $\times 10^8$ M$_{\sun}$. The upper limit of the mass calculated by using the broader component of Pa$\beta$ line is 3.4 $\times 10^7$ M$_{\sun}$. Being upper limits, all the values obtained here are compatible with the value of the inner mass calculated from the H$_2$ velocity curve (see section~\ref{sec6}), which is of the order of 10$^7$ M$_{\sun}$, assuming that the inclination angle of this gas rotation is not close to zero. These values are also compatible with those obtained by \citet{woo} and \citet{smajic}. If we assume that the black hole mass is 8.3 $\times 10^6$ M$_{\sun}$ \citep{woo}, the BLR radius is between 1 and 15 light-days, which is within the upper limit of 20 light-days found by \citet{alloin85}.

\section{Gas kinematics} \label{sec6}

\subsection{Highly ionized gas in the optical}

\begin{figure}
\begin{center}
 \includegraphics[scale=0.32]{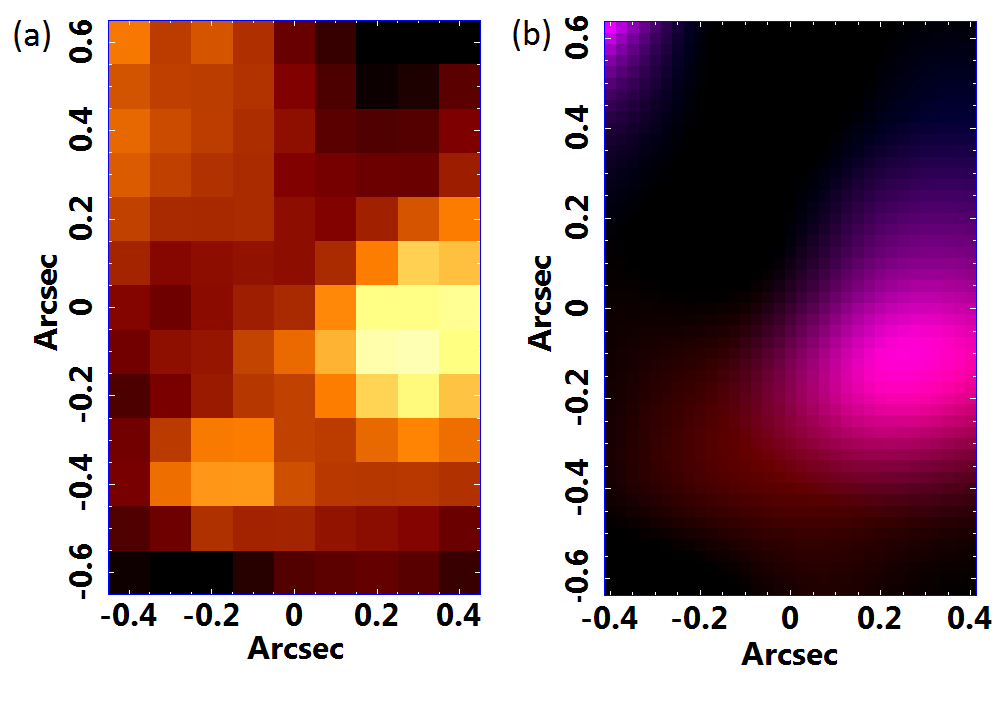}
  \caption{(a) Image of the H$\alpha$/F555W ratio cut in the H \textsc{ii} region. (b) RB composition of the \textit{J}-band SINFONI data cube cut on the H \textsc{ii} region: in red, the image of the red wing of the narrow component of Pa$\beta$ and, in blue, the image of the blue wing of the same component. The orientation is north up and east to the left, as for the GMOS/IFU data cube. \label{kinregiaoHII}}
\end{center}
\end{figure}

\begin{figure*}
\begin{center}

  \includegraphics[angle=0,scale=0.35]{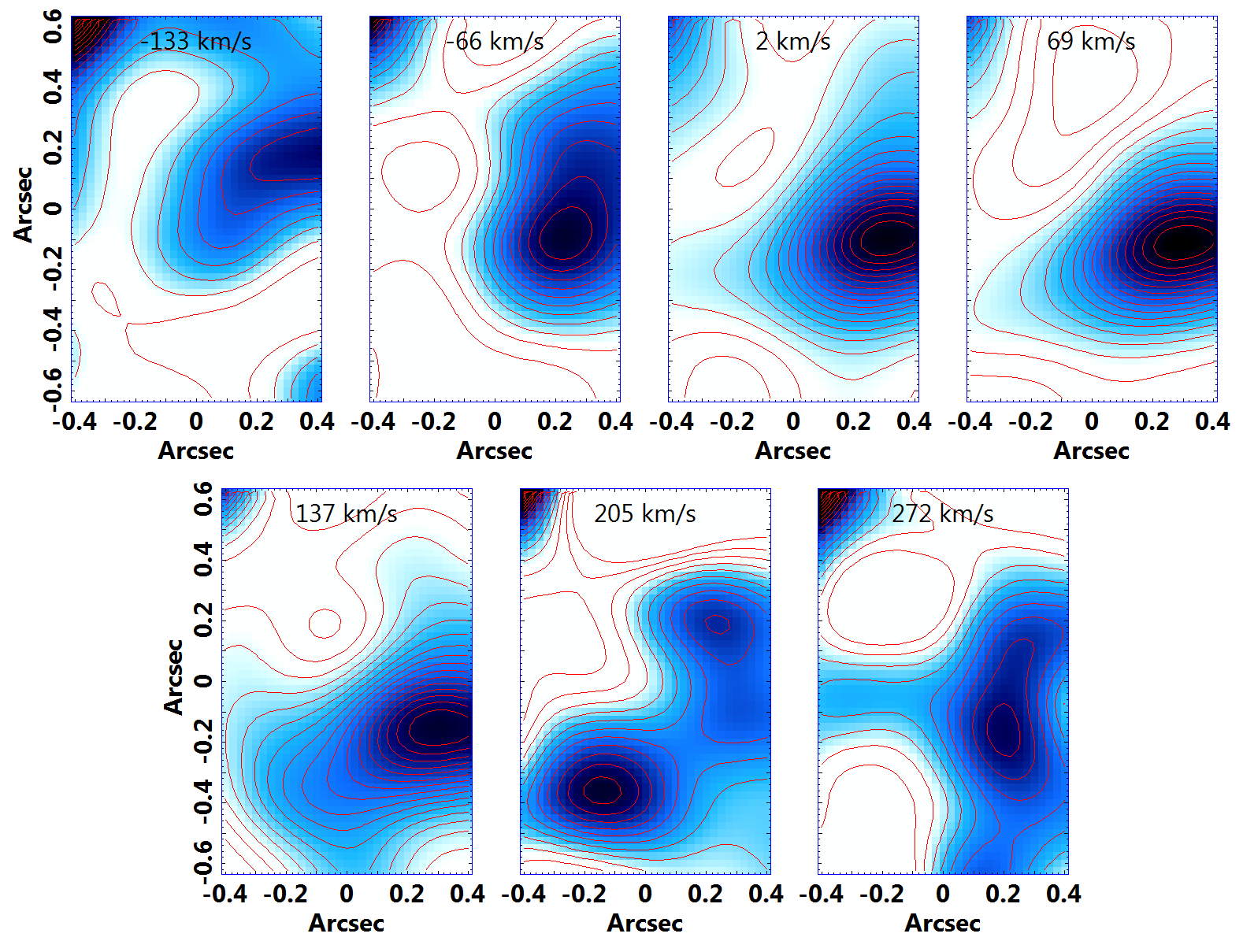}
  \caption{Channel maps of the Pa$\beta$ line in the area corresponding to the H \textsc{ii} region data cube. The isocontours of the positive values (higher than zero) are shown in red. The images were made at intervals of 1.45\AA\ and the velocity values were obtained from the mean wavelength of the intervals relative to the Pa$\beta$ central wavelength of the nuclear spectrum. \label{channelmapregiaoHII}}
  
\end{center}
\end{figure*}

\begin{figure*}
\begin{center}

  \includegraphics[angle=90,scale=0.37]{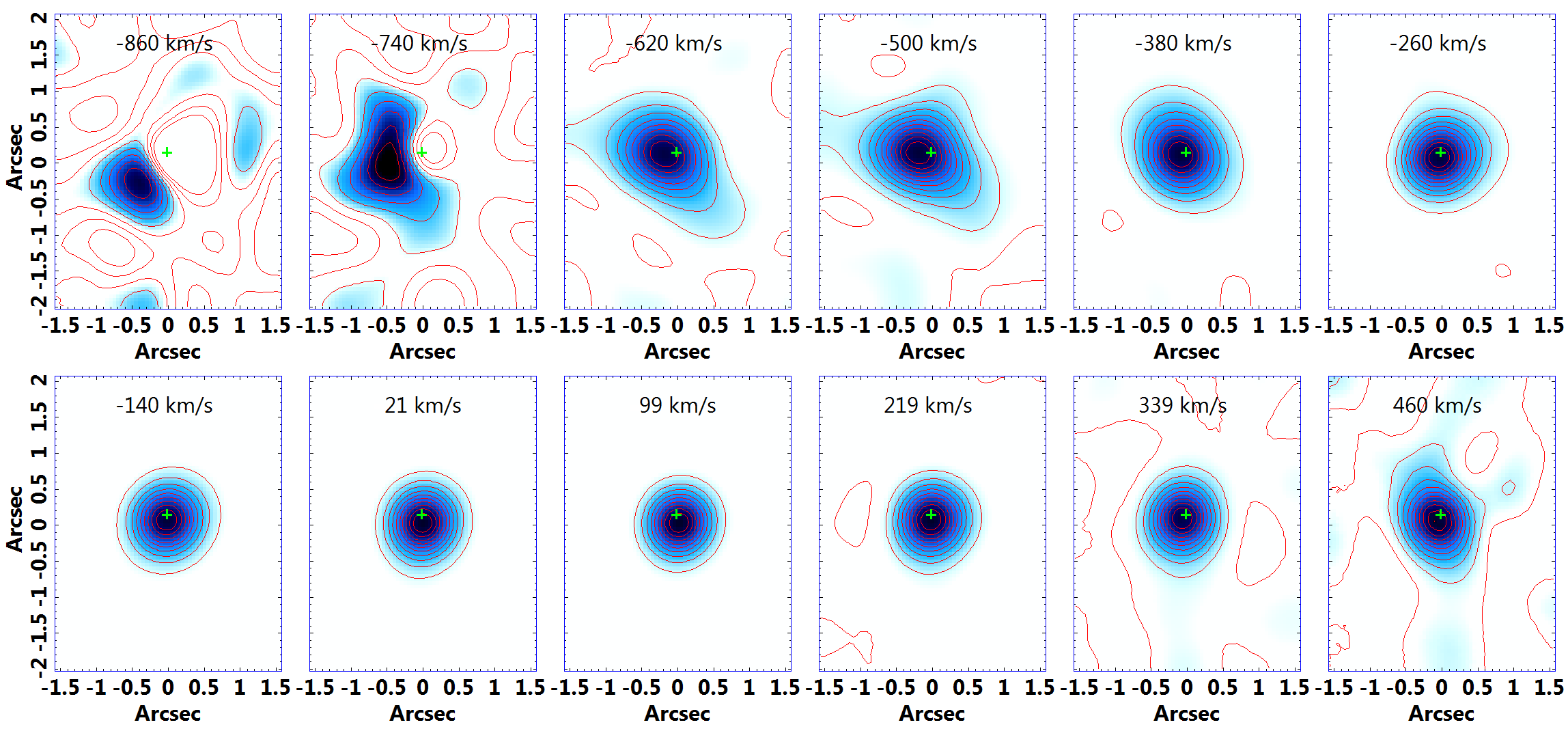}
  \caption{[O \textsc{iii}]$\lambda$5007 channel maps. The isocontours of the positive values are shown in red. The green cross indicates the AGN position and its size represents the 3$\sigma$ uncertainty. The images were taken at intervals of 2\AA\ and the velocity values were obtained from the mean wavelengths of the intervals relative to the rest wavelength of [O \textsc{iii}]$\lambda$5007. \label{channelmapoiii}}
  
\end{center}
\end{figure*}

\begin{figure*}
\begin{center}

  \includegraphics[angle=90,scale=0.38]{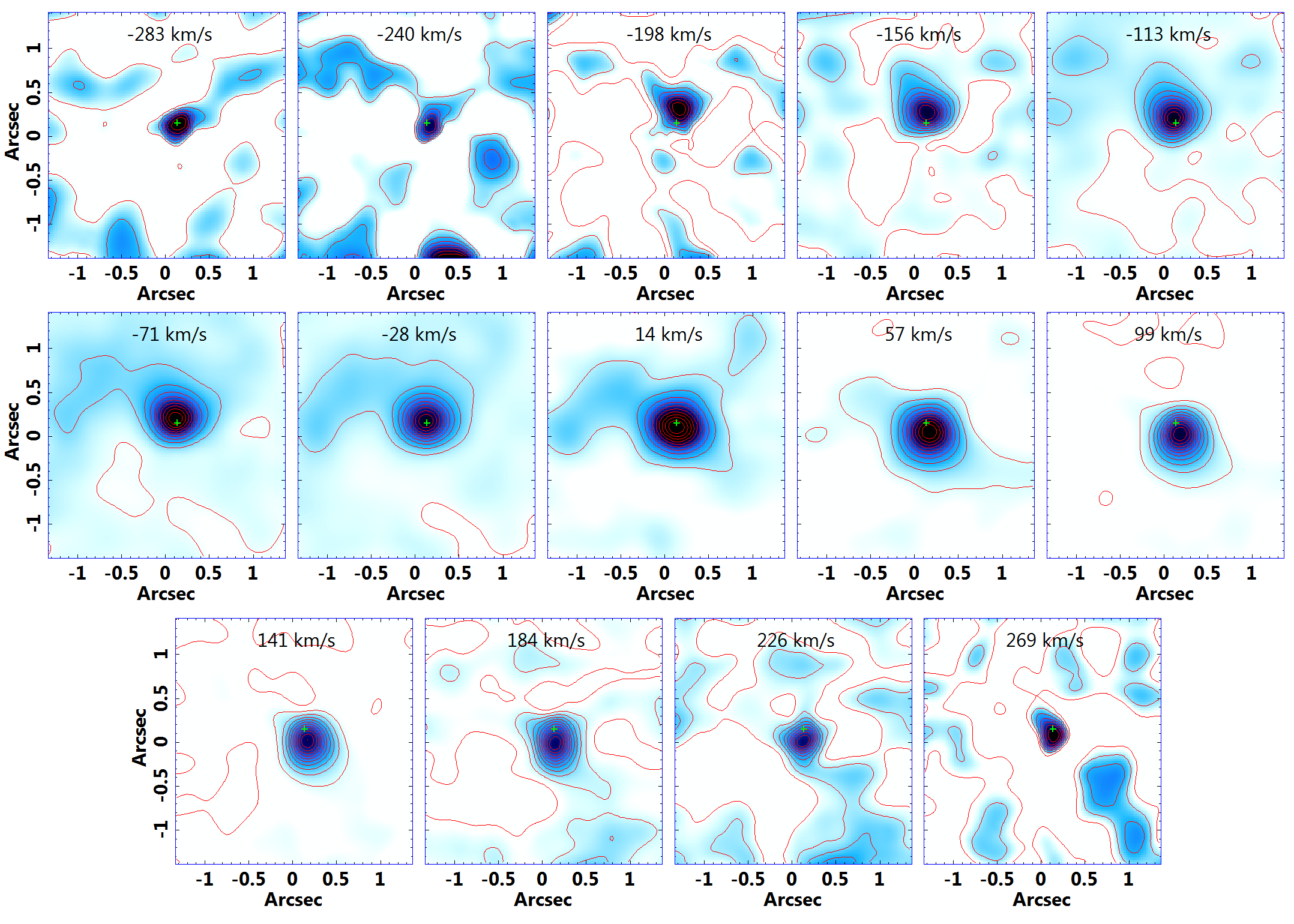}
  \caption {$H_2\lambda$21218 channel maps. The isocontours of the positive values are shown in red. The green cross indicates the position of the AGN and its size represents the 3$\sigma$ uncertainty. The images were taken at intervals of 2\AA\ and the velocities were obtained from the interval's mean wavelength relative to the central wavelength of H$_2\lambda$21218 of the nuclear spectrum. \label{channelmaph2}}
  
\end{center}
\end{figure*}

\begin{figure*}
\begin{center}
 \includegraphics[scale=0.4]{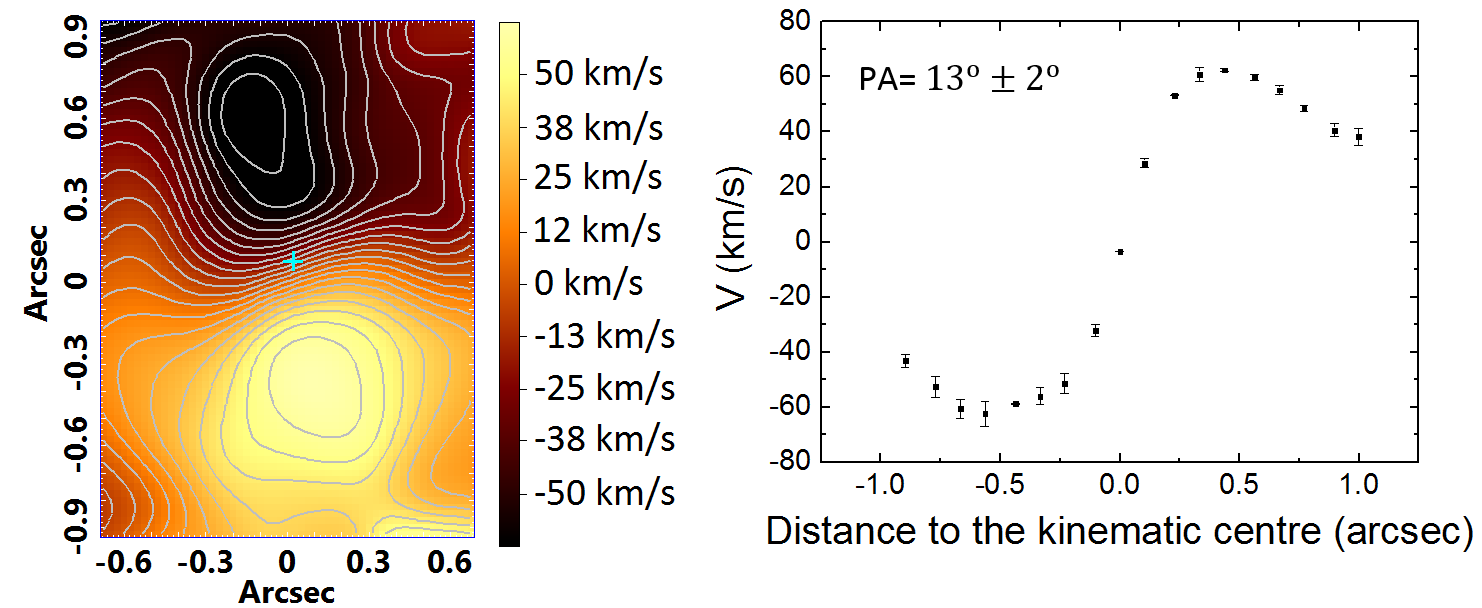}
  \caption{Velocity map of H$_2$ with its velocity curve. The cyan cross indicates the position of the AGN, which was determined from the image of Br$\gamma$. The cross size represents the uncertainty of 3$\sigma$. \label{mapavelh2}}
  
   \end{center}
\end{figure*}
  
  \begin{figure}
\begin{center}
  
  \includegraphics[scale=0.4]{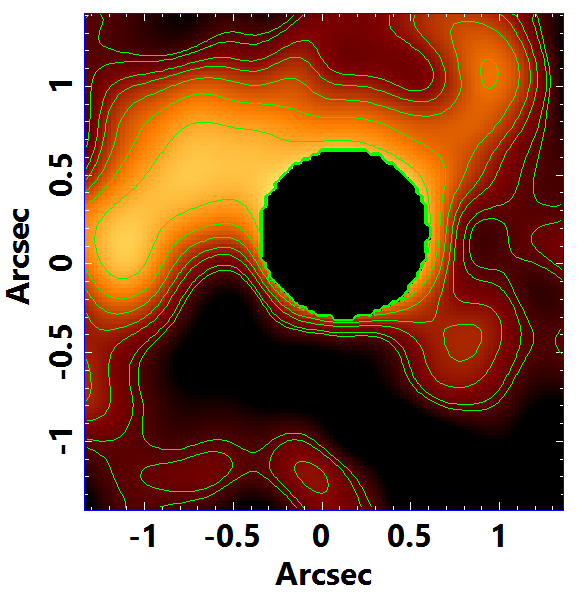}
  \caption{Image of the sum of the three channels with velocities close to zero (-71, -28 and 14 km s$^{-1}$). The spiral is better seen in images of low velocities and slightly blueshifted. To a better visualization and adjustment of LUT (look-up table), the values of the central region were masked and isocontours of positive values are shown in green.\label{espiralh2}}
 
  \end{center}
\end{figure}

\begin{figure*}
\begin{center}

  \includegraphics[scale=0.4]
  {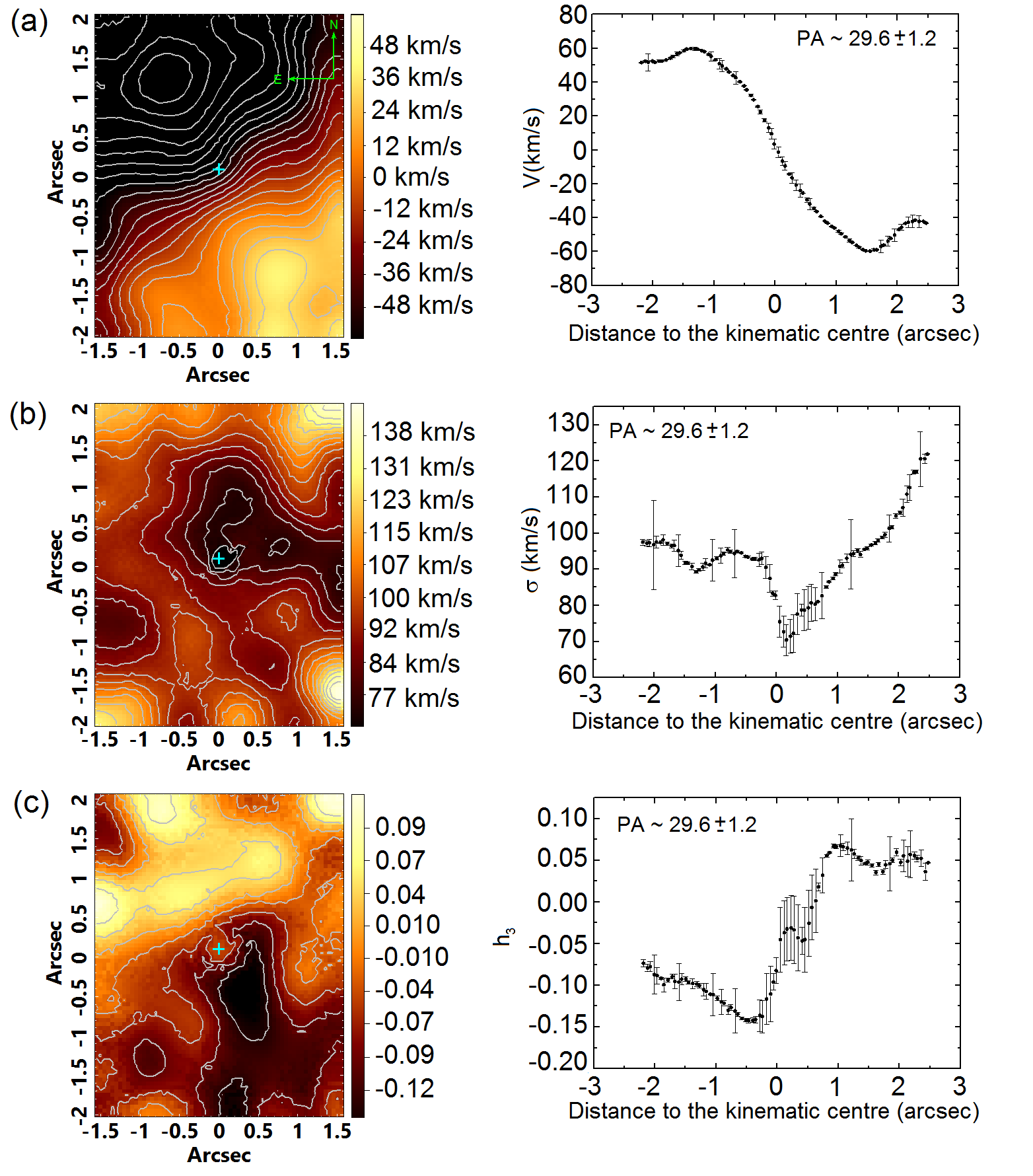}
  \caption{pPXF results: (a) $V_*$ map, (b) $\sigma_*$ map and (c) $h_3$ map, which measures the deviation of the stellar absorption profile relative to a Gaussian profile.(a) also shows the orientation of the observations and all the figures have isocontours and a light blue cross, which indicates the position of the AGN, the size of which depicts the uncertainty of 3$\sigma$ of the position. The radial curve of each map was extracted along the kinematic axis, whose PA is indicated.\label{ppxf}}
  
\end{center}
\end{figure*}

By observing the line profiles of [O \textsc{iii}]$\lambda$5007 from the AGN spectrum and also from the H \textsc{ii} region spectrum (see Fig.~\ref{fig8}), one may notice an asymmetry, with a more pronounced blue wing. In order to analyse in greater detail this line's kinematics, we made the channel maps shown in Fig.~\ref{channelmapoiii}. It is easy to see the existence of an extended emission, with a maximum at southeast from the nucleus, with velocities between -860 and -500 km s$^{-1}$. The emission from clouds with more positive velocities is considerably more compact and centred approximately on the nucleus. A possible explanation for this blueshifted extended emission is that it is associated with an outflow coming from the AGN. The emission associated with this apparent outflow also seems to extend towards Region 2, which suggests a possible contamination of the H \textsc{ii} region emission by the outflow. In this case, the prominent blue wing of the [O \textsc{iii}]$\lambda$5007 line of the H \textsc{ii} region spectrum could be due to this contamination by the outflow. The average position angle (PA) of the first two channels is 122$\degr$ $\pm$ 16$\degr$. We will see next that this value suggests that the molecular gas disc is approximately perpendicular to the outflow.

\subsection{H II Region}

The previous analysis showed that the detected H \textsc{ii} region in the optical data cube of NGC 1566 is in redshift relative to the central AGN. Furthermore, an RB composition obtained from the images of the red and blue wings of the Pa$\beta$ line in the \textit{J}-band data cube (Fig.~\ref{kinregiaoHII}b) shows that this H \textsc{ii} region also seems to have a velocity gradient. The north-west portion of this area (shown in blue in  Fig.~\ref{kinregiaoHII}b) presents lower radial velocities than the southeast portion (shown in red).

In order to perform a more detailed analysis of this apparent velocity gradient, we compared this result with the H$\alpha$/F555W image obtained with \textit{HST} (Fig.~\ref{kinregiaoHII}a). We immediately notice that the two brightest components of this H \textsc{ii} region have different radial velocities (with the brightest component having lower radial velocities than the other one), originating the observed gradient. We also made channel maps of the Pa$\beta$ line in the area corresponding to the H \textsc{ii} region (Fig.~\ref{channelmapregiaoHII}). The result shows with more detail the pattern already observed in the RB composition in Fig. ~\ref{kinregiaoHII}(b). The observed lower velocities (in the brightest component of the H \textsc{ii} region) were $\sim$ -133 km s$^{-1}$, while the highest velocities (in the other component of the H \textsc{ii} region) were $\sim$ 205 km s$^{-1}$. 

\subsection{NIR molecular gas}

The \textit{K}-band data cube revealed a significant emission of the H$_2\lambda$21218 line. In order to analyse the molecular gas kinematics in the central region of NGC 1566, we made a radial velocity map, obtained from the Gaussian fits of the H$_2\lambda$21218 line of each spectrum of the data cube. We verified that an amplitude/noise (\textit{A/N}) ratio greater than 5 was necessary to obtain reliable values of the radial velocity from the Gaussian fit. However, the values of the \textit{A/N} ratio were lower than 5 in a large part of the FOV. So, we re-sized the obtained velocity map, keeping only areas where the \textit{A/N} ratio of the H$_2\lambda$21218 line was greater than 5. We assumed that the systemic velocity is equal to the arithmetic mean of the minimum and maximum velocities of the velocity map. The centre of the velocity map was taken as being equal to the point along the line of nodes where the measured velocity was equal to the systemic velocity, which was subtracted from the map. The result of this subtraction is shown in Fig.~\ref{mapavelh2}, together with the velocity curve extracted along the line of nodes. One may notice that the observed pattern is consistent with a rotation around the kinematic centre, whose position is compatible, at 1$\sigma$ level, with the AGN position (estimated from the image of the broad wing of Br$\gamma$). The PA of the line of nodes of the velocity map is 13$\degr$ $\pm$ 2$\degr$.

The velocity uncertainties were obtained by a Monte Carlo simulation. For that, a representative Gaussian distribution of the spectral noise was estimated. We then created different Gaussian distributions of random noise with the same width of the initial noise distribution. These distributions were added to the initial Gaussian fitted to the line, and new Gaussian fits were made sequentially. At last, the final uncertainty was taken as the standard deviation of the values found for the velocity in all the obtained fits.  

The inner mass calculated from the H$_2$ velocity curve is $M_{int}= (2.67 \pm 0.08)/\sin i \times 10^7 M_{\sun}$, where $i$ is the inclination of the molecular gas rotating disc. 

We also made channel maps of the H$_2\lambda$21218 line, shown in Fig.~\ref{channelmaph2}. The images provide a more detailed visualization of the rotational pattern revealed by the velocity map. Significant emission of H$_2$ was detected at velocities between -283 and 269 km s$^{-1}$. We notice that the channels with velocities close to zero show a pattern similar to a spiral, that is better seen in Fig.~\ref{espiralh2}, which is the sum of the three channels with velocities centred in -71, -28 and 14 km s$^{-1}$. The morphology of this spiral is very similar to the one observed by \citet{comb14} and by \citet{smajic}, who observed the molecular gas spiral from the CO(3-2) and H$_2$ emission, respectively.

\section{Stellar kinematics} \label{sec7}

In order to describe the stellar kinematics of the central region of NGC 1566, we applied the pPXF procedure \citep{cappellari} to the data cube with the emission lines masked, that is, to the data cube containing only the stellar continuum. pPXF is a method that uses a combination of template spectra of a given base, convolved with a Gauss-Hermite expansion, in order to reproduce the stellar spectrum of an object. In this case, a stellar population spectral base created from the Medium-resolution Isaac Newton Telescope Library of Empirical Spectra (MILES, \citealt{blazquez}) was used. This procedure allows one to obtain the stellar radial velocity ($V_*$), the stellar velocity dispersion ($\sigma_*$) and the Gauss-Hermite coefficient $h_3$, which reveals the profile asymmetries of the stellar absorption lines relative to  Gaussian profiles. As this method was applied to each spectrum of the data cube, we obtained maps of all the parameters mentioned above. 

Similarly to what was done in the case of the H$_2$ radial velocity map, the systemic $V_*$ was taken as being equal to the arithmetic mean of the maximum and minimum velocities of the $V_*$ map. We also assumed that the centre of the $V_*$ map was the point along the line of nodes in which the measured velocity is equal to the systemic velocity. The $V_*$ map, after the subtraction of the systemic velocity, together with the $\sigma_*$ and $h_3$ maps and also their extracted curves along the line of nodes of the $V_*$ map, are shown in Fig.~\ref{ppxf}.

The obtained $V_*$ map (Fig.~\ref{ppxf}a) shows a bipolar velocity distribution, with negative values at northeast from the nucleus and positive values at southwest. As observed for the molecular gas, this pattern is consistent with a stellar rotation around the kinematic centre, which is compatible with the AGN position (estimated from the image of the broad wing of H$\alpha$), at 1$\sigma$ level. The PA found for the line of nodes is 29.6$\degr$ $\pm$ 1.2$\degr$. 

The $\sigma_*$ map (Fig.~\ref{ppxf}b) shows an abrupt decrease of the values in the central region. We believe that such decrease is a consequence of imprecisions in the fit caused by the obfuscation of the absorption lines in this area due to the featureless continuum emitted by the AGN. In addition the map shows a generalized decrease of $\sigma_*$ towards the nucleus. The minimum value of $\sigma_*$ is found approximately at the position of the AGN.

The $h_3$ map (Fig.~\ref{ppxf}c) shows an anticorrelation with the $V_*$ map, which is more evident in the extracted curve. This behaviour is typical of a stellar rotation superposed to a stellar background with approximately null radial velocities. The uncertainties are higher at the central region due to, possibly, the obfuscation caused by the featureless continuum, which may have hampered the fits obtained with the pPXF procedure.

The uncertainties of the kinematic parameters obtained with the pPXF were estimated using a Monte Carlo simulation, in a similar way to what was done for the H$_2$ velocity map. First, we subtracted the synthetic spectrum obtained with the pPXF from the original spectrum, for each spaxel of the data cube. We then estimated a representative Gaussian distribution of the noise of the obtained residual spectrum. After that, we created different Gaussian noise distributions with the same width of the initial noise distribution. These distributions were added to the initial synthetic spectrum and the pPXF was applied sequentially to each resulting spectrum. Finally, for each kinematic parameter, the final uncertainty was taken as the standard deviation of the obtained values. 

\section{Discussion} \label{sec8}

NGC 1566 has a well-studied nucleus in the literature, since \citet{devdev61}. We know that this nucleus has a variable activity and we found that, at the time of the observation reported here, it has a Seyfert type 1 activity. When we compare the line profiles of the optical and NIR spectra of the AGN analysed here, we see that there is a significant variation, which we attributed to the variability of this object, since the observations were taken at different epochs. As evidence of this, while the H$\alpha$, H$\beta$ and Br$\gamma$ lines (see Figs~\ref{fig7} and \ref{ajustepb}b) can be decomposed with only one broad component, the Pa$\beta$ line decomposition requires two broad components (Fig.~\ref{ajustepb}a). Another evidence of this variability concerns the relation between the FWHM and the redshift, as shown in the graph of Fig.~\ref{zfwhm}. A change in the AGN activity could result in a variation of the values of the emitting radii of the BLR lines. A simple toy model of a ring of gas rotating around the central black hole, assuming different values for $z$ (implying different emitting radii) and $i$, was capable of reproducing the observed relation between the FWHM and the redshift. All the results indicate low values of $i$, that is, an almost face-on ring. More sophisticated models are needed to characterize in greater detail the detected broad component emitting region. However, such modelling is beyond the scope of this paper.

From the broad components of the H$\alpha$ and H$\beta$ lines isolated with the aid of Gaussian fits (see Fig.~\ref{hahblarga}), we verified that the ratio between them is compatible with 2.81, which is the value corresponding to case B of recombination, with \textit{T}=10$^4$ K and density equals 10$^6$ cm$^{-3}$. This suggests that there is no inner extinction in the BLR. Furthermore, we noticed that the profiles of these lines are compatible; thus, the statement of the discrepancy between the line profiles of H$\alpha$ and H$\beta$, made by \citet{osmer74}, is not consistent with what we see here. This inconsistency is probably due to the fact that the authors worked with lower resolution and data quality. 

The position of the AGN (estimated from the image of the broad component of the H$\alpha$ line) is compatible, at the 1$\sigma$ level, with the stellar bulge centre (determined from the synthetic data cube obtained with the spectral synthesis fits) - see Fig.~\ref{cuboestelar}.

The featureless continuum emission of this AGN is very strong and is concentrated in the central region of the data cube, that is, there is no evidence of scattering (see Appendix \ref{sinteseespectral} and section \ref{secfeatcont}). This emission was so relevant in the data cube that, in the results of the PCA Tomography of the data cube after the deconvolution, eigenvector E2 (which is the second most relevant eigenvector obtained with PCA Tomography) revealed an anticorrelation with the stellar absorption lines in the central region. In view of this result, we subtracted data cubes containing various power laws (with different spectral indexes and PSFs) describing this featureless continuum from the original data cube, in order to nullify this effect in the results provided by PCA Tomography. The best result was obtained with a power law with a spectral index of 1.7 and with a PSF with a FWHM, at the wavelength of H$\alpha$, slightly greater ($\sim$ 0$\arcsec\!\!$.75) than the one corresponding to the image of the broad component of H$\alpha$, as mentioned before. The featureless continuum map provided by the spectral synthesis with the \textsc{starlight} software (see Appendix \ref{sinteseespectral}: Figs~\ref{mapas2_starlight}d and ~\ref{perfilradfeat}) also revealed a central source with a FWHM ($\sim$ 0$\arcsec\!\!$.94) larger than the one obtained with the H$\alpha$ broad component image. A possible explanation is that part of the emission detected with PCA Tomography and with the spectral synthesis comes from hot and young stars, whose spectra (with few absorption lines), in certain circumstances, may resemble a power law. The fact that the FWHM of the central source detected with the spectral synthesis is larger than the one revealed by PCA Tomography suggests that the spectral synthesis was more affected by the resemblance between the featureless continuum and the young stellar population spectra than PCA Tomography. The featureless continuum emission is so important in this object that it compromises the pPXF and \textsc{starlight} (see Appendix \ref{sinteseespectral}) fits in the most central region, having obfuscated the stellar absorption lines and made it impossible both to identify stellar populations and to determine precisely the kinematic parameters. 

\begin{figure}
\begin{center}

  \includegraphics[scale=0.4]{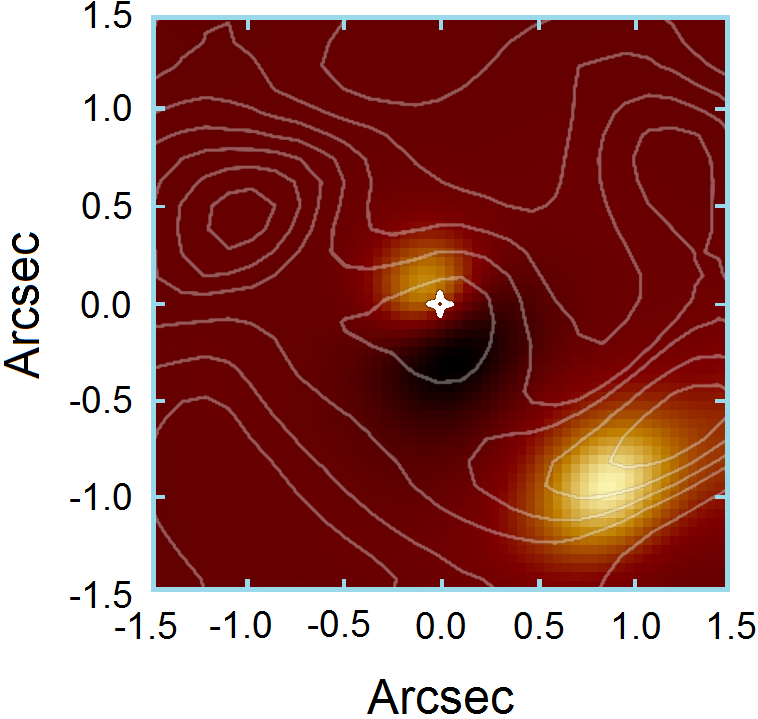}
  \caption{Isocontours of the integrated flux of CO(3-2) obtained by \citealt{comb14} (the white cross is the AGN position, according to the authors) superposed to tomogram 3, which shows the H \textsc{ii} region centred on $\Delta_x =$ 1$\arcsec$ and $\Delta_y=$ -1$\arcsec$. \label{t3_co}}
   
\end{center}
\end{figure}

An H \textsc{ii} region was observed near the central AGN (at a projected distance of 73 pc). This region was also detected by \citet{comb14} and \citet{smajic}. The spectrum of this H \textsc{ii} region seems to present a contamination associated with the AGN emission, which manifests itself as a prominent blue wing of the [O \textsc{iii}]$\lambda$5007 line. Fig.~\ref{t3_co} shows a superposition between the map of the integrated flux of CO(3-2) (molecular gas), obtained by \citet{comb14}, and tomogram 3 (Fig.~\ref{fig2}c), which shows that the H \textsc{ii} region is centred on $\Delta_x =$ 1$\arcsec$ and $\Delta_y=$ -1$\arcsec$. We notice that the H \textsc{ii} region is very close, or even aligned, to one of the concentrations of molecular gas associated with the arms of a spiral in the central region of NGC 1566. The image of H$\alpha$ obtained with the \textit{HST} (Figs~\ref{hst_ha}a and b) revealed that the H \textsc{ii} region detected in the optical data cube is actually formed by many components located along an emitting structure that is also related to a probable spiral arm in the central region of this galaxy. A velocity gradient was observed along this H \textsc{ii} region, which is probably due to this spiral arm's own kinematics (Fig.~\ref{kinregiaoHII}). However, we cannot discard the possibility that an eventual influence of the AGN's outflow (detected over the H \textsc{ii} region as well) is associated with the observed velocity gradient (which can also be the responsible for the prominent blue wing observed in the [O \textsc{iii}]$\lambda$5007 line of this region). When we compared the H$\alpha$/F555W and the velocity gradient images (Figs~\ref{kinregiaoHII}a and b), we saw that there are possibly two regions of gas with different velocities, the superior portion having lower radial velocity values than the inferior portion. The PCA Tomography of the optical data cube showed that the H \textsc{ii} region, as a whole, is in redshift relative to the nucleus, with a radial velocity, also relative to the nucleus, of \textit{V}= 36 $\pm$ 8 km s$^{-1}$. This value was obtained from the analysis of the nucleus and of the H \textsc{ii} region spectra (involving Gaussian fits) extracted from the optical data cube. 

When analysing the \textit{HST} F555W-F814W image (equivalent to \textit{V-I}), one may notice that the areas with redder spectra in the central region present a morphology similar to a spiral (as can be seen in Fig.~\ref{hst_vi}). When we compare the images of Fig.~\ref{hst_vi} with the extinction image provided by the spectral synthesis (see Fig.~\ref{mapas2_starlight}a), we see that the extinction pattern is very similar to the pattern of this spiral, with redder spectra. So we may say that this morphology detected through analysis of the \textit{HST} images is probably due to extinction, possibly by dust, in the areas along the spiral. Similar results were obtained by \citet{mezcua} and \citet{comb14}. As mentioned before, spiral patterns associated with the emission of ionized gas (Fig.~\ref{hst_ha}a) and of molecular gas \citep{comb14} were also observed in the central region of this galaxy.

The H$_2$ velocity map, obtained from the \textit{K}-band data cube (Fig.~\ref{mapavelh2}), revealed the presence of a molecular gas disc around the AGN, as the kinematic centre of this disc is compatible with the AGN position, at 1$\sigma$ level. The existence of a molecular gas disc in the central region of NGC 1566 has already been verified in previous studies. Using images of the CO(3-2) emission obtained with ALMA, \citet{comb14} detected a molecular gas disc, with radius of 3$\arcsec$, which presents a spiral pattern. The same disc was also analysed by \citet{smajic}, using, besides ALMA data, measurements of the H$_2\lambda$21218 line of a \textit{K}-band data cube observed with SINFONI, obtained with a fore-optics with FOV of 8$\arcsec$ $\times$ 8$\arcsec$. 

The PA values of the line of nodes of the molecular gas disc determined by \citet{comb14} and by \citet{smajic} were $\sim 44\degr$ and $\sim 45\degr$, respectively. These values are not compatible with the one obtained in this present work (PA = $13\degr \pm 2\degr$), even at the 3$\sigma$ level. The reason of this discrepancy probably concerns the fact that the H$_2$ velocity map analysed here has a smaller FOV than the velocity maps analysed by \citet{comb14} and \citet{smajic}. Therefore, only the central part of the molecular gas disc analysed in these works was seen here. This result reveals that the PA of the line of nodes of this molecular disc varies with the distance from the nucleus. In fact, \citet{smajic} had already detected an apparent variation in the PA of the line of nodes in the inner parts of the molecular disc, but the greater spatial resolution of the data analysed here allows a better visualization of the kinematics of those regions.\citet{smajic} also say that the apparent variation of the PA can be due to the bar or to an spiral density wave. The authors, indeed, observed a molecular gas spiral in the central region and this structure leads to a perturbation in the molecular gas rotation pattern. We detected a very similar spiral (see Fig.~\ref{espiralh2}), which \citet{smajic} observed in a slightly large scale, and, based on what has been said, the kinematic perturbation caused by this spiral can explain the variations in the PA of the molecular disc line of nodes. Another possible cause for this variation is the observed outflow, but the PA associated with this outflow ($122\degr \pm 16\degr$) does not indicate any relation to the molecular gas kinematics in the central region, which makes this scenario improbable. A molecular gas flux through a bar can also change the PA of the line of nodes of the molecular disc, however, the observed molecular gas kinematic pattern in the central region of this galaxy seems much more compatible with an spiral wave (\citealt{comb14}; \citealt{smajic}) than with a coherent flux of gas through a bar. Thus, the hypothesis of the bar influence is also unlikely. Finally, one cannot discard the possibility of that PA variation being due to a warped disc. Nevertheless, redoing this observation with full adaptive optics, to obtain a better resolution, would be necessary to find out if this emission of molecular gas has some relation with the torus.

The emitting region of ionized gas (with radial velocity, relative to the nucleus, between -860 and -740 km s$^{-1}$) southeast from the central AGN, detected by the channel maps of the [O \textsc{iii}]$\lambda$ 5007 line (Fig.~\ref{channelmapoiii}), is probably associated with an outflow from the AGN. The apparent direction of this possible outflow (with a mean PA of 122$\degr$ $\pm$ 16$\degr$) is approximately perpendicular to the line of nodes of the radial velocity map obtained from the  H$_2\lambda$21218 line. This direction also seems to be aligned with the extended emission detected in the H$\alpha$/F555W image (Fig.~\ref{hst_ha}). This same extended emission, probably associated with the NLR of the AGN, was observed by \citet{schmitt}. A possible scenario to explain these observations is the following: the detected molecular disc corresponds to an extension of a type torus/disc structure more compact and near the AGN. This structure collimates the emission (and also the outflows) from the AGN, originating the observed morphology. This scenario is compatible with the  Unified Model. Although the probable outflow has been detected mainly southeast from the AGN, as mentioned before, the outflow also seems to extend towards Region 2 (the H \textsc{ii} region). This may indicate some influence of this phenomenon on the H \textsc{ii} region and, consequently, could explain the prominent blue wing of the [O \textsc{iii}]$\lambda$5007 line, detected in the spectrum of this area, which was also observed in many emission lines of Region 1 spectrum, associated with the AGN. 

The $V_*$ and $\sigma_*$ maps obtained from the optical data cube (Fig.~\ref{ppxf}) are consistent with what was found by \citet{smajic}. However, the $\sigma_*$ map analysed here seems to be slightly less noisy and allows a better visualization of the $\sigma_*$ drop, which was not discussed by \citet{smajic}. $\sigma_*$ drops in galactic nuclei have been observed with increasing frequency \citep[e.g.,][]{ems01,zeu02,mar03,sha03,fal06,gan06}. The most accepted model to explain such behaviour in the values of $\sigma_*$ assumes that the stars in the nuclear region of the galaxy were formed from circumnuclear cold gas. The recently born stars remain with the velocity dispersion values of their progenitor gas clouds, thus giving rise to stellar populations with low values of $\sigma_*$. The observed stellar spectrum and, consequently, the calculated values of $\sigma_*$ are dominated by the younger stars, originating the \textit{$\sigma_*$-drop} phenomenon \citep{woz03,all05,all06}. Although part of the observed \textit{$\sigma_*$ drop} in the map of Fig.~\ref{ppxf} may be caused by the obfuscation of the absorption lines due to the featureless continuum emission from the AGN, it is probable that the model described above is applicable to the central region of NGC 1566. It is also possible that these same young stars are responsible for part of the featureless continuum emission, since its PSF, as mentioned before, is larger than the PSF obtained with the image of the broad component of H$\alpha$ (see Fig.~\ref{perfilradfeat} and section \ref{secfeatcont}). The difference  between the PA values obtained for the $V_*$ and H$_2\lambda$21218 velocity maps suggests that the molecular gas and the stellar discs are not coplanar.

\section{Conclusions}\label{sec9}

We analysed an optical data cube of the central region of NGC 1566 (obtained with the IFU of GMOS) and also the NIR data cubes in the \textit{J} and \textit{K} bands (retrieved from the SINFONI archive), in order to evaluate the properties of the emission-line spectrum, of the featureless continuum, of the stellar kinematics and of the ionized and molecular gas kinematics. The main conclusions of our analysis are the following:

(1) Two emitting regions were found: the central region, whose spectrum is that of a Seyfert 1, and the region centred on $\Delta_x =$ 1$\arcsec$ and $\Delta_y=$ -1$\arcsec$, with a spectrum typical of an H \textsc{ii} region. The projected distance between these two regions is 73.3 $\pm$ 1.9 pc. Furthermore, the H \textsc{ii} region is in redshift relative to the nucleus, with a radial velocity (also relative to the nucleus) of 36 $\pm$ 8 km s$^{-1}$.

(2) Analysis of \textit{HST} images revealed that the H \textsc{ii} region is composed of many substructures and is located along an apparent spiral arm. Although it is, as a whole, in redshift, the H \textsc{ii} region has a velocity gradient, which may be due to the kinematics of this probable spiral arm, although one cannot discard the possibility of an influence of the AGN outflow on the H \textsc{ii} region.

(3) An outflow from the AGN was detected with the analysis of the [O \textsc{iii}]$\lambda$5007 line kinematics, and it seems to extend towards the H \textsc{ii} region, which explains the prominent blue wing that was seen in this region's spectrum. The mean PA found for this outflow was 122$\degr$ $\pm$ 16$\degr$. The \textit{HST} image of H$\alpha$/F555W suggests the existence of an elongated emission that seems to coincide with the position of the outflow.

(4) PCA Tomography revealed an obfuscation of the nuclear absorption lines of this object by the emission of the featureless continuum coming from the AGN. An analysis, also made with PCA Tomography, revealed that this featureless continuum may be represented by a power law with a spectral index of 1.7. The FWHM of the PSF, at the wavelength of H$\alpha$, corresponding to the featureless continuum emission, $\sim$ 0$\arcsec\!\!$.75, is larger than the FWHM of the PSF of the data cube, 0$\arcsec\!\!$.66, estimated from the image of the broad component of H$\alpha$, suggesting that part of this emission is associated not only with the AGN, but also with hot and young stars in the central region, whose spectrum can be similar to a power-law. The power-law map corresponding to the featureless continuum obtained with the spectral synthesis revealed a central source with a FWHM of $\sim$ 0$\arcsec\!\!$.94, also larger than the FWHM of the PSF of the data cube.

(5) The kinematic maps provided by the pPXF revealed the presence of a stellar disc rotating around the AGN (the position of the kinematic centre is compatible, at 2$\sigma$ level, with the position of the AGN). One may notice a decrease in the velocity dispersion towards the nucleus, which may be due to the presence of stars formed from cold gas clouds that inherited the low values of velocity dispersion of this gas. Possibly, these young stars are the same ones responsible for the emission of the fake featureless continuum mentioned before. 

(6) An H$_2$ rotating disc was identified in the \textit{K}-band data cube. This disc is rotating around the central region and the PA of its line of nodes  is $\sim$ 13$\degr$ $\pm$ 2$\degr$. This PA is not compatible with the values obtained for the same disc (in larger scales) by \citet{comb14} and \citet{smajic}. This may indicate that the PA of the line of nodes of this disc varies with the distance from the nucleus and is perhaps due to the spiral molecular gas perturbation in the inner central region.

(7) We detected a slightly blueshifted molecular gas spiral from the images of H$_2\lambda$21218.

(8) The line of nodes observed in the H$_2$ velocity map is approximately perpendicular to the probable outflow of the AGN, which may indicate that this molecular disc corresponds to the extension of a more compact structure of a torus/disc type, which collimates the outflow of the AGN. This scenario is consistent with the Unified Model. 

(9) From the analysis of the BLR emission lines in the AGN spectra of the optical and NIR data cubes, we saw first that there is a variation in the optical and NIR line profiles. There is also a relevant difference between Pa$\beta$ and all others line profiles, and it was possible to identify a second broad component in Pa$\beta$. As the observations were made at different epochs, it is possible that the difference in the profiles is due to the variability of the AGN activity. In addition, we also see that the FWHMs of the lines increase with the redshift. This relation can be explained overall by a simple toy model of a gas ring rotating around a central black hole, with small inclination angles, assuming the influence of the gravitational redshift and a variation of the distances of the emitting clouds from the centre, the latter caused by the variability of the AGN. If the upper limit of the BLR size is 0.01 pc, as determined by \citet{alloin85}, we conclude that the upper limit for the black hole mass is between 3.4 $\times 10^7$ and 4 $\times 10^8$ M$_{\sun}$.

\section*{Acknowledgements}

This work is based on observations obtained at the Gemini Observatory (processed using the Gemini \textsc{iraf} package), which is operated
by the Association of Universities for Research in Astronomy, Inc., under a cooperative agreement with the NSF on behalf of the Gemini partnership: the National Science Foundation (United States), the National Research Council (Canada), CONICYT (Chile), the Australian Research Council (Australia), Minist\'erio da Ci\^encia, Tecnologia e Inova\c{c}\~ao (Brazil) and Ministerio de Ciencia, Tecnolog\'ia e Innovaci\'on Productiva (Argentina). A special thanks to Dr. D. May for constructing Fig.\ref{t3_co} and to CNPq (Conselho Nacional de Desenvolvimento Cient\'ifico e Tecnol\'ogico), CAPES (Coordena\c{c}\~ao de Aperfei\c{c}oamento de Pessoal de N\'ivel Superior) and FAPESP (Funda\c{c}\~ao de Amparo \`a Pesquisa do Estado de S\~ao Paulo), under grant 2011/51680-6, for supporting this work.




\bibliographystyle{mnras}
\bibliography{references} 

\begin{thebibliography}{}
\makeatletter
\relax
\def\mn@urlcharsother{\let\do\@makeother \do\$\do\&\do\#\do\^\do\_\do\%\do\~}
\def\mn@doi{\begingroup\mn@urlcharsother \@ifnextchar [ {\mn@doi@}
  {\mn@doi@[]}}
\def\mn@doi@[#1]#2{\def\@tempa{#1}\ifx\@tempa\@empty \href
  {http://dx.doi.org/#2} {doi:#2}\else \href {http://dx.doi.org/#2} {#1}\fi
  \endgroup}
\def\mn@eprint#1#2{\mn@eprint@#1:#2::\@nil}
\def\mn@eprint@arXiv#1{\href {http://arxiv.org/abs/#1} {{\tt arXiv:#1}}}
\def\mn@eprint@dblp#1{\href {http://dblp.uni-trier.de/rec/bibtex/#1.xml}
  {dblp:#1}}
\def\mn@eprint@#1:#2:#3:#4\@nil{\def\@tempa {#1}\def\@tempb {#2}\def\@tempc
  {#3}\ifx \@tempc \@empty \let \@tempc \@tempb \let \@tempb \@tempa \fi \ifx
  \@tempb \@empty \def\@tempb {arXiv}\fi \@ifundefined
  {mn@eprint@\@tempb}{\@tempb:\@tempc}{\expandafter \expandafter \csname
  mn@eprint@\@tempb\endcsname \expandafter{\@tempc}}}

\bibitem[\protect\citeauthoryear{{Ag{\"u}ero}, {D{\'{\i}}az}  \&
  {Bajaja}}{{Ag{\"u}ero} et~al.}{2004}]{ag04}
{Ag{\"u}ero} E.~L.,  {D{\'{\i}}az} R.~J.,   {Bajaja} E.,  2004, \mn@doi [\aap]
  {10.1051/0004-6361:20031644}, \href
  {http://adsabs.harvard.edu/abs/2004A%26A...414..453A} {414, 453}

\bibitem[\protect\citeauthoryear{{Allard}, {Peletier}  \& {Knapen}}{{Allard}
  et~al.}{2005}]{all05}
{Allard} E.~L.,  {Peletier} R.~F.,   {Knapen} J.~H.,  2005, \mn@doi [\apjl]
  {10.1086/498264}, \href {http://adsabs.harvard.edu/abs/2005ApJ...633L..25A}
  {633, L25}

\bibitem[\protect\citeauthoryear{{Allard}, {Knapen}, {Peletier}  \&
  {Sarzi}}{{Allard} et~al.}{2006}]{all06}
{Allard} E.~L.,  {Knapen} J.~H.,  {Peletier} R.~F.,   {Sarzi} M.,  2006,
  \mn@doi [\mnras] {10.1111/j.1365-2966.2006.10751.x}, \href
  {http://adsabs.harvard.edu/abs/2006MNRAS.371.1087A} {371, 1087}

\bibitem[\protect\citeauthoryear{{Alloin}, {Pelat}, {Phillips}  \&
  {Whittle}}{{Alloin} et~al.}{1985}]{alloin85}
{Alloin} D.,  {Pelat} D.,  {Phillips} M.,   {Whittle} M.,  1985, \mn@doi [\apj]
  {10.1086/162783}, \href {http://adsabs.harvard.edu/abs/1985ApJ...288..205A}
  {288, 205}

\bibitem[\protect\citeauthoryear{{Alloin}, {Pelat}, {Phillips}, {Fosbury}  \&
  {Freeman}}{{Alloin} et~al.}{1986}]{alloin}
{Alloin} D.,  {Pelat} D.,  {Phillips} M.~M.,  {Fosbury} R.~A.~E.,   {Freeman}
  K.,  1986, \mn@doi [\apj] {10.1086/164475}, \href
  {http://adsabs.harvard.edu/abs/1986ApJ...308...23A} {308, 23}

\bibitem[\protect\citeauthoryear{{Beckman}, {Bransgrove}  \&
  {Phillips}}{{Beckman} et~al.}{1986}]{beckman}
{Beckman} J.~E.,  {Bransgrove} S.~G.,   {Phillips} J.~P.,  1986, \aap, \href
  {http://adsabs.harvard.edu/abs/1986A%26A...157...49B} {157, 49}

\bibitem[\protect\citeauthoryear{{Bon}, {Bon}, {Marziani}  \&
  {Jovanovi{\'c}}}{{Bon} et~al.}{2015}]{bon}
{Bon} N.,  {Bon} E.,  {Marziani} P.,   {Jovanovi{\'c}} P.,  2015, \mn@doi
  [\apss] {10.1007/s10509-015-2555-5}, \href
  {http://adsabs.harvard.edu/abs/2015Ap%26SS.360...41B} {360, 7}

\bibitem[\protect\citeauthoryear{{Bottema}}{{Bottema}}{1993}]{bottema}
{Bottema} R.,  1993, \aap, \href
  {http://adsabs.harvard.edu/abs/1993A%26A...275...16B} {275, 16}

\bibitem[\protect\citeauthoryear{{Cappellari} \& {Emsellem}}{{Cappellari} \&
  {Emsellem}}{2004}]{cappellari}
{Cappellari} M.,  {Emsellem} E.,  2004, \mn@doi [\pasp] {10.1086/381875}, \href
  {http://adsabs.harvard.edu/abs/2004PASP..116..138C} {116, 138}

\bibitem[\protect\citeauthoryear{{Cardelli}, {Clayton}  \& {Mathis}}{{Cardelli}
  et~al.}{1989}]{cardelli}
{Cardelli} J.~A.,  {Clayton} G.~C.,   {Mathis} J.~S.,  1989, \mn@doi [\apj]
  {10.1086/167900}, \href {http://adsabs.harvard.edu/abs/1989ApJ...345..245C}
  {345, 245}

\bibitem[\protect\citeauthoryear{{Clavel} \& {Joly}}{{Clavel} \&
  {Joly}}{1984}]{joly}
{Clavel} J.,  {Joly} M.,  1984, \aap, \href
  {http://adsabs.harvard.edu/abs/1984A%26A...131...87C} {131, 87}

\bibitem[\protect\citeauthoryear{{Clavel} et~al.,}{{Clavel}
  et~al.}{2000}]{clavel}
{Clavel} J.,  et~al., 2000, \aap, \href
  {http://adsabs.harvard.edu/abs/2000A%26A...357..839C} {357, 839}

\bibitem[\protect\citeauthoryear{{Combes} et~al.,}{{Combes}
  et~al.}{2014}]{comb14}
{Combes} F.,  et~al., 2014, \mn@doi [\aap] {10.1051/0004-6361/201423433}, \href
  {http://adsabs.harvard.edu/abs/2014A%26A...565A..97C} {565, A97}

\bibitem[\protect\citeauthoryear{{Comer{\'o}n}, {Knapen}, {Beckman},
  {Laurikainen}, {Salo}, {Mart{\'{\i}}nez-Valpuesta}  \& {Buta}}{{Comer{\'o}n}
  et~al.}{2010}]{comeron}
{Comer{\'o}n} S.,  {Knapen} J.~H.,  {Beckman} J.~E.,  {Laurikainen} E.,  {Salo}
  H.,  {Mart{\'{\i}}nez-Valpuesta} I.,   {Buta} R.~J.,  2010, \mn@doi [\mnras]
  {10.1111/j.1365-2966.2009.16057.x}, \href
  {http://adsabs.harvard.edu/abs/2010MNRAS.402.2462C} {402, 2462}

\bibitem[\protect\citeauthoryear{{Comte} \& {Duquennoy}}{{Comte} \&
  {Duquennoy}}{1982}]{comte82}
{Comte} G.,  {Duquennoy} A.,  1982, \aap, \href
  {http://adsabs.harvard.edu/abs/1982A%26A...114....7C} {114, 7}

\bibitem[\protect\citeauthoryear{{Davies} et~al.,}{{Davies}
  et~al.}{2016}]{davies}
{Davies} R.~L.,  et~al., 2016, \mn@doi [\apj] {10.3847/0004-637X/824/1/50},
  \href {http://adsabs.harvard.edu/abs/2016ApJ...824...50D} {824, 50}

\bibitem[\protect\citeauthoryear{{Ehle}, {Beck}, {Haynes}, {Vogler}, {Pietsch},
  {Elmouttie}  \& {Ryder}}{{Ehle} et~al.}{1996}]{ehle}
{Ehle} M.,  {Beck} R.,  {Haynes} R.~F.,  {Vogler} A.,  {Pietsch} W.,
  {Elmouttie} M.,   {Ryder} S.,  1996, \aap, \href
  {http://adsabs.harvard.edu/abs/1996A%26A...306...73E} {306, 73}

\bibitem[\protect\citeauthoryear{{Elvis}, {Fassnacht}, {Wilson}  \&
  {Briel}}{{Elvis} et~al.}{1990}]{elvis}
{Elvis} M.,  {Fassnacht} C.,  {Wilson} A.~S.,   {Briel} U.,  1990, \mn@doi
  [\apj] {10.1086/169210}, \href
  {http://adsabs.harvard.edu/abs/1990ApJ...361..459E} {361, 459}

\bibitem[\protect\citeauthoryear{{Emsellem}, {Greusard}, {Combes}, {Friedli},
  {Leon}, {P{\'e}contal}  \& {Wozniak}}{{Emsellem} et~al.}{2001}]{ems01}
{Emsellem} E.,  {Greusard} D.,  {Combes} F.,  {Friedli} D.,  {Leon} S.,
  {P{\'e}contal} E.,   {Wozniak} H.,  2001, \mn@doi [\aap]
  {10.1051/0004-6361:20000523}, \href
  {http://adsabs.harvard.edu/abs/2001A%26A...368...52E} {368, 52}

\bibitem[\protect\citeauthoryear{{Erwin}}{{Erwin}}{2004}]{erwin}
{Erwin} P.,  2004, \mn@doi [\aap] {10.1051/0004-6361:20034408}, \href
  {http://adsabs.harvard.edu/abs/2004A%26A...415..941E} {415, 941}

\bibitem[\protect\citeauthoryear{{Falc{\'o}n-Barroso}
  et~al.,}{{Falc{\'o}n-Barroso} et~al.}{2006}]{fal06}
{Falc{\'o}n-Barroso} J.,  et~al., 2006, \mn@doi [\mnras]
  {10.1111/j.1365-2966.2006.10261.x}, \href
  {http://adsabs.harvard.edu/abs/2006MNRAS.369..529F} {369, 529}

\bibitem[\protect\citeauthoryear{{Ganda}, {Falc{\'o}n-Barroso}, {Peletier},
  {Cappellari}, {Emsellem}, {McDermid}, {de Zeeuw}  \& {Carollo}}{{Ganda}
  et~al.}{2006}]{gan06}
{Ganda} K.,  {Falc{\'o}n-Barroso} J.,  {Peletier} R.~F.,  {Cappellari} M.,
  {Emsellem} E.,  {McDermid} R.~M.,  {de Zeeuw} P.~T.,   {Carollo} C.~M.,
  2006, \mn@doi [\mnras] {10.1111/j.1365-2966.2005.09977.x}, \href
  {http://adsabs.harvard.edu/abs/2006MNRAS.367...46G} {367, 46}

\bibitem[\protect\citeauthoryear{{Garrison} \& {Walborn}}{{Garrison} \&
  {Walborn}}{1974}]{gw74}
{Garrison} R.~F.,  {Walborn} N.~R.,  1974, \jrasc, \href
  {http://adsabs.harvard.edu/abs/1974JRASC..68..117G} {68, 117}

\bibitem[\protect\citeauthoryear{{Georgiev} \& {B{\"o}ker}}{{Georgiev} \&
  {B{\"o}ker}}{2014}]{georgiev}
{Georgiev} I.~Y.,  {B{\"o}ker} T.,  2014, \mn@doi [\mnras]
  {10.1093/mnras/stu797}, \href
  {http://adsabs.harvard.edu/abs/2014MNRAS.441.3570G} {441, 3570}

\bibitem[\protect\citeauthoryear{{Gonzalez} \& {Woods}}{{Gonzalez} \&
  {Woods}}{2002}]{gwoods}
{Gonzalez} R.~C.,  {Woods} R.~E.,  2002, {Digital image processing}

\bibitem[\protect\citeauthoryear{{Hackwell} \& {Schweizer}}{{Hackwell} \&
  {Schweizer}}{1983}]{hackwell}
{Hackwell} J.~A.,  {Schweizer} F.,  1983, \mn@doi [\apj] {10.1086/160710},
  \href {http://adsabs.harvard.edu/abs/1983ApJ...265..643H} {265, 643}

\bibitem[\protect\citeauthoryear{{Hawley} \& {Phillips}}{{Hawley} \&
  {Phillips}}{1980}]{hp}
{Hawley} S.~A.,  {Phillips} M.~M.,  1980, \mn@doi [\apj] {10.1086/157681},
  \href {http://adsabs.harvard.edu/abs/1980ApJ...235..783H} {235, 783}

\bibitem[\protect\citeauthoryear{{Ho}, {Filippenko}  \& {Sargent}}{{Ho}
  et~al.}{1997}]{ho}
{Ho} L.~C.,  {Filippenko} A.~V.,   {Sargent} W.~L.~W.,  1997, \mn@doi [\apjs]
  {10.1086/313041}, \href {http://adsabs.harvard.edu/abs/1997ApJS..112..315H}
  {112, 315}

\bibitem[\protect\citeauthoryear{{Kauffmann} et~al.,}{{Kauffmann}
  et~al.}{2003}]{kauffmann}
{Kauffmann} G.,  et~al., 2003, \mn@doi [\mnras]
  {10.1111/j.1365-2966.2003.07154.x}, \href
  {http://adsabs.harvard.edu/abs/2003MNRAS.346.1055K} {346, 1055}

\bibitem[\protect\citeauthoryear{{Kawamuro}, {Ueda}, {Tazaki}  \&
  {Terashima}}{{Kawamuro} et~al.}{2013}]{kawa}
{Kawamuro} T.,  {Ueda} Y.,  {Tazaki} F.,   {Terashima} Y.,  2013, \mn@doi
  [\apj] {10.1088/0004-637X/770/2/157}, \href
  {http://adsabs.harvard.edu/abs/2013ApJ...770..157K} {770, 157}

\bibitem[\protect\citeauthoryear{{Kewley}, {Dopita}, {Sutherland}, {Heisler}
  \& {Trevena}}{{Kewley} et~al.}{2001}]{kewley1}
{Kewley} L.~J.,  {Dopita} M.~A.,  {Sutherland} R.~S.,  {Heisler} C.~A.,
  {Trevena} J.,  2001, \mn@doi [\apj] {10.1086/321545}, \href
  {http://adsabs.harvard.edu/abs/2001ApJ...556..121K} {556, 121}

\bibitem[\protect\citeauthoryear{{Kewley}, {Groves}, {Kauffmann}  \&
  {Heckman}}{{Kewley} et~al.}{2006}]{kewley2}
{Kewley} L.~J.,  {Groves} B.,  {Kauffmann} G.,   {Heckman} T.,  2006, \mn@doi
  [\mnras] {10.1111/j.1365-2966.2006.10859.x}, \href
  {http://adsabs.harvard.edu/abs/2006MNRAS.372..961K} {372, 961}

\bibitem[\protect\citeauthoryear{{Lucy}}{{Lucy}}{1974}]{lucy}
{Lucy} L.~B.,  1974, \mn@doi [\aj] {10.1086/111605}, \href
  {http://adsabs.harvard.edu/abs/1974AJ.....79..745L} {79, 745}

\bibitem[\protect\citeauthoryear{{M{\'a}rquez}, {Masegosa}, {Durret},
  {Gonz{\'a}lez Delgado}, {Moles}, {Maza}, {P{\'e}rez}  \&
  {Roth}}{{M{\'a}rquez} et~al.}{2003}]{mar03}
{M{\'a}rquez} I.,  {Masegosa} J.,  {Durret} F.,  {Gonz{\'a}lez Delgado} R.~M.,
  {Moles} M.,  {Maza} J.,  {P{\'e}rez} E.,   {Roth} M.,  2003, \mn@doi [\aap]
  {10.1051/0004-6361:20031059}, \href
  {http://adsabs.harvard.edu/abs/2003A%26A...409..459M} {409, 459}

\bibitem[\protect\citeauthoryear{{Martin}}{{Martin}}{1974}]{martin74}
{Martin} W.~L.,  1974, \mnras, \href
  {http://adsabs.harvard.edu/abs/1974MNRAS.168..109M} {168, 109}

\bibitem[\protect\citeauthoryear{{Menezes}, {Steiner}  \& {Ricci}}{{Menezes}
  et~al.}{2013}]{menezes13}
{Menezes} R.~B.,  {Steiner} J.~E.,   {Ricci} T.~V.,  2013, \mn@doi [\apjl]
  {10.1088/2041-8205/765/2/L40}, \href
  {http://adsabs.harvard.edu/abs/2013ApJ...765L..40M} {765, L40}

\bibitem[\protect\citeauthoryear{{Menezes}, {Steiner}  \& {Ricci}}{{Menezes}
  et~al.}{2014}]{rob1}
{Menezes} R.~B.,  {Steiner} J.~E.,   {Ricci} T.~V.,  2014, \mn@doi [\mnras]
  {10.1093/mnras/stt2381}, \href
  {http://adsabs.harvard.edu/abs/2014MNRAS.438.2597M} {438, 2597}

\bibitem[\protect\citeauthoryear{{Menezes}, {da Silva}, {Ricci}, {Steiner},
  {May}  \& {Borges}}{{Menezes} et~al.}{2015}]{rob2}
{Menezes} R.~B.,  {da Silva} P.,  {Ricci} T.~V.,  {Steiner} J.~E.,  {May} D.,
  {Borges} B.~W.,  2015, \mn@doi [\mnras] {10.1093/mnras/stv629}, \href
  {http://adsabs.harvard.edu/abs/2015MNRAS.450..369M} {450, 369}

\bibitem[\protect\citeauthoryear{{Mezcua}, {Prieto}, {Fern{\'a}ndez-Ontiveros},
  {Tristram}, {Neumayer}  \& {Kotilainen}}{{Mezcua} et~al.}{2015}]{mezcua}
{Mezcua} M.,  {Prieto} M.~A.,  {Fern{\'a}ndez-Ontiveros} J.~A.,  {Tristram} K.,
   {Neumayer} N.,   {Kotilainen} J.~K.,  2015, \mn@doi [\mnras]
  {10.1093/mnras/stv1408}, \href
  {http://adsabs.harvard.edu/abs/2015MNRAS.452.4128M} {452, 4128}

\bibitem[\protect\citeauthoryear{{Morganti}, {Tsvetanov}, {Gallimore}  \&
  {Allen}}{{Morganti} et~al.}{1999}]{morg99}
{Morganti} R.,  {Tsvetanov} Z.~I.,  {Gallimore} J.,   {Allen} M.~G.,  1999,
  \mn@doi [\aaps] {10.1051/aas:1999258}, \href
  {http://adsabs.harvard.edu/abs/1999A%26AS..137..457M} {137, 457}

\bibitem[\protect\citeauthoryear{{Nelson} \& {Whittle}}{{Nelson} \&
  {Whittle}}{1995}]{nelson}
{Nelson} C.~H.,  {Whittle} M.,  1995, \mn@doi [\apjs] {10.1086/192179}, \href
  {http://adsabs.harvard.edu/abs/1995ApJS...99...67N} {99, 67}

\bibitem[\protect\citeauthoryear{{Netzer}}{{Netzer}}{1977}]{netzer}
{Netzer} H.,  1977, \mn@doi [\mnras] {10.1093/mnras/181.1.89P}, \href
  {http://adsabs.harvard.edu/abs/1977MNRAS.181P..89N} {181, 89P}

\bibitem[\protect\citeauthoryear{{Osmer}, {Smith}  \& {Weedman}}{{Osmer}
  et~al.}{1974}]{osmer74}
{Osmer} P.~S.,  {Smith} M.~G.,   {Weedman} D.~W.,  1974, \mn@doi [\apj]
  {10.1086/152787}, \href {http://adsabs.harvard.edu/abs/1974ApJ...189..187O}
  {189, 187}

\bibitem[\protect\citeauthoryear{{Pastoriza} \& {Gerola}}{{Pastoriza} \&
  {Gerola}}{1970}]{pg70}
{Pastoriza} M.,  {Gerola} H.,  1970, \aplett, \href
  {http://adsabs.harvard.edu/abs/1970ApL.....6..155P} {6, 155}

\bibitem[\protect\citeauthoryear{{Ricci}, {Steiner}  \& {Menezes}}{{Ricci}
  et~al.}{2011}]{ricci11}
{Ricci} T.~V.,  {Steiner} J.~E.,   {Menezes} R.~B.,  2011, \mn@doi [\apjl]
  {10.1088/2041-8205/734/1/L10}, \href
  {http://adsabs.harvard.edu/abs/2011ApJ...734L..10R} {734, L10}

\bibitem[\protect\citeauthoryear{{Ricci}, {Steiner}  \& {Menezes}}{{Ricci}
  et~al.}{2014}]{ricci14}
{Ricci} T.~V.,  {Steiner} J.~E.,   {Menezes} R.~B.,  2014, \mn@doi [\mnras]
  {10.1093/mnras/stu441}, \href
  {http://adsabs.harvard.edu/abs/2014MNRAS.440.2419R} {440, 2419}

\bibitem[\protect\citeauthoryear{{Richardson}}{{Richardson}}{1972}]{rich}
{Richardson} W.~H.,  1972, Journal of the Optical Society of America
  (1917-1983), \href {http://adsabs.harvard.edu/abs/1972JOSA...62...55R} {62,
  55}

\bibitem[\protect\citeauthoryear{{S{\'a}nchez-Bl{\'a}zquez}
  et~al.,}{{S{\'a}nchez-Bl{\'a}zquez} et~al.}{2006}]{blazquez}
{S{\'a}nchez-Bl{\'a}zquez} P.,  et~al., 2006, \mn@doi [\mnras]
  {10.1111/j.1365-2966.2006.10699.x}, \href
  {http://adsabs.harvard.edu/abs/2006MNRAS.371..703S} {371, 703}

\bibitem[\protect\citeauthoryear{{Schmitt} \& {Kinney}}{{Schmitt} \&
  {Kinney}}{1996}]{schmitt}
{Schmitt} H.~R.,  {Kinney} A.~L.,  1996, \mn@doi [\apj] {10.1086/177264}, \href
  {http://adsabs.harvard.edu/abs/1996ApJ...463..498S} {463, 498}

\bibitem[\protect\citeauthoryear{{Shapiro}, {Gerssen}  \& {van der
  Marel}}{{Shapiro} et~al.}{2003}]{sha03}
{Shapiro} K.~L.,  {Gerssen} J.,   {van der Marel} R.~P.,  2003, \mn@doi [\aj]
  {10.1086/379306}, \href {http://adsabs.harvard.edu/abs/2003AJ....126.2707S}
  {126, 2707}

\bibitem[\protect\citeauthoryear{{Shobbrook}}{{Shobbrook}}{1966}]{shobb}
{Shobbrook} R.~R.,  1966, \mn@doi [\mnras] {10.1093/mnras/131.2.293}, \href
  {http://adsabs.harvard.edu/abs/1966MNRAS.131..293S} {131, 293}

\bibitem[\protect\citeauthoryear{{Smaji{\'c}}, {Moser}, {Eckart}, {Busch},
  {Combes}, {Garc{\'{\i}}a-Burillo}, {Valencia-S.}  \& {Horrobin}}{{Smaji{\'c}}
  et~al.}{2015}]{smajic}
{Smaji{\'c}} S.,  {Moser} L.,  {Eckart} A.,  {Busch} G.,  {Combes} F.,
  {Garc{\'{\i}}a-Burillo} S.,  {Valencia-S.} M.,   {Horrobin} M.,  2015,
  \mn@doi [\aap] {10.1051/0004-6361/201424850}, \href
  {http://adsabs.harvard.edu/abs/2015A%26A...583A.104S} {583, A104}

\bibitem[\protect\citeauthoryear{{Steiner}, {Menezes}, {Ricci}  \&
  {Oliveira}}{{Steiner} et~al.}{2009}]{steiner}
{Steiner} J.~E.,  {Menezes} R.~B.,  {Ricci} T.~V.,   {Oliveira} A.~S.,  2009,
  \mn@doi [\mnras] {10.1111/j.1365-2966.2009.14530.x}, \href
  {http://adsabs.harvard.edu/abs/2009MNRAS.395...64S} {395, 64}

\bibitem[\protect\citeauthoryear{{Thompson}, {Levenson}, {Uddin}  \&
  {Sirocky}}{{Thompson} et~al.}{2009}]{Tomp09}
{Thompson} G.~D.,  {Levenson} N.~A.,  {Uddin} S.~A.,   {Sirocky} M.~M.,  2009,
  \mn@doi [\apj] {10.1088/0004-637X/697/1/182}, \href
  {http://adsabs.harvard.edu/abs/2009ApJ...697..182T} {697, 182}

\bibitem[\protect\citeauthoryear{{Woo} \& {Urry}}{{Woo} \& {Urry}}{2002}]{woo}
{Woo} J.-H.,  {Urry} C.~M.,  2002, \mn@doi [\apj] {10.1086/342878}, \href
  {http://adsabs.harvard.edu/abs/2002ApJ...579..530W} {579, 530}

\bibitem[\protect\citeauthoryear{{Wozniak}, {Combes}, {Emsellem}  \&
  {Friedli}}{{Wozniak} et~al.}{2003}]{woz03}
{Wozniak} H.,  {Combes} F.,  {Emsellem} E.,   {Friedli} D.,  2003, in {Collin}
  S.,  {Combes} F.,   {Shlosman} I.,  eds,  Astronomical Society of the Pacific
  Conference Series Vol. 290, Active Galactic Nuclei: From Central Engine to
  Host Galaxy. p.~563

\bibitem[\protect\citeauthoryear{{de Vaucouleurs}}{{de
  Vaucouleurs}}{1973}]{dev73}
{de Vaucouleurs} G.,  1973, \mn@doi [\apj] {10.1086/152028}, \href
  {http://adsabs.harvard.edu/abs/1973ApJ...181...31D} {181, 31}

\bibitem[\protect\citeauthoryear{{de Vaucouleurs} \& {de Vaucouleurs}}{{de
  Vaucouleurs} \& {de Vaucouleurs}}{1961}]{devdev61}
{de Vaucouleurs} G.,  {de Vaucouleurs} A.,  1961, \memras, \href
  {http://adsabs.harvard.edu/abs/1961MmRAS..68...69D} {68, 69}

\bibitem[\protect\citeauthoryear{{de Zeeuw} et~al.,}{{de Zeeuw}
  et~al.}{2002}]{zeu02}
{de Zeeuw} P.~T.,  et~al., 2002, \mn@doi [\mnras]
  {10.1046/j.1365-8711.2002.05059.x}, \href
  {http://adsabs.harvard.edu/abs/2002MNRAS.329..513D} {329, 513}

\bibitem[\protect\citeauthoryear{{van Dokkum}}{{van Dokkum}}{2001}]{van01}
{van Dokkum} P.~G.,  2001, \mn@doi [\pasp] {10.1086/323894}, \href
  {http://adsabs.harvard.edu/abs/2001PASP..113.1420V} {113, 1420}

\makeatother
\end{thebibliography}



\appendix

\section{Spectral synthesis} \label{sinteseespectral}

The spectral synthesis aims to fit, with a linear combination of stellar population spectra of a base (convolved with Gaussian functions), the stellar continuum of a given object. In order to apply this method to the optical data cube of NGC 1566, we used the \textsc{starlight} software and a stellar population base created from MILES \citep{blazquez}, the same one used to apply the pPXF procedure. We added to this base a spectrum with a power law representing the featureless continuum emission from the AGN, in this case with a spectral index of 1.7.

\begin{figure*}
\begin{center}

  \includegraphics[scale=0.5]
  {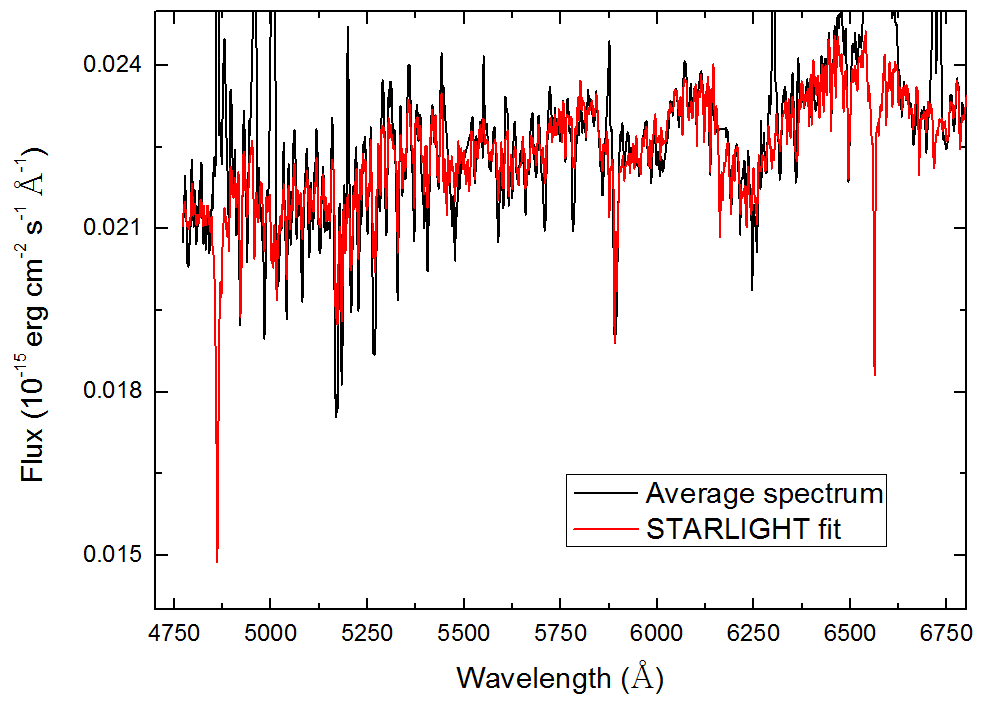}
  \caption{Result of the \textsc{starlight} fit applied to the average spectrum of NGC 1566. This gives an idea of the quality of the fits obtained with this method. \label{fitespectromedio}}
  
\end{center}
\end{figure*}

When we apply the spectral synthesis to a data cube, we obtain, for each spectrum of the data cube, the flux fractions attributed to the stellar populations used in the fit and, as we have a spectrum for each spaxel, we also obtain flux maps of these populations. Likewise, it is possible to draw a histogram of the flux fractions of the stellar populations found, in terms of metallicity and age, considering the entire FOV of the data cube. The spectral synthesis also calculates the interstellar extinction, attributed to each spectrum, resulting in a map of the extinction at the observed object. Other maps resulting from the spectral synthesis have the values of $\chi^2$ of the fits and of the S/N ratio of the spectra. As we added to the base a power law representing the featureless continuum emission, we also obtain a flux map of this emission. Fig.\ref{fitespectromedio} shows the fit made with \textsc{starlight} on the mean spectrum of NGC 1566 data cube.

\begin{figure}
\begin{center}

  \includegraphics[scale=0.52]
  {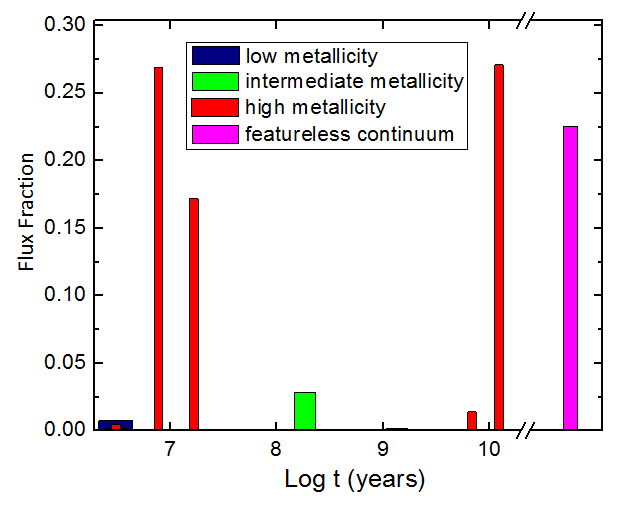}
  \caption{Histogram of the flux fractions of the stellar populations detected with the spectral synthesis. In red, the populations with high metallicity (0.02 and 0.05), whose flux fractions are higher for ages of about $10^{7}$ and $10^{10}$ yr. In blue, the populations with low metallicity ($10^{-4}$ and $10^{-4}$ and $4\times10^{-4}$), which have flux fractions too low to be identified, in this case. In green, stellar populations with intermediate metallicity ($4\times10^{-3}$ and $8\times10^{-3}$), with higher fractions at ages of about $10^{8}$ yr. In magenta, the flux fraction attributed to the power law with spectral index of 1.7, corresponding to the featureless continuum emission of the AGN in this galaxy. \label{fig.hist}}
  
\end{center}
\end{figure}

\begin{figure*}
\begin{center}

  \includegraphics[scale=0.5]
  {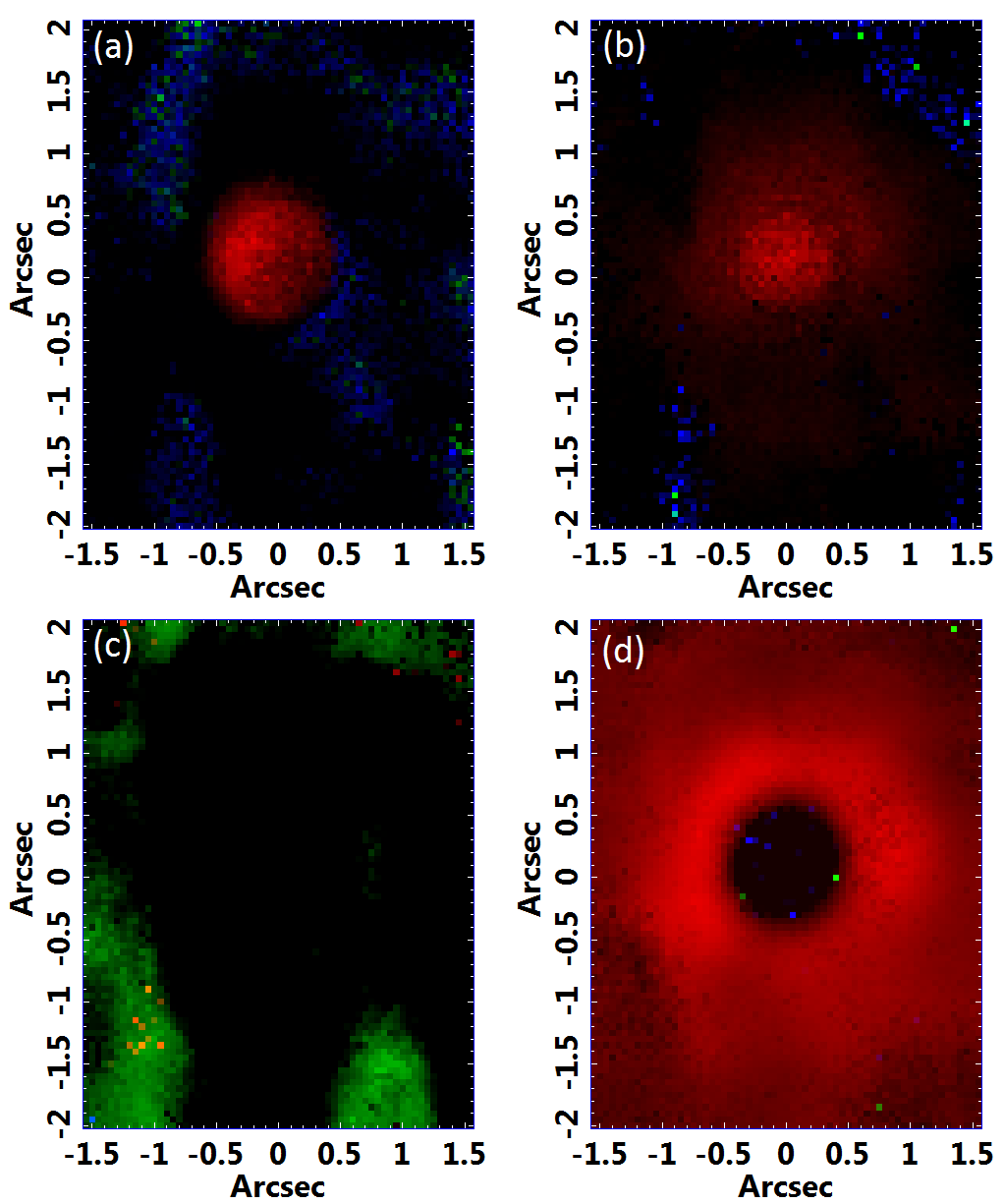}
  \caption{Flux maps associated with the stellar populations detected with the spectral synthesis. In red, populations with high metallicity (0.02 and 0.05); in blue, populations with low metallicity ($10^{-4}$ and $10^{-4}$); and, in green, stellar populations with intermediate metallicity ($4\times10^{-3}$ and $8\times10^{-3}$). (a) represents populations with $10^{6}$ yr, (b) $10^{7}$ yr, (c) $10^{8}$ yr, and (d) $10^{10}$ yr.  \label{mapas_starlight}}
  
\end{center}
\end{figure*}

\begin{figure*}
\begin{center}

  \includegraphics[scale=0.5]
  {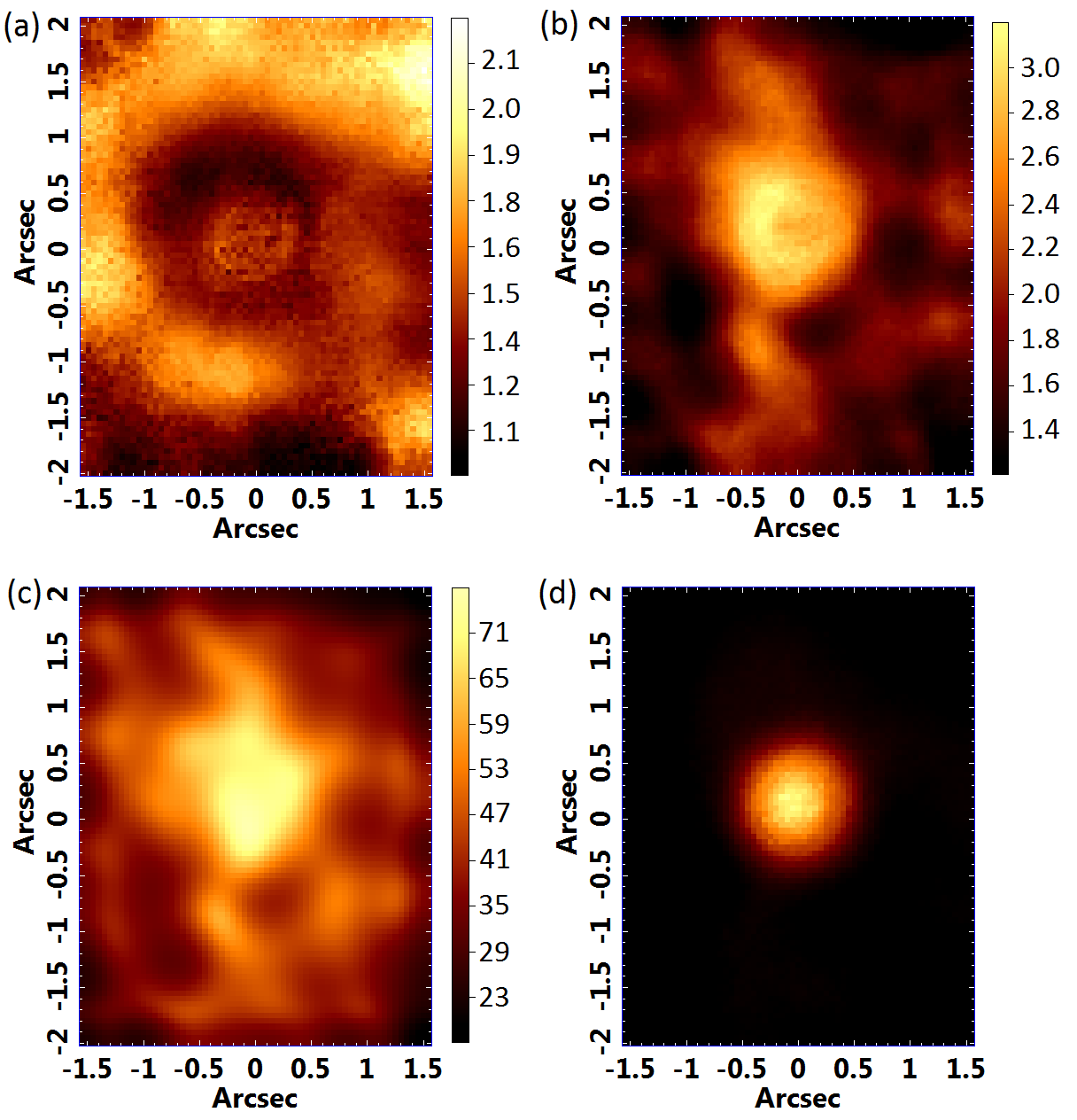}
  \caption{Maps of the parameters provided by the spectral synthesis. (a) shows the extinction map of NGC 1566, (b) is the $\chi^2$ map of the \textsc{starlight} fits, (c) is the S/N map and (d) is the map of the flux associated with the power law with spectral index of 1.7, representing the featureless continuum emission.  \label{mapas2_starlight}}
  
\end{center}
\end{figure*}

\begin{figure}
\begin{center}

  \includegraphics[scale=0.3]
  {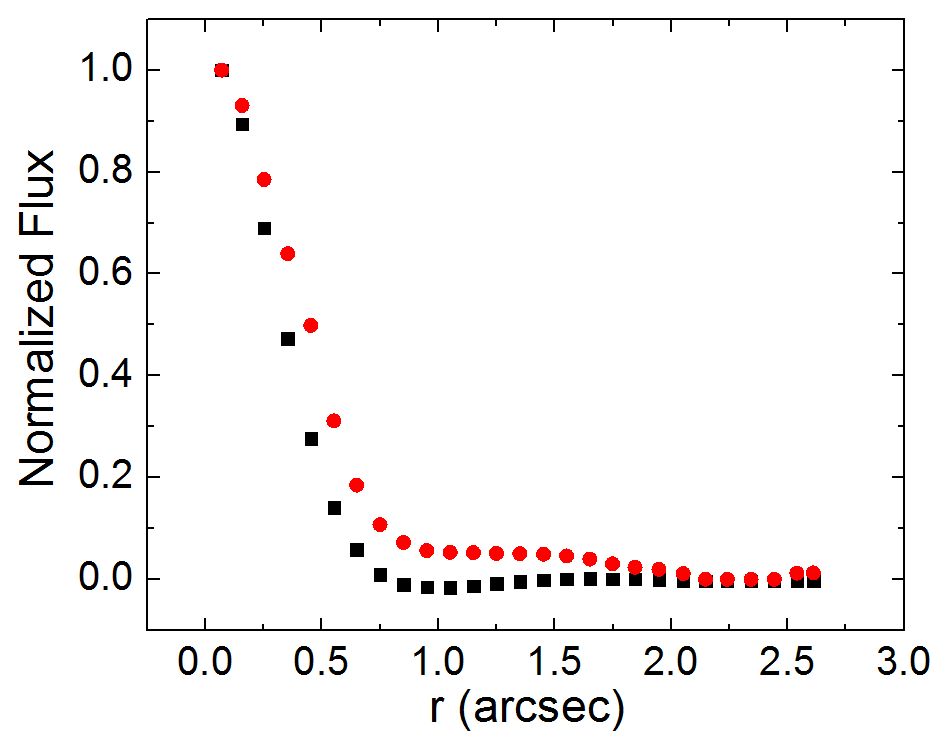}
  \caption{In red, radial profile of the featureless continuum image and, in black, radial profile of the image representing the red wing of the broad component of H$\alpha$, used to determine the PSF of the optical data cube. \label{perfilradfeat}}
  
\end{center}
\end{figure}

The histogram of Fig. \ref{fig.hist} shows that the highest flux fractions are from the young ($10^{7}$ yr), $\sim$43 per cent of total, and old ($10^{10}$ yr), $\sim$28 per cent of total, stellar populations with high metallicity (0.02 and 0.05; 0.02 being the solar metallicity). There is also a large flux fraction attributed to the featureless continuum, $\sim$22 per cent of total, which is a very relevant emission in this galaxy. 

Figs~\ref{mapas_starlight}(a) and (b) show the flux maps representing the stellar populations with $10^{6}$ and $10^{7}$ yr, respectively. Both maps show a central concentration of stellar populations with high metallicity.

A very low flux fraction was detected, $\sim$3 per cent, associated with intermediate metallicity. As we can see in Fig.~\ref{mapas_starlight}(c), these populations are scattered throughout the circumnuclear regions. 

The representative flux map of stellar populations with $10^{10}$ yr is shown in Fig.~\ref{mapas_starlight}(d). It is possible to notice that these populations, most of which having high metallicity, are scattered along  the entire FOV, but were not detected in the central region. The absence of flux from these populations in this region may derive from a possible obfuscation of their absorption lines due to featureless continuum emission. 

The extinction map (Fig.~\ref{mapas2_starlight}a) shows a pattern that, at first, resembles a ring-like emission. However, as indicated in section \ref{sec4}, when we compare this map to the \textit{V-I} image convolved with the PSF of GMOS data cube, we clearly see that this pattern is a spiral composed of regions with redder spectra.

Great part of the spectral synthesis maps show possible contamination from the featureless continuum emission. This contamination is visible in the extinction map and in the flux maps of populations with $10^6$, $10^7$ and $10^{10}$ yr. Possible damages to the fits caused by the obfuscation due to the featureless continuum may be one of the reasons for the values of $\chi^2$ to be higher in the central region (see Fig.~\ref{mapas2_starlight}b). Another reason is that the uncertainties of the flux values in the spectra were estimated from the \textit{rms} of these spectra (in a given wavelength interval); so, as the spectra of the central region have higher  S/N (and, consequently, lower values of uncertainty), and the $\chi^2$ is inversely proportional to the square of the uncertainty, we obtained higher values of $\chi^2$ in the central region. The S/N map (Fig.~\ref{mapas2_starlight}c) shows that S/N is higher in the central region and decays towards the edges of the FOV. 

The flux map of the power law with spectral index of 1.7, representing the featureless continuum of the AGN in this galaxy, is shown in Fig.~\ref{mapas2_starlight}(d). The featureless continuum emission is central and compact, and its profile can be well fitted by a Gaussian function with FWHM = 0.86$\arcsec$. When comparing the radial profile of the featureless continuum map with the radial profile of the image corresponding to the red wing of the broad component of H$\alpha$ (Fig.~\ref{perfilradfeat}), we see what we previously mentioned: the featureless continuum emission area is larger than that of the PSF. So, as there is no evidence of scattering of the featureless continuum emission, it is probable that part of this detected emission is, actually, the continuum emitted by young stars, wrongly identified by the \textsc{starlight} software.

\begin{figure}
\begin{center}

  \includegraphics[scale=0.6]{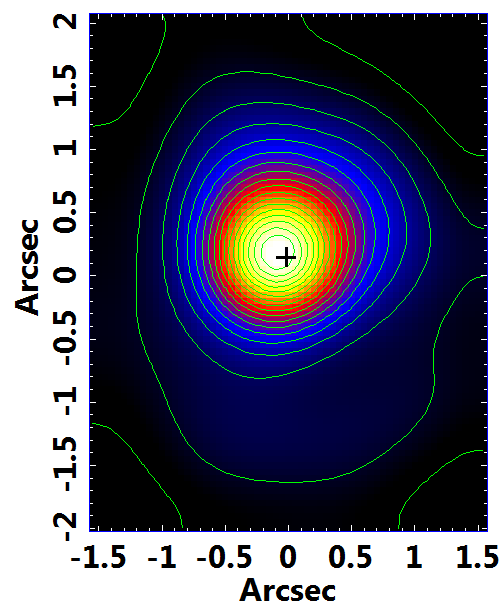}
  \caption{Collapsed image of the stellar synthetic data cube, obtained from the spectral synthesis, with the cross representing the position of the AGN. The cross size is equivalent to 3$\sigma$ of the uncertainty of this position, obtained from the image representing the red wing of the broad component of H$\alpha$. \label{cuboestelar}}
  
\end{center}
\end{figure}

When we observe Fig.~\ref{cuboestelar}, which is the collapsed image of the stellar synthetic data cube (obtained from the spectral synthesis fits), we see that the centre of the stellar emission does not coincide exactly with the position of the AGN, but is compatible with it at 1$\sigma$ level.
 
In order to have an idea of the uncertainty of the parameters obtained with the spectral synthesis, we performed the following procedure: first, the spectral synthesis was applied to the mean spectrum of the data cube of NGC 1566. The resulting synthetic stellar spectrum was subtracted from the mean spectrum. We then calculated a Gaussian distribution representing the residual spectral noise obtained. After that, we made different Gaussian distributions of random noise, with the same width of the initial Gaussian distribution. These different noise distributions were added to the synthetic stellar spectrum, obtained from the data cube mean spectrum fit, and the spectral synthesis was applied to each resulting spectrum. The mean age uncertainty (given by the weighted mean of the stellar population ages based on the flux fraction associated with each one of them) and the uncertainty of the interstellar extinction at the observed object were taken as the standard deviation of the values obtained with all those applications of the spectral synthesis. The obtained values are 0.11 dex for the mean age and 0.07 mag for the interstellar extinction.

\subsection{Spectral synthesis of the nuclear and circumnuclear regions}

\begin{figure}
\begin{center}

  \includegraphics[scale=0.3]
  {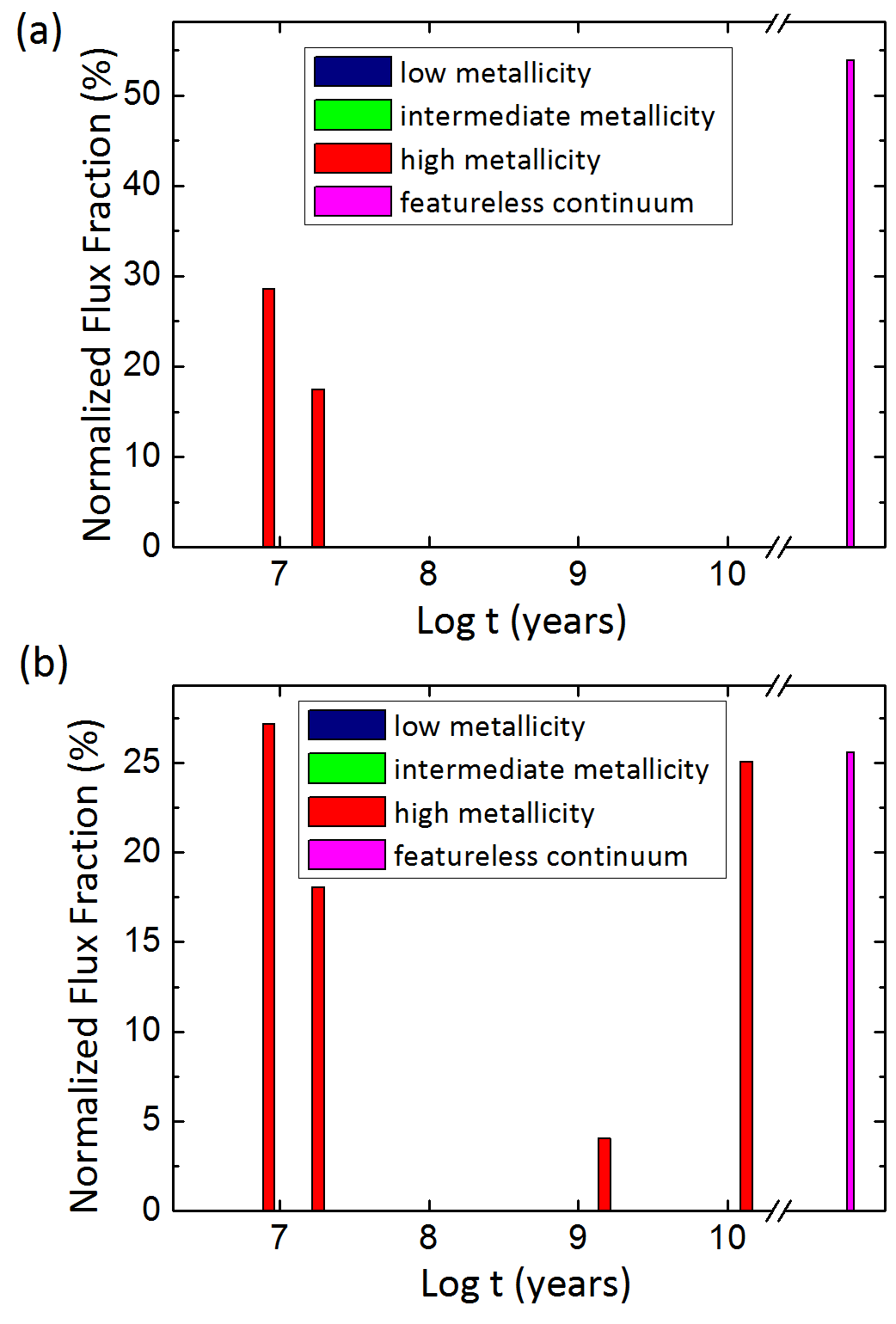}
  \caption{Histogram of the flux fractions of the stellar populations detected with the spectral synthesis (a) at the central region and (b) at the circumnuclear regions. In red, stellar populations with high metallicity (0.02 and 0.05); in blue, populations with low metallicity ($10^{-4}$ and $10^{-4}$ and $4\times10^{-4}$); in green, stellar populations with intermediate metallicity ($4\times10^{-3}$ and $8\times10^{-3}$); and, in magenta, the flux fraction attributed to the power law with spectral index of 1.7, representing the featureless continuum emission of the AGN of this galaxy.\label{hist2}}
  
\end{center}
\end{figure}

We extracted two spectra from the same data cube used to apply the spectral synthesis: one from a circular region, centred on the nucleus, with a radius of 0$\arcsec\!\!$.3, and the other from the entire FOV. We then subtracted the nuclear spectrum from that with the total emission, in order to obtain a spectrum having only the circumnuclear emission. At last, we applied the spectral synthesis to the two spectra, obtaining the two histograms shown in Fig.~\ref{hist2}.

The histogram of the spectral synthesis applied to the nuclear region shows that the highest flux fraction is due to the featureless continuum emission, $\sim$54 per cent of total. There is only one other relevant flux fraction in this region, comprised of young stellar populations ($10^6$ and $10^7$ yr) with high metallicity (0.02 and 0.05): $\sim$46 per cent of total. 

In the circumnuclear region, the highest flux fractions are due to the young ($10^6$ and $10^7$ yr), $\sim$45 per cent of total, and old stellar populations ($10^9$ and $10^{10}$yr), $\sim$29 per cent of total, with high metallicity. 

The greatest difference between the results of the spectral synthesis of these two regions is the presence of the old stellar populations only in the circumnuclear region. This may be explained by a possible obfuscation of the absorption lines of these stars in the nuclear region due to the featureless continuum emission. It is also important to mention that at least part of the featureless continuum detected with these two applications of the spectral synthesis can, in fact, be associated with the continuum emitted by young stellar populations, wrongly identified by the \textsc{starlight} software, as already mentioned before.


\bsp	
\label{lastpage}
\end{document}